\documentclass[a4paper,11pt]{article}
\pdfoutput=1 

\usepackage{jheppub} 
\usepackage{amsmath}
\allowdisplaybreaks[4]        
\usepackage{amssymb}
\usepackage{euscript}     
\usepackage{color}         
\usepackage{tensor}        
\usepackage{amsthm}
\usepackage{bbm}
\usepackage{float}
\usepackage{pgfplots}
\usepackage{graphicx,subcaption}
\usepackage{mathtools}

\usepackage{hyperref}
\hypersetup{colorlinks=true,allcolors=blue}

\usepackage[most]{tcolorbox}
\tcbuselibrary{skins, breakable, theorems}

\usepackage[header,title,page,titletoc]{appendix}  
\usepackage[T1]{fontenc} 
\usepackage[numbers]{natbib}

\newcommand{\eg}{{\it e.g.,}\ }
\newcommand{\ie}{{\it i.e.,}\ }
\newcommand{\viz}{{\it viz,}\ }
\newcommand{\mt}[1]{\textrm{\tiny #1}}

\renewcommand{\(}{\left(}
\renewcommand{\)}{\right)}
\renewcommand{\[}{\left[}

\newcommand{\talpha}{\tilde{\alpha}}

\newcommand{\csch}{\text{csch}}

\definecolor{browna}{rgb}{0.76,0.72,0.65}
\definecolor{brownb}{rgb}{0.71,0.69,0.65}
\definecolor{SpringGreen}{rgb}{0.95,0.97,0.95}
\definecolor{OliverGreen}{rgb}{0.09,0.34,0.09}
\definecolor{LeftGreen}{rgb}{0.13,0.54,0.13}

\newtcbtheorem[no counter]{theorem}{Theorem}{
	enhanced,
	sharp corners,
	attach boxed title to top left={
		yshifttext=-1mm
	},
	colback=white,
	colframe=browna,
	fonttitle=\bfseries,
	boxed title style={
		sharp corners,
		size=small,
		colback=browna,
		colframe=browna,
	} 
}{thm}

\newtcbtheorem[no counter]{Example}{Example}{
	enhanced,
	sharp corners,
	attach boxed title to top left={
		yshifttext=-1mm
	},
	colback=white,
	colframe=brownb,
	fonttitle=\bfseries,
	coltitle=white,
	boxed title style={
		sharp corners,
		size=small,
		colback=brownb,
		colframe=brownb,
	} 
}{exm}

\newtcbtheorem{myclaim}{Claim -- }
{    coltitle=OliverGreen,    
	colback=SpringGreen,    colframe=LeftGreen,    detach title,    boxrule=0pt,    leftrule=2pt,    attach title to upper,    sharp corners,    left=1mm,}
{claim}

\definecolor{browna}{rgb}{0.76,0.72,0.65}

\subheader{\today}

\title{Non-perturbative Overlaps in JT Gravity:
\vskip 0.1in
\Large{\it From Spectral Form Factor to Generating Functions of Complexity}}

\author{Masamichi Miyaji$^1$,~Shan-Ming Ruan$^{2,3,4}$,~Shono Shibuya$^5$,~Kazuyoshi Yano$^5$}

\affiliation{$^1$Yukawa Institute for Theoretical Physics, Kyoto University, Kyoto, 606-8267, Japan}
\affiliation{$^2$School of Physics, Peking University, Beijing 100871, China}
\affiliation{$^3$Center for High Energy Physics, Peking University, Beijing 100871, China}
\affiliation{$^4$Theoretische Natuurkunde, Vrije Universiteit Brussel and The International Solvay Institutes, \\
Pleinlaan 2, B-1050 Brussels, Belgium}
\affiliation{$^5$Department of Physics, Nagoya University,
Nagoya, Aichi 464-8602, Japan}
\emailAdd{masamichi.miyaji@gmail.com, Shan-Ming.Ruan@vub.be, shibuya.shono.n8@s.mail.nagoya-u.ac.jp,
yano.kazuyoshi.h8@s.mail.nagoya-u.ac.jp}

\abstract{The interplay between black hole interior dynamics and quantum chaos provides a crucial framework for probing quantum effects in quantum gravity. In this work, we investigate non-perturbative overlaps in Jackiw-Teitelboim (JT) gravity to uncover universal signatures of quantum chaos and quantum complexity. Taking advantage of universal spectral correlators from random matrix theory, we compute the overlaps between the thermofield double (TFD) state and two distinct classes of states: fixed-length states, which encode maximal volume slices, and time-shifted TFD states. The squared overlaps naturally define probability distributions that quantify the expectation values of gravitational observables. Central to our results is the introduction of generating functions for quantum complexity measures, such as $\langle e^{-\alpha \ell} \rangle$. The time evolution of these generating functions exhibits the universal slope-ramp-plateau structure, mirroring the behavior of the spectral form factor (SFF). Using generating functions, we further demonstrate that the universal time evolution of complexity for chaotic systems, which is characterized by a linear growth followed by a late-time plateau, arises from the disappearance of the linear ramp as the regularization parameter $\alpha$ decreases. With regard to the time-shifted TFD state, we derive a surprising result: the expectation value of the time shift, which classically grows linearly, vanishes when non-perturbative quantum corrections are incorporated. This cancellation highlights a fundamental distinction between semiclassical and quantum gravitational descriptions of the black hole interior. All our findings establish generating functions as powerful probes of quantum complexity and chaos in gravitational and quantum systems.
}

\newcommand{\imineq}[2]{\vcenter{\hbox{\includegraphics[height=#2ex]{#1}}}}

\numberwithin{equation}{section}

\begin{document}
\begin{flushright}
YITP-25-28
\\
\end{flushright}
	\maketitle
	\flushbottom

	
\section{Introduction}

\begin{figure}[t]
	\centering
	\includegraphics[width=5.5in]{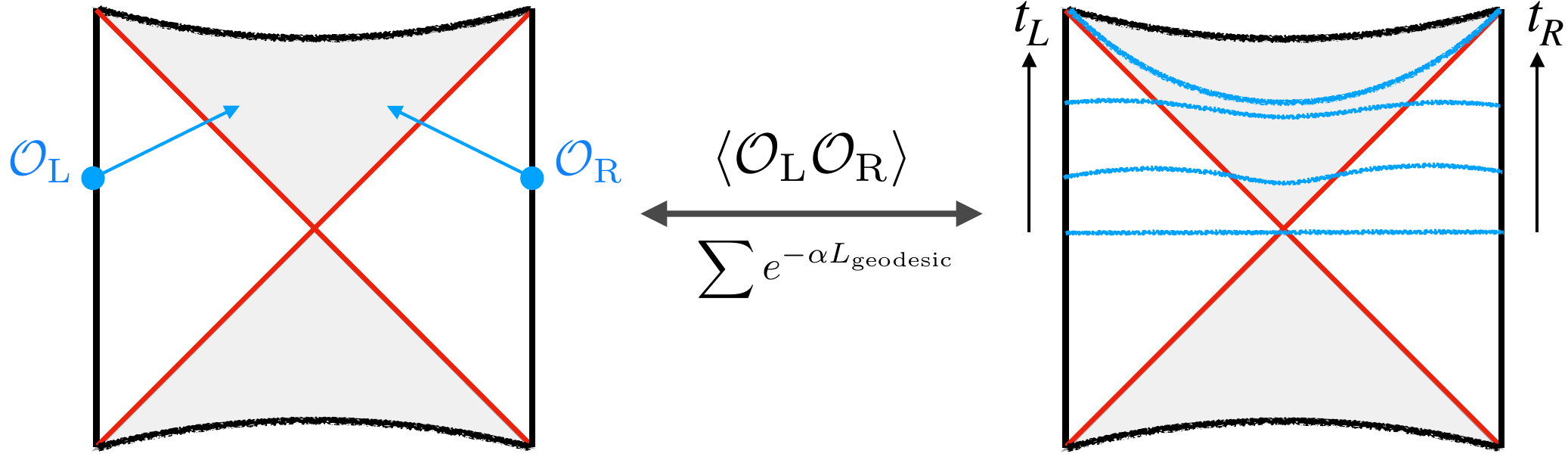}
    \caption{\textbf{Classical Gravitational Picture.} \textit{Left:} Two simple operators, $\mathcal{O}_{\mt{L}}$ and $\mathcal{O}_{\mt{R}}$, are inserted on the left and right boundaries. The holographic correlation function, computed from classical black hole geometry, exhibits an exponential decay with time. \textit{Right:} The time evolution of the maximal-volume hypersurface with the increase of boundary time $t_{\mt{L}} + t_{\mt{R}}$. The linear growth of this maximal volume has been conjectured to correspond to the growth of quantum complexity of the boundary theory. These two distinct semiclassical problems—(i) the infinite decay of thermal correlators and (ii) the infinite linear growth of the wormhole size (such as the maximal volume) in the AdS black hole background—are closely related. Their relationship can be understood through the calculation of holographic correlators using the geodesic approximation.}\label{fig:SFFComplexityClassical}
\end{figure}  

The black hole information problem has long been a fundamental challenge in quantum gravity, raising concerns about the compatibility of black hole physics with the fundamental principles of quantum mechanics. A particularly elegant yet profound formulation of this problem was articulated by Maldacena in the context of the AdS/CFT correspondence \cite{Maldacena:2001kr}, where it was observed that the long-time behavior of holographic correlation functions in semiclassical gravity appears to be in stark contrast to the expectations from a finite-dimensional quantum system. In this context, the classical two-sided black hole in AdS is one of the most extensively studied solutions, known to be dual to the thermofield double (TFD) state $|\text{TFD} \rangle$ in the boundary CFT \cite{Maldacena:2001kr}. This state remains invariant under the generator $H_{\mt{L}} - H_{\mt{R}}$ while evolving non-trivially under $H_{\mt{L}} + H_{\mt{R}}$.
From the perspective of the dual boundary theory, one can consider the insertion of simple operators $\mathcal{O}$ either on the same boundary or on opposite boundaries, as illustrated in figure \ref{fig:SFFComplexityClassical}. For instance, the two-sided two-point function takes the form \cite{Saad:2019pqd}
\begin{equation}\label{eq:twopointfuncion}
\langle \mathrm{TFD}_\beta|  \mathcal{O}(t_{\mt{L}}) \mathcal{O}(t_{\mt{R}}) | \mathrm{TFD}_\beta \rangle = \frac{1}{Z} \sum_{i, j} e^{-\frac{\beta}{2}\left(E_i+E_j\right)} e^{-i T\left(E_i-E_j\right)}\left|\left\langle E_i\right| \mathcal{O}\right| E_j\rangle|^2 \,,
\end{equation}
where the time shift is given by $T= t_{\mt{L}} + t_{\mt{R}}$,\footnote{The time coordinates $t_{\mt{L}}$ and $t_{\mt{R}}$ on the left and right boundaries both increase upwards.} $Z = \mathrm{Tr} \, e^{-\beta H}$ denotes the partition function, and the sum extends over energy eigenstates. The corresponding Euclidean representation of this two-sided correlator is equivalent to the thermal correlator $\langle\mathcal{O}(\frac{\beta}{2}+i T) \mathcal{O}(0)\rangle_\beta := \mathrm{Tr}\left[ e^{-\beta H}\mathcal{O}(\frac{\beta}{2}+i T) \mathcal{O}(0) \right]$. On the gravity side, the initial decay of this correlator is governed by the relaxation of quasinormal modes \cite{Horowitz:1999jd}, leading to an exponential suppression of the correlation function over time. From the dual boundary perspective, the decay of thermal correlators is a signature of chaotic thermalization. In the semiclassical limit, this holographic correlator decays indefinitely, reflecting the intuitive expectation that excitations outside the horizon inevitably fall into the black hole. However, in a unitary quantum system with a discrete spectrum, correlation functions cannot decay indefinitely; instead, they must eventually reach an exponentially small but nonzero average value and exhibit erratic fluctuations over time. The discrepancy between semiclassical expectations and the constraints of quantum mechanics signals the necessity of non-perturbative corrections in the gravitational path integral, which restore unitarity and modify the late-time behavior of correlators.

\begin{figure}[t]
	\centering
	\includegraphics[width=5.5in]{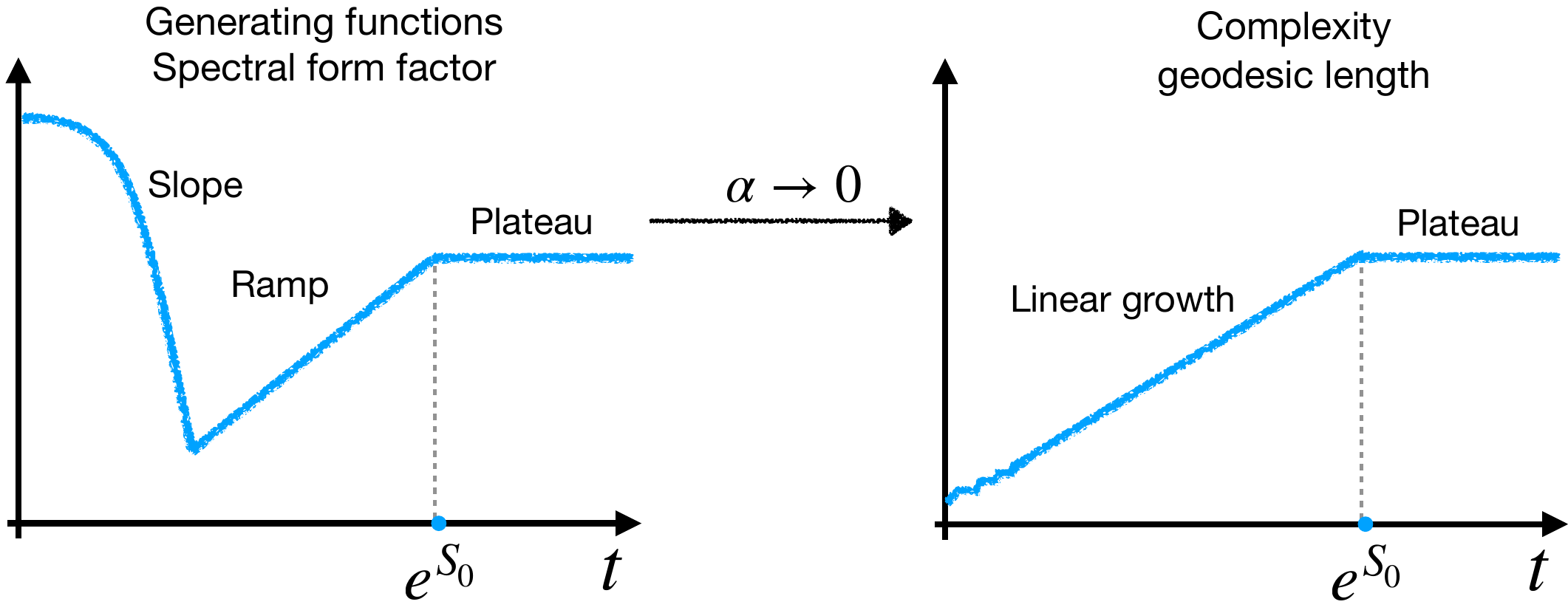}
    \caption{The time evolution of the spectral form factor or generating functions of complexity shows the well-known slope-ramp-plateau structure. The time evolution of quantum complexity is a result of the disappearance of the linear ramp region of its generating function at $\alpha \to 0$ limit, \ie quantum complexity grows linearly up to the time scale at $e^{S_0}$ and then reaches a plateau.}\label{fig:SFFvsC}
\end{figure}

Significant progress has been made in recent years toward resolving this tension by demonstrating that the late-time behavior of correlators is governed by random matrix universality. The simplest example of this phenomenon is the so-called spectral form factor (SFF). As illustrated in figure \ref{fig:SFFvsC}, ensemble-averaged correlators as well as the spectral form factor typically exhibit a characteristic slope-ramp-plateau structure \cite{Saad:2018bqo,Saad:2019lba, Saad:2019pqd,Okuyama:2020ncd,Blommaert:2022lbh,Saad:2022kfe,Okuyama:2023pio,Griguolo:2023jyy}. The ramp corresponds to a phase of linear growth, which finally transitions into a plateau at a time scale exponentially large in the system’s entropy, \ie $e^{S_0}$. These late-time features are invisible in perturbative gravity regime but emerge naturally when non-perturbative effects, such as Euclidean wormholes, are taken into account. The realization that quantum gravitational corrections give rise to this universal behavior provides a resolution to Maldacena’s version of the black hole information problem: rather than decaying forever, quantum correlators in gravitational systems obey the same spectral statistics, which govern quantum chaotic systems.

On the other hand, a simple yet remarkable feature of the two-sided AdS black hole geometry is that the size of the wormhole—\ie the black hole interior or the Einstein-Rosen bridge—exhibits linear growth at late times, as illustrated in figure \ref{fig:SFFComplexityClassical}. Motivated by this geometric observation, Susskind proposed that a new quantum information measure beyond entanglement entropy \cite{Susskind:2014moa}, namely, quantum complexity\footnote{We refer to the complexity measures for boundary field theory as quantum complexity. The well-studied complexity measure in the high energy community is the so-called quantum circuit complexity \cite{Nielsen:2006cea,Dowling:2006tnk}, \ie the minimal cost of the quantum circuit. Other closely related quantum measures are the spread complexity \cite{Balasubramanian:2022tpr} or the Krylov complexity \cite{Parker:2018yvk}, which capture the complexity of state/operator spreading. We cannot specify the boundary dual of holographic complexity, but later we will provide more precise definitions of quantum complexity measures using the spectral representation.}, is necessary to encode the growth of the wormhole \cite{Susskind:2014rva,Stanford:2014jda}. This proposal draws inspiration from the behavior of random quantum circuits, whose circuit complexity increases linearly with time \cite{Brown:2017jil,Susskind:2018fmx,Haferkamp:2021uxo}, mirroring the expansion of the black hole interior. To understand the growth of the black hole interior from a boundary perspective, numerous holographic conjectures and proposals have been put forth regarding the relationship between wormhole growth and quantum complexity \cite{Susskind:2014rva,Stanford:2014jda, Brown:2015bva,Brown:2015lvg,Caputa:2017urj,Caputa:2017yrh,Brown:2019rox,Belin:2021bga,Belin:2022xmt}. Additionally, related studies have explored the connection to the growth of the quantum Fisher information metric \cite{Miyaji:2015woj,Miyaji:2016fse,Belin:2018bpg} and, more recently, to the evolution of Krylov complexity \cite{Jian:2020qpp,Balasubramanian:2022tpr,Balasubramanian:2023kwd,Rabinovici:2023yex,Erdmenger:2023wjg,Balasubramanian:2024lqk,Heller:2024ldz}. 

The first and simplest version is the complexity=volume conjecture \cite{Susskind:2014rva,Stanford:2014jda}, which states that the holographic complexity is dual to the maximal volume of a hypersurface anchored at the boundary time slice. A closely related fundamental tension, reminiscent of Maldacena’s black hole information problem, arises in the context of wormhole growth: in the semiclassical limit, the wormhole size increases linearly without bound. However, just as correlation functions in a finite-dimensional quantum system cannot decay forever, the linear growth of the wormhole size must eventually saturate due to the finite dimensionality of the Hilbert space in a fully quantum gravitational description. Susskind and Brown \cite{Brown:2017jil,Susskind:2018fmx} conjectured that the characteristic time evolution of quantum complexity in a chaotic system consists of an long period of linear growth, followed by a plateau after a time scale of order $e^{S_0}$, as depicted in figure \ref{fig:SFFvsC}. Consequently, the infinite growth of the wormhole size represents another striking discrepancy between semiclassical expectations and those of quantum gravity. To achieve late-time saturation of wormhole size or holographic complexity, non-perturbative quantum gravity effects must be incorporated to suppress this indefinite growth.

A unifying perspective on Maldacena’s black hole information problem and Susskind’s wormhole growth paradox emerges from the geodesic approximation used to evaluate holographic correlators. As a fundamental aspect of the AdS/CFT correspondence, holographic two-point functions for {\it heavy} operators at leading order (in the semiclassical regime) can be computed using the saddle-point approximation, \ie the geodesic approximation \cite{Balasubramanian:1999zv,Louko:2000tp,Aparicio:2011zy,Balasubramanian:2012tu}:
\begin{equation}\label{eq:geodesic}
\left\langle\mathcal{O}\left(X_1\right) \mathcal{O}\left(X_2\right)\right\rangle \sim \sum_{\text{all geodesics }} e^{- \alpha L_{\rm{geodesic}}} \,,
\end{equation}
where the parameter $\alpha$ denotes the conformal dimension of the operator $\mathcal{O}(X_i)$ located at the bulk point $X_i$, and $L_{\rm{geodesic}}$ represents the length of the bulk geodesic connecting the operator insertion points. Notably, in two-dimensional gravity, the maximal volume is equivalent to the geodesic length. Considering the two-point function in eq.~\eqref{eq:twopointfuncion}, this approximation naturally illustrates the exponential decay of correlation functions, as the geodesic length connecting the left and right boundaries increases linearly. Similarly, in the case of the Einstein-Rosen bridge, the maximal volume slice—which quantifies the growth of the wormhole—is also governed by the bulk geodesic structure. As depicted in figure \ref{fig:SFFComplexityClassical}, it becomes evident that the black hole information problem and the wormhole growth paradox are equivalent in the classical AdS black hole spacetime.

Recent studies incorporating quantum corrections from Euclidean wormholes and the universal properties of random matrix theory have provided a resolution to the problem of infinitely decaying correlation functions. Given the correspondence between these two paradoxes at the semiclassical level, it is natural to conjecture that a similar connection persists at the quantum level: just as quantum corrections are necessary to prevent the indefinite decay of correlation functions, analogous quantum gravitational effects must regulate the unbounded growth of wormhole size. Ultimately, both problems point to the necessity of non-perturbative gravitational corrections to ensure compatibility with the principles of quantum mechanics and unitary evolution. 

This correspondence, as depicted in figure \ref{fig:SFFvsC} and table \ref{table:GvsC}, represents one of the central conclusions of this paper. We will explicitly demonstrate that the resolution to both the correlator decay problem and the wormhole growth paradox follows from the same underlying mechanism. The key ingredient is the introduction of what we term {\it generating functions of complexity}, which exhibit a characteristic slope-ramp-plateau structure, analogous to that observed in averaged correlators and the spectral form factor. The time evolution of complexity is thus dictated by the behavior of the corresponding generating function in the $\alpha \to 0$ limit where the linear ramp disappears. This explains why the linear growth of wormhole size derived from classical spacetime could still be dominant up to $e^{S_0}$. Notably, the same mechanism that governs the ramp-to-plateau transition in the spectral form factor also leads to the saturation of wormhole growth at timescales of order $e^{S_0}$. 
This conclusion can also be extended to other {\it infinite complexity measures} within the framework of the complexity=anything proposal \cite{Belin:2021bga,Belin:2022xmt,Jorstad:2023kmq}. Our results suggest that quantum complexity, like spectral correlations, obeys universal spectral statistics in chaotic systems, highlighting profound connections between quantum gravity, quantum chaos and quantum complexity theory.

To quantitatively understand the late-time evolution and non-perturbative behavior of wormhole size, we focus on studying TFD states in Jackiw-Teitelboim (JT) gravity \cite{Jackiw:1984je,Teitelboim:1983ux,Louis-Martinez:1993bge,Almheiri:2014cka,Maldacena:2016upp,Engelsoy:2016xyb,Kitaev:2017awl,Yang:2018gdb,Saad:2018bqo,Saad:2019lba,Johnson:2019eik}, where the maximal volume corresponds to the geodesic length. While the classical geodesic length grows linearly in time, it has been found that in the presence of Euclidean wormholes \cite{Coleman:1988cy, Giddings:1987cg}, baby universe emission significantly affects the linear growth of geodesic length at late times \cite{Saad:2019pqd, Iliesiu:2021ari}. Although the gravitational path integral with Euclidean wormholes suggests an averaging mechanism—such as ensemble averaging \cite{Maldacena:2004rf, Saad:2019lba, Bousso:2019ykv, Pollack:2020gfa, Bousso:2020kmy, VanRaamsdonk:2020tlr, Stanford:2020wkf} or coarse-graining \cite{Langhoff:2020jqa, Chandra:2022fwi}—which could lead to non-factorization, signatures of a fine-grained, factorized Hilbert space can still be recovered through a refined use of the gravitational path integral \cite{Marolf:2020xie, Marolf:2020rpm, Saad:2021uzi, Blommaert:2021fob, Blommaert:2022ucs, Bousso:2023efc, Miyaji:2023wcf,Boruch:2024kvv, Balasubramanian:2024yxk,Banerjee:2024fmh}. Crucially, the effects of baby universe emission transform a black hole with an expanding Einstein-Rosen bridge into a white hole with a contracting Einstein-Rosen bridge \cite{Stanford:2022fdt, Iliesiu:2024cnh, Blommaert:2024ftn, Miyaji:2024ity, Balasubramanian:2024lqk, Cui:2024ibh}, finally leading to an equal mixture of black holes and white holes at very late times, \ie the so-called {\it gray hole} \cite{Susskind:2015toa, Susskind:2020wwe}. This phenomenon has been suggested \cite{Stanford:2022fdt} to be related to the firewall paradox \cite{Almheiri:2012rt,Almheiri:2013hfa}.

\begin{table}[h]
    \centering
    \renewcommand{\arraystretch}{1.5}
    \setlength{\tabcolsep}{10pt}
    \begin{tabular}{|c|}
        \hline
        \textbf{Generating Function} \\
        \eg $\langle e^{-\alpha \ell} \rangle, 
        \mathrm{G}_{\mt{SC}}(\alpha, t)$ \\
        \hline
        Slope \\
        \hline
        Ramp \\
        \hline
        Plateau \\
        \hline
    \end{tabular}
    \begin{tabular}{c}
       $\xRightarrow[\eg \langle e^{-\alpha \ell} \rangle \rightarrow \langle \hat{\ell} \rangle]{\makebox[2.5cm]{$\alpha \to 0$}} $ 
    \end{tabular}
    \begin{tabular}{|c|}
        \hline
        \textbf{Quantum Complexity} \\
        \eg $\langle \hat{\ell} \rangle$, spectral complexity \\
        \hline
        Linear growth \\
        \hline
        $\emptyset$\\
        \hline
        Plateau \\
        \hline
    \end{tabular}
\caption{The correspondence between the time evolution of the quantum complexity of a chaotic system and the slope-ramp-plateau structure of the corresponding generating functions. It is important to note that the ramp part of the generating function may not disappear completely, resulting in the residual ramp becoming the decreasing part from the peak of complexity to the plateau. A more explicit illustration of this phenomenon can be found in section \ref{section:comment}.}\label{table:GvsC}
\end{table}

\subsection{Outline and Summary}

The fundamental building block for computing the expectation value of wormhole size and other gravitational observables is the non-perturbative overlaps between the TFD state and eigenstates of relevant operators, as the squared overlap determines the probability distribution. The quantum state $|\ell \rangle$, characterized by a fixed geodesic length $\ell$, is defined through a generalized version of the Hartle-Hawking prescription \cite{Hartle:1983ai,Miyaji:2024ity}:
\begin{equation}
    \langle \ell |E\rangle = \psi_E(\ell)=~\imineq{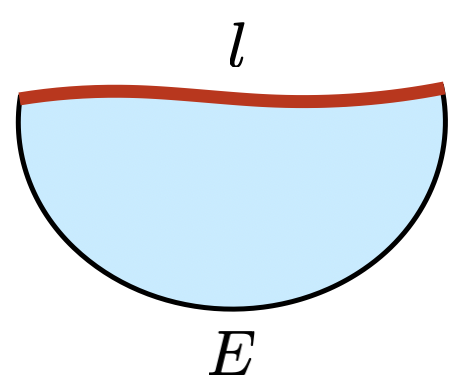}{13} \,, 
\end{equation}
where the right-hand side represents the JT gravity partition function on a disk with a fixed two-sided energy $E = E_{\mt{L}} + E_{\mt{R}}$ (with $E_{\mt{L}} = E_{\mt{R}}$) and a boundary given by a geodesic of length $\ell$. The squared overlap with the TFD state is then defined as 
\begin{equation}
    P(\ell,t):=\langle \text{TFD}(t)|\ell \rangle\langle \ell |\text{TFD}(t)\rangle \,. 
\end{equation}
Classically, $P(\ell,t)$ is sharply peaked at the classical value $\ell = \ell_{\text{classical}}(t)$. However, incorporating contributions from Euclidean wormholes reveals a nonzero overlap between the TFD state and all fixed-length states due to baby universe emission and absorption. This substantially modifies the behavior of $P(\ell,t)$, as illustrated in figures \ref{fig:Ptotal01} and \ref{fig:Ptotal02}. Notably, $P(\ell,t)$ becomes constant for large $\ell$, implying that a classical Einstein-Rosen bridge can transition into a fixed-length state of arbitrarily large $\ell$ by emitting and absorbing baby universes. Consequently, $P(\ell,t)$ fails to define a proper probability distribution, as the total ``probability'' and the corresponding expectation value, \ie the ``length expectation value'' \cite{Iliesiu:2021ari, Iliesiu:2024cnh}, \ie 
\begin{equation}\label{eq:length}
    \langle \, \hat{\ell}  \,\rangle:=\int_{-\infty}^{\infty}d\ell ~P(\ell,t) \,\ell \,,
\end{equation}
both diverge. This divergence arises because fixed-length states are not mutually orthogonal, leading to an infinite total contribution, $\int_{-\infty}^{\infty}d\ell~P(\ell,t)  \to  \infty$. 

Instead, a well-defined quantity associated with the geodesic length is given by the generating function:
\begin{equation}\label{eq:lengthgen}
    \langle e^{-\alpha \ell}\rangle :=\int_{-\infty}^{\infty}d\ell~P(\ell,t)e^{-\alpha \ell} \,, 
\end{equation}
where the factor $e^{-\alpha \ell}$ also serves as a regulator. Specifically, in the limit $\alpha \to 0$, we obtain the regularized length expectation value, \ie
\begin{equation}\label{eq:alphato0}
\lim_{\alpha \to 0} \left(  - \partial_\alpha \langle e^{-\alpha \ell } \rangle   \right) = \text{Divergent Constant} + \langle \ell \rangle_{\rm reg} \,.
\end{equation}
Interestingly, the generating function $\langle e^{-\alpha \ell }\rangle $ typically exhibits a characteristic slope-ramp-plateau structure, analogous to that observed in the spectral form factor, with a transition occurring at the Heisenberg time 
\begin{equation}
    T_{\mt{H}}:=2\pi e^{S_0}D(E_0) \,.
\end{equation}
Here, $D(E_0)$ denotes the density of states. This timescale is the same as the ramp-to-plateau transition of the spectral form factor. However, as $\alpha \to 0$, the linear ramp regime disappears. 
The disappearance of linear ramp just explains the time evolution of maximal volume: it initially grows as the inverse of the slope of its generating function and then transits to the plateau at the Heisenberg time $T_{\mt{H}}$. This behavior is illustrated in figures \ref{fig:expal02} and \ref{fig:Lvalues}. Using the spectral representation of the generating function $\langle e^{-\alpha \ell}\rangle$, we identify the universal component of the microscopic description of generating functions of complexity in the microcanonical ensemble as 
\begin{equation}
G_{\mt{SC}}(\alpha,t) = 
  \sum_{E_1, E_2} \frac{ \alpha  \cos \left( (E_1 -E_2)t \right) }{(E_1-E_2)^2 + 2(E_1 +E_2) \alpha^2}  \,.  
\end{equation}
Following the same procedure as in eq.~\eqref{eq:alphato0}, we recover the so-called {\it spectral complexity} \eqref{eq:SC} proposed in \cite{Iliesiu:2020qvm}. The correspondence between quantum complexity and its generating function, in terms of characteristic time evolution, is summarized in table \ref{table:GvsC}.  

Similar to fixed-length states, we consider another class of quantum states by fixing the relative time shift $\delta$ between the two boundaries. In the study of two-sided black holes, the time shift $\delta$ plays a fundamental role, as it is canonically conjugate to the boundary Hamiltonian \cite{Harlow:2018tqv}. The corresponding eigenstate $|\delta\rangle$ is equivalent to the time-evolved TFD state $| \mathrm{TFD}(\delta) \rangle$. Following the same approach as for fixed-length states, we compute the squared overlap between the TFD state and $|\delta\rangle$, 
\begin{equation}
    P^{\mt{TFD}}(\delta,t):=\langle \text{TFD}(t)|\delta \rangle\langle \delta|\text{TFD}(t)\rangle,
\end{equation}
which corresponds to the microcanonical spectral form factor evaluated at time $|t-\delta|$. Classically, $P(\delta,t)$ is sharply peaked at $\delta = t$, but becomes nonzero when quantum corrections are included. Applying the explicit form of $P^{\mt{TFD}}(\delta,t)$ derived in eq.~\eqref{eq:PTFD}, we can compute the generating function associated with the absolute value of the time shift:  
\begin{equation}
\begin{split}
 \langle e^{-\alpha |\delta|} \rangle   
:=\int_{-\infty}^{\infty} P^{\mt{TFD}}(\delta,t) e^{-\alpha |\delta |} \, d \delta  \,, \\
\end{split}
\end{equation}
or alternatively, $\langle e^{\mp \alpha \delta_\pm} \rangle$ \eqref{eq:gendeltapm} for positive/negative time shifts. These generating functions typically display the slope-ramp-plateau structure with time evolution, as illustrated in figures \ref{fig:expadelta} and \ref{fig:expalphadeltapm}. Similarly, we can still find that the linear ramp part would disappear as $\alpha$ approaches zero. However, a surprising result is that the (regularized) expectation value of the time shift vanishes, namely, 
\begin{equation} 
\begin{split} 
\langle  \hat{\delta} \rangle_{\rm reg } := - \lim_{\alpha\to0} \frac{\partial}{\partial \alpha} \left( \langle e^{-\alpha\delta_+}\rangle - \langle e^{\alpha\delta_-}\rangle \right)= 0 + \mathcal{O}(\Delta E) \,. 
\end{split} 
\end{equation} 
This cancellation arises because classical contributions from disk geometry and quantum corrections from Euclidean wormholes exactly offset each other at all time regimes. The spectral representation of these generating functions closely parallels that of spectral complexity. For instance, the spectral representation of the generating function $\langle e^{-\alpha |\delta|} \rangle$ can be derived by the Fourier transform,
\begin{equation} 
\mathrm{G}_{|\delta|} (\alpha, t) = \sum_{E_1, E_2} \frac{ 2\alpha }{(E_1-E_2)^2 +\alpha^2} e^{- i(E_1 -E_2)t}\,, 
\end{equation} 
whose derivative in the $\alpha \to 0$ limit also reproduces spectral complexity.

The organization of this paper is as follows. In sections \ref{section:lengthstate} and \ref{section:timeshiftstate}, we investigate two classes of quantum states: $|\ell\rangle$ with a fixed geodesic length and $|\delta \rangle$ with a fixed time shift. We compute their non-perturbative overlaps with TFD states. In section \ref{section:probe}, we introduce generating functions for the expectation values of the geodesic length and time shift, demonstrating analytically that they exhibit a characteristic slope-ramp-plateau behavior. Furthermore, we study the $\alpha \to 0$ limit to derive the time evolution of the geodesic length and time shift including quantum corrections. We also identify the spectral quantities associated with these generating functions. Finally, in section \ref{section:comment}, we conclude with remarks and discuss possible generalizations.

\section{Overlap between Fixed-Length State and TFD State}\label{section:lengthstate}

We begin by introducing the fixed-length state within the context of JT gravity. Consider an energy eigenstate of the Hamiltonian, denoted by $|E\rangle$.\footnote{We omit the subscripts for left/right eigenmodes, \ie
\begin{equation}
\frac{\hat{H}_{\mt{R}}+ \hat{H}_{\mt{L}}}{2} |E_n \rangle_{\mt{L}} \otimes |E_n \rangle_{\mt{R}} =\hat{H} | E_n \rangle  = E_n  | E_n \rangle  \,.
\end{equation} 
} 
The inner products and completeness relations for these states are given by 
\begin{equation} 
\langle E | E' \rangle = \frac{\delta(E - E')}{e^{S_0} D(E)} \,, \qquad e^{S_0} \int_{E_0 - \Delta E/2}^{E_0 + \Delta E/2} dE \, D(E) \, |E\rangle \langle E| = \hat{\mathbbm{1}} \,, 
\end{equation}
where $D(E)$ denotes the density of states in JT gravity. The leading-order density of states corresponds to the disk-level contribution, 
\begin{equation}\label{eq:Ddisk} 
 D(E) \approx D_{\rm{Disk}}(E) = \frac{\sinh(2\pi \sqrt{E})}{4\pi^2} \,. 
\end{equation} 
The microcanonical Hilbert subspace consists of independent microstates defined within a narrow energy window, 
\begin{equation} 
E \in \left[E_0 - \frac{\Delta E}{2}, E_0 + \frac{\Delta E}{2}\right] \,, \quad \text{with} \quad \Delta E \ll E_0 \,. 
\end{equation}
The total number of microstates, \ie the microcanonical partition function, is finite and given by 
\begin{equation}\label{eq:N}
    Z \equiv e^{S_0} \int_{E_0-\Delta E/2}^{E_0+\Delta E/2}  D(E) dE \approx  e^{S_0}D_{\text{Disk}}(E_0)\Delta E\,.
\end{equation}
One key focus in this work is the microcanonical TFD state, which serves as the dual of a two-sided black hole and is defined as
\begin{equation}\label{eq:TFD}
\begin{split}
    |\text{TFD}(t)\rangle
    &=\frac{e^{S_0}}{\sqrt{Z}}\int_{E_0-\Delta E/2}^{E_0+\Delta E/2}dE~  D(E) e^{-iEt} \, |E\rangle \,.
\end{split}
\end{equation}
The normalization factor $N$ is the same as the total number of states as defined in eq.~\eqref{eq:N}. 

On the other hand, the fixed-length state, characterized by a regularized geodesic length $\ell$ is defined as \cite{Iliesiu:2024cnh}
\begin{equation}\label{eq:lengthstate}
|\ell \rangle = e^{S_0} \int_{E_0 - \Delta E/2}^{E_0 + \Delta E/2} dE \, D(E) \psi_{E}(\ell) |E\rangle \,, 
\end{equation}
where the wavefunctions in the eigenenergy basis are given by 
\begin{equation}\label{eq:psidisk}
\psi_{E}(\ell) = \sqrt{8 e^{-S_0}} K_{i2\sqrt{E}}(2e^{-\ell/2})\,.
\end{equation} 
For later calculations, we note that applying the Kontorovich–Lebedev transform\footnote{Using the Kontorovich–Lebedev transform and inversion formulas, one can derive the conjugate integrals:
\begin{equation}
\int_0^{\infty} \frac{d z}{z} K_{i a}(z) K_{i b}(z)=\frac{\pi}{2}|\Gamma(i a)|^2 \delta(a-b), \quad a, b>0 \,,
\end{equation}
with $|\Gamma(i a)|^2 = \frac{\pi}{a \sinh (a \pi )}$, and
\begin{equation}\label{eq:KLtransformation}
  \int^\infty_0  \frac{y}{\pi^2} \sinh (\pi y) K_{iy}(x') K_{iy}(x) d y  = \frac{x}{2} \delta (x- x') \,.
\end{equation}
} yields the following explicit integrals:
\begin{equation}\label{eq:integrals01}
  e^{S_0} \int_{0}^{\infty}dE~  D_{\rm{Disk}}(E) \psi_{E}(\ell)\psi_{E}(\ell')
 =\delta (\ell- \ell') \,, 
\end{equation}
and
\begin{equation}\label{eq:integrals02}
  \int_{-\infty}^{\infty} d \ell  \, \psi_{E}(\ell)\psi_{E'}(\ell) = \frac{\delta(E- E')}{e^{S_0} D_{\rm{Disk}}(E) }  = \langle E | E' \rangle  \,.
\end{equation}
From eq.~\eqref{eq:integrals02}, we find that the fixed-length states defined in eq.~\eqref{eq:lengthstate} are not yet normalized, with a normalization factor given by 
\begin{equation}
\begin{split}
 \mathcal{N}_\ell 
 \equiv\langle \ell  |\ell \rangle  &=
  e^{S_0} \int_{E_0 - \frac{\Delta E}{2}}^{E_0 + \frac{\Delta E}{2}}dE~  D_{\rm{Disk}}(E) \psi_{E}(\ell)\psi_{E}(\ell) \,.  \\
 &\approx   
 \begin{cases}
     \frac{\Delta E}{2 \pi E_0} \,, \qquad \qquad  E_0 e^{\ell} \gg 1 \,,\\
    2\pi \Delta E e^{-4 e^{\ell/2}} \,,  \quad e^{-\ell} \gg E_0 \,.\\
 \end{cases}
\end{split}
\end{equation}
Note that the current normalization of the fixed-length state is different from that in the canonical ensemble (see Appendix \ref{sec:canonical} for more details on the canonical ensemble case). More importantly, two distinct fixed-length states are not orthogonal even at the classical level, \ie  
\begin{equation}\label{eq:overlapllp}
\langle \ell  | \ell'  \rangle  = e^{S_0} \int_{E_0 - \Delta E/2}^{E_0+\Delta E/2} dE~  D_{\rm{Disk}}(E) \psi_{E}(\ell)\psi_{E}(\ell') \neq \delta (\ell - \ell') \,,
\end{equation}
due to the finite energy window corresponding to the microcanonical ensemble. This result highlights a key distinction from the canonical ensemble.

On the other hand, the fixed-length states $| \ell \rangle$ can be interpreted as forming a different choice of basis. These states naturally define the geodesic length operator, denoted as $\hat{\ell}$, given by\footnote{For a canonical ensemble, the fixed-length state corresponds to the eigenstate of geodesic length operator $\hat{\ell}_{\rm can}$ with satisfying  
\begin{equation}
\hat{\ell}_{\rm can}  \,  |\ell_{\rm can} \rangle  = \ell_{\rm can} \, |\ell_{\rm can} \rangle \,.
\end{equation}
For more details about the results in canonical ensemble, see Appendix \ref{sec:app}. However, we note that a similar eigen equation does {\it not} hold in the microcanonical ensemble even at the classical level, \ie 
\begin{equation}
\hat{\ell}  \,  |\ell \rangle  \ne \ell \, |\ell \rangle \,,
\end{equation}
due to the absence of orthogonality as shown in eq.~\eqref{eq:overlapllp}.
}
\begin{equation}\label{eq:loperator01}
\hat{\ell} = \int_{-\infty}^{\infty} \ell\, |\ell\rangle\langle \ell |  \, d \ell \,.
\end{equation}
However, the geodesic length operator $\hat{\ell}$ is not well-defined at this stage because the fixed-length states form an over-complete basis. A well-defined construction of the geodesic length operator has been proposed in \cite{Miyaji:2024ity}, where Gram-Schmidt orthogonalization is applied starting from shorter length states. In this approach, the refined fixed-length states become orthogonal, preventing divergence, and the length spectrum terminates at a value proportional to the Heisenberg time $T_{\mt{H}}$. Using the explicit integrals from eq.~\eqref{eq:integrals02}, we can show that the completeness relation of the energy eigenbasis implies a corresponding completeness relation in terms of fixed-length states, namely 
\begin{equation}\label{eq:suml}
\begin{split}
  \left(\int_{-\infty}^{\infty} d \ell  \, | \ell \rangle   \langle \ell |  \right) \bigg|_{\mathrm{classical}} =  e^{S_0} \int_{E_0-\Delta E/2}^{E_0+\Delta E/2}dE~  D_{\rm{Disk}}(E) \, |E\rangle\langle E|=\hat{\mathbbm{1}}  \,. 
\end{split}
\end{equation}
It is important to note that this completeness relation holds only at the classical level, as it relies on the factorized spectral correlation function $D(E_i)D(E_j) \sim D_{\rm{Disk}}(E_i) D_{\rm{Disk}}(E_j)$. 

By construction, the fixed-length states satisfy 
\begin{equation}\label{eq:definepsiEl}
    \langle E|\ell\rangle= \psi_{E}(\ell) \,,
\end{equation} 
where $|E\rangle$ is an energy eigenstate in the microcanonical ensemble, and $\psi_{E}(\ell)$ represents the wavefunction corresponding to a geodesic boundary with a regularized length $\ell$. In the remainder of this paper, we focus on the regime where the geodesic length is not too small, such that $E_0 e^\ell \gg 1$. In this regime, the wavefunction can be approximated as (see Appendix \ref{sec:app} for details): 
\begin{equation}\label{eq:psiDiskapp}
\begin{split}
 \psi_{E}(\ell) 
 &\approx  \frac{\sqrt{8 \pi } e^{-\pi  \sqrt{E}-S_0/2} \sin \left(-2 e^{-\frac{\ell}{2}} \sqrt{E e^\ell-1}+2 \sqrt{E} \log \left(\sqrt{E e^\ell-1}+\sqrt{E e^\ell}\right)+\frac{\pi }{4}\right)}{e^{-\frac{\ell}{4}} (E e^\ell-1)^{\frac{1}{4}}}\,, \\
&\approx 
\frac{\sqrt{8\pi}e^{-\pi\sqrt{E}-S_0/2}}{E^{1/4}}
\left(\cos\left(\sqrt{E}(\ell+\log 4E - 2)-\frac{\pi}{4} \right)   + \mathcal{O}\left( 
\frac{e^{-\ell}}{2\sqrt{E}} \right)  \right)\,,
\end{split}
\end{equation}
where the subleading term is suppressed for either $E \gg 1$ or $e^{-\ell} \ll 1$.\footnote{This implies that we neglect higher-order terms of the form $\frac{(e^{-\ell})^p}{(\sqrt{E})^q}$ with $p, q\geq 1$.} Since $\ell$ denotes the renormalized geodesic length, its range covers all real values $\ell \in (- \infty, +\infty)$. However, contributions from negative values of $\ell$ are doubly exponentially suppressed, as demonstrated by the approximate wavefunction for $E e^{\ell} < 1$:  
\begin{equation}\label{eq:psinegative}
   \psi_{E}(\ell)  \approx \frac{\sqrt{2 \pi e^{-S_0}} e^{-2 \left(\sqrt{e^{-\ell}-E}+\sqrt{E} \arcsin \left(\sqrt{E e^{\ell}} \right)\right)}}{(e^{-\ell}-E)^{1/4}} \,.   
\end{equation}
Therefore, in most of the calculations in this paper, we neglect contributions from the negative length regime.

To investigate the overlaps of quantum states in JT gravity, we start by exploring the overlap between the fixed-length state and TFD state, namely 
\begin{equation}\label{eq:TFDloverlap}
\begin{split}
    \langle \ell |\text{TFD}(t)\rangle
    &=
   \frac{e^{S_0}}{\sqrt{Z}}
    \int_{E_0-\Delta E/2}^{E_0+\Delta E/2}dE~D(E)\psi_{E}(\ell)e^{-iEt} \,.
\end{split}
\end{equation}
The squared overlap, which defines the probability distribution, is formulated as
\begin{equation}
 P(\ell,t) \equiv \langle \rm{TFD}(t)| \ell\rangle\langle \ell|\rm{TFD}(t)\rangle  \,. 
\end{equation}
In general, the squared overlap between (normalized) states can be interpreted as the transition probability\footnote{Since we consider continuous variables in this paper, a more accurate term would be probability density.}. By applying the completeness relation of fixed-length states, we can find that summing the probability distribution over all fixed-length states yields unity:
\begin{equation}
\begin{split}
  \left(\int_{-\infty}^{\infty} P(\ell,t) \, d \ell  \right) \bigg|_{\mathrm{classical}}&= \langle \rm{TFD}(t)| \left(  \int_{-\infty}^{\infty} |\ell\rangle\langle \ell | d \ell \right)|\rm{TFD}(t)\rangle  \\
 &= \langle \rm{TFD}(t)|\rm{TFD}(t)\rangle= 1  \,. 
\end{split}
\end{equation}
This result follows from the normalization of the TFD state as defined in eq.~\eqref{eq:TFD}. While one might expect $P(\ell,t)$ to define a proper probability density since it is simply the squared overlap, we will later demonstrate that after including quantum corrections, the sum of all probability distributions leads to a divergence. Nevertheless, for the remainder of this paper, we continue to refer to $P(\ell,t)$ as a probability distribution to avoid confusion.

\subsection{Classical Contribution}

\begin{figure}[t]
	\centering
	\includegraphics[width=4in]{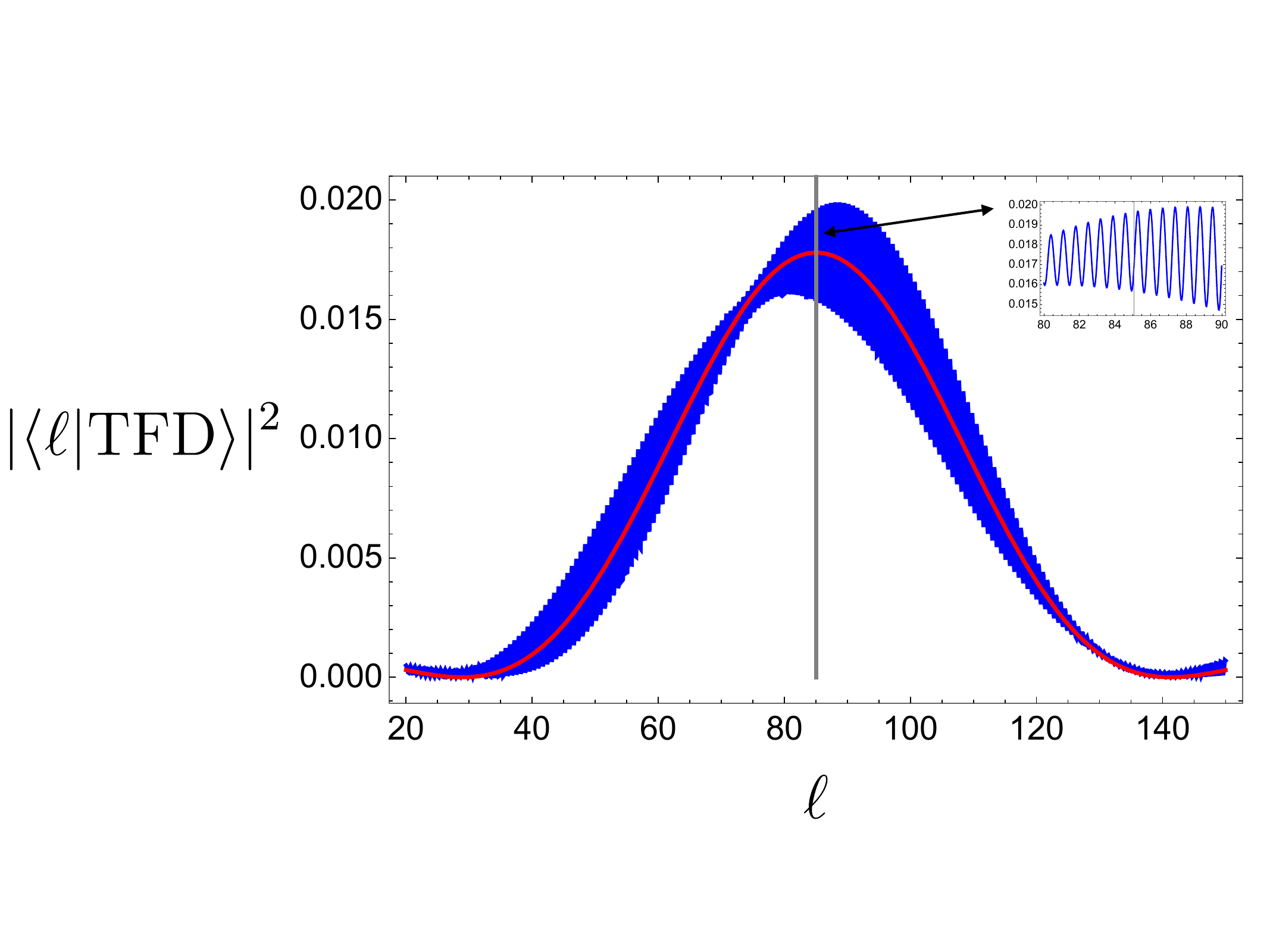}
    \caption{The squared overlap $P_{\rm{classical}}(\ell,t)$ is shown as the blue curve with oscillatory behavior. The red smooth curve represents the leading analytical approximation in eq.~\eqref{eq:Pclassical}, while the gray straight line corresponds to the classical geodesic length with $t_{\ell
    }=t$, \ie $\ell_{\rm{classical}}$ derived in eq.~\eqref{eq:classicall}. This numerical plot is generated by using 
    $t=200$, and $ E_0 =1,  \Delta E=\frac{1}{20}$.}\label{fig:overlap}
\end{figure}  

To proceed, we first focus on the classical contribution, denoted as $P_{\rm{classical}}(\ell,t)$. Specifically, this term is associated with the factorized spectral two-point function and defined by 
\begin{equation}
    \begin{split}
    P_{\text{classical}}(\ell,t) &= | \langle \rm{TFD}(t)| \ell\rangle  |^2 \\
    &= \frac{e^{2S_0}}{Z}\int_{E_0-\Delta E/2}^{E_0+\Delta E/2}dE_1dE_2~e^{-i(E_1-E_2)t}
    \psi_{E_1}(\ell) \psi_{E_2}(\ell)
     D_{\rm{Disk}}(E_1)D_{\rm{Disk}}(E_2) \,. \\
\end{split}
\end{equation}
Because of the $Z_2$ symmetry (\ie the invariance under the exchange $E_1 \leftrightarrow E_2$), the imaginary part of the expression cancels out, as expected. For a state characterized by fixed $E_0$ and $\Delta E$, deriving an analytical expression for this integral is challenging. However, numerical results can be obtained using the explicit wavefunction form in eq.~\eqref{eq:psidisk}, as illustrated in figure \ref{fig:overlap}. In addition, the normalization of the probability distribution $P_{\rm{classical}}(\ell,t)$ can be easily verified:
\begin{equation}
 \int_{-\infty}^\infty  P_{\text{classical}}(\ell,t)  d \ell =\frac{e^{S_0}}{Z}\int_{E_0-\Delta E/2}^{E_0+\Delta E/2}dE~
     D_{\rm{Disk}}(E) =1 \,.
\end{equation}

Since our focus is on the regime where $\Delta E / E_0 \ll 1$, we approximate the wavefunction $\psi_{E}(\ell)$ by expanding around the central energy $E_0$, \ie
\begin{equation}\label{eq:appwave}
\begin{split}
 \psi_{E}(\ell)
    \approx  \frac{  \sqrt{8 \pi} e^{-\pi\sqrt{E_0}-S_0/2}}{E_0^{1/4}}
    \cos\left(\sqrt{E_0}(\ell+\log(4E_0)-2)-\frac{\pi}{4} +t_{\ell}(E-E_0) \right)  \,,
    \end{split}
\end{equation}
where we define a characteristic length scale\footnote{Since $\ell$ is arbitrary, the subleading term can be neglected provided that $(t_{\ell}-E_0^{-1/2})\frac{(\Delta E)^2}{E_0} \ll 1$.}
\begin{equation}
t_{\ell}: = \frac{\log (4 E_0 e^{\ell})}{2\sqrt{E_0}} =\frac{\ell+\log (4E_0)}{2\sqrt{E_0}} \,.
\end{equation}
associated with the characteristic energy scale $E_0$ and the geodesic length $\ell$. At leading order, the overlap in eq.~\eqref{eq:TFDloverlap} simplifies to
\begin{equation}
\begin{split}
    \langle \ell |\text{TFD}(t)\rangle
    &\approx
    e^{-iE_0t}
    \frac{1}{\pi^{1/2}E_0^{1/4}(\Delta E)^{1/2}}
    \Bigg[
    e^{i(\sqrt{E_0}(\ell+\log(4E_0)-2)-\frac{\pi}{4})}
    \frac{\sin\big((t_{\ell}-t)\frac{\Delta E}{2}\big)}{t_{\ell}-t} \\
    &\quad +
    e^{-i(\sqrt{E_0}(\ell+\log(4E_0)-2)-\frac{\pi}{4})}    
    \frac{\sin(t_{\ell}+t)\frac{\Delta E}{2})
    }{t_{\ell}+t}
    \Bigg].
\end{split}
\end{equation}
The squared overlap, \ie the probability distribution, is thus given by
\begin{equation}\label{eq:Pclassical}
    \begin{split}
    & P_{\text{classical}}(\ell,t) \approx 
   \frac{1}{\pi\sqrt{E_0}\Delta E}
   \left(  \frac{\sin^2\left((t_{\ell}-t)\frac{\Delta E}{2}\right)}{\big(t_{\ell}-t\big)^2}  +  \right. \\
   &\quad \left. +\frac{\sin^2\left((t_{\ell}+t)\frac{\Delta E}{2}\right)}{\big(t_{\ell}+t\big)^2} + 2 \sin ( 4(E_0 t_\ell -\sqrt{E_0}) )\frac{\sin\left((t_{\ell}-t)\frac{\Delta E}{2}\right)\sin\left((t_{\ell}+t)\frac{\Delta E}{2}\right)}{(t_{\ell}-t)(t_{\ell}+t)}  
      \right) \,, 
    \end{split}
\end{equation}
where the first term dominates in the late-time regime when $(t_{\ell} +t) \gg 1$. As shown in figure \ref{fig:overlap}, it is evident that the squared overlap concentrates around
\begin{equation}
P_{\text{classical}} \big|_{\rm max}  = \frac{\Delta E}{4\pi \sqrt{E_0}} \,, \quad \text{with}  \quad t_{\ell} = t  \,. 
\end{equation}
In terms of $\ell$, this peak corresponds to the classical geodesic length of a two-sided black hole, \ie 
\begin{equation}\label{eq:classicall}
    \ell =  \ell_{\rm{classical}} \equiv 2\sqrt{E_0} \, t-\log (4E_0).
\end{equation}

\subsection{Quantum Contributions}

\begin{figure}[t]
	\centering
	\includegraphics[width=4in]{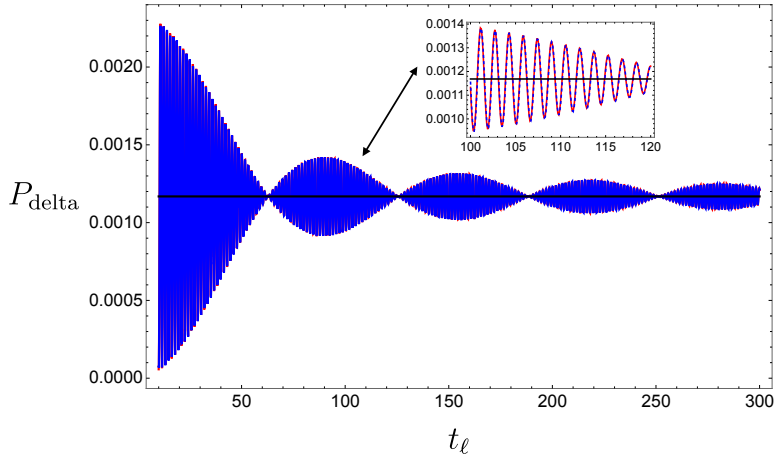}
	\caption{The quantum correction $P_{\rm{delta}}(\ell)$ derived from the contact term in the spectral correlation. The leading term, \ie the constant $\frac{1}{T_{\mt{H}}\sqrt{E_0}}$, is represented by the black line. The analytical approximation \eqref{eq:Pdeltaapp} is shown as the red curve, while numerical results are indicated by the blue curve with the same parameter values as in other figures, \ie $S_0=3, E_0=1$, and $ \Delta E=\frac{1}{20}$.}
	\label{fig:Pdelta}
\end{figure}  

The classical contribution $P_{\rm{classical}}(\ell,t)$ is expected to dominate in the limit $e^{S_0} \to \infty$. However, for a finite-dimensional Hilbert space, quantum corrections become necessary. Starting from the basic definition, the probability distribution can be expressed as  
\begin{equation}\label{eq:Plt}
    \begin{split}
    P(\ell, t)
    &= \frac{e^{2S_0}}{Z}\int dE_1\int dE_2~e^{-i(E_1-E_2)t}
    \psi_{E_1}(\ell)  \psi_{E_2}(\ell)
    \langle D(E_1)D(E_2) \rangle \,, \\
\end{split}
\end{equation}
where $\langle D(E_1)D(E_2) \rangle$ represents the spectral two-point function.\footnote{Since eq.~\eqref{eq:Plt} involves the ensemble average of a positive quantity, it is manifestly positive. However, when approximating the wavefunction and extrapolating beyond its valid parameter region, the expression may yield negative values, such as when $t_\ell\sim 0$.} 
Since we work within the microcanonical ensemble, we restrict the energy to a narrow window $\Delta E$, ensuring that $\Delta E / E_i \ll 1$, which allows us to approximate $E_i \approx E_0$. In the limit $E_i \to E_j$, the spectral correlations $\langle D(E_i)D(E_j) \rangle$ in JT gravity \cite{Saad:2019lba} are well approximated by the universal expression:
\begin{equation}\label{eq:sine}
    \begin{split}
    D_{\text{Disk}}(E_i)D_{\text{Disk}}(E_j)    
    +
    e^{-S_0}\delta(E_i-E_j)D_{\text{Disk}}(E_i)-\frac{\sin^2(\pi e^{S_0}D_{\text{Disk}}(\bar{E})(E_i-E_j))}{e^{2S_0}\pi^2(E_i-E_j)^2} \,, 
    \end{split}
\end{equation}
where $\bar{E}= \frac{E_i +E_j}{2}$. The delta function term represents a contact contribution, while the sine kernel captures non-perturbative effects. Notably, the sine kernel is a signature of the Gaussian Unitary Ensemble (GUE) in random matrix theory, indicating level repulsion and spectral rigidity.

The factorized two-point term yields the classical distribution $P_{\rm{classical}}(\ell,t)$, as discussed in the previous subsection. The remaining two contributions, arising from the delta function and the sine kernel, are denoted as $P_{\rm{delta}}(\ell, t)$ and $P_{\rm{sine}}(\ell, t)$, respectively. It is straightforward to show that $P_{\rm{delta}}(\ell, t)$ simplifies to a time-independent constant:
\begin{equation}\label{eq:Pdelta}
\begin{split}
 P_{\rm{delta}}(\ell,t) &=P_{\rm{delta}}(\ell) =\frac{e^{S_0}}{Z}\int_{E_0 -\frac{\Delta E}{2}}^{E_0+\frac{E_0}{2}} dE\, 
    ( \psi_{E}(\ell))^2 D_{\rm{Disk}}(E)\,. 
\end{split}    
\end{equation}
To leading order, this reduces to 
\begin{equation}\label{eq:Pdeltaapp}
\begin{split}
 P_{\rm{delta}}(\ell)   &\approx \frac{1}{\Delta E}\int_{E_0 -\frac{\Delta E}{2}}^{E_0+\frac{E_0}{2}} dE\, 
    \left( \psi_{E}(\ell)\right)^2 \,, \\ 
    &\approx \frac{1}{T_{\mt{H}} \sqrt{E_0}}  \left(1 +
    \frac{\sin\left(4( E_0 t_{\ell} -\sqrt{E_0}) \right)  \sin(t_{\ell}\Delta E)}{t_{\ell}\Delta E}
    \right)  +\mathcal{O}(\Delta E) \,, \\
\end{split}    
\end{equation}
where we have used the fact that the normalization factor of the wavefunction associated with eigenvalue $E_0$ is formulated as
\begin{equation}
 \left(  \psi_{E_0}(\ell)  \right)^2 \sim \frac{1}{\sqrt{E_0}}\frac{4\pi e^{-S_0}}{ \sinh(2\sqrt{E_0} \pi) } =  \frac{2}{T_{\mt{H}} \sqrt{E_0}} \,.
\end{equation}
As a result, $P_{\rm{delta}}(\ell, t)$, expressed in terms of the geodesic length $\ell$, oscillates around the constant $1/(T_{\mt{H}} \sqrt{E_0})$. The deviation is suppressed by a factor of $t_{\ell}$, indicating that $P_{\rm{delta}}(\ell, t)$ approaches this constant for a fixed-length state with a large geodesic length. Moreover, it is worth noting that compared to the classical contribution $P_{\rm{classical}}(\ell, t)$, the quantum correction $P_{\rm{delta}}(\ell, t)$ is suppressed by the dimension of the Hilbert space, \ie   $e^{S_0}$, since $1/T_{\mt{H}} \sim e^{-S_0}$. This suppression highlights the subleading nature of quantum corrections. 

\begin{figure}[t]
	\centering
	\includegraphics[width=2.95in]{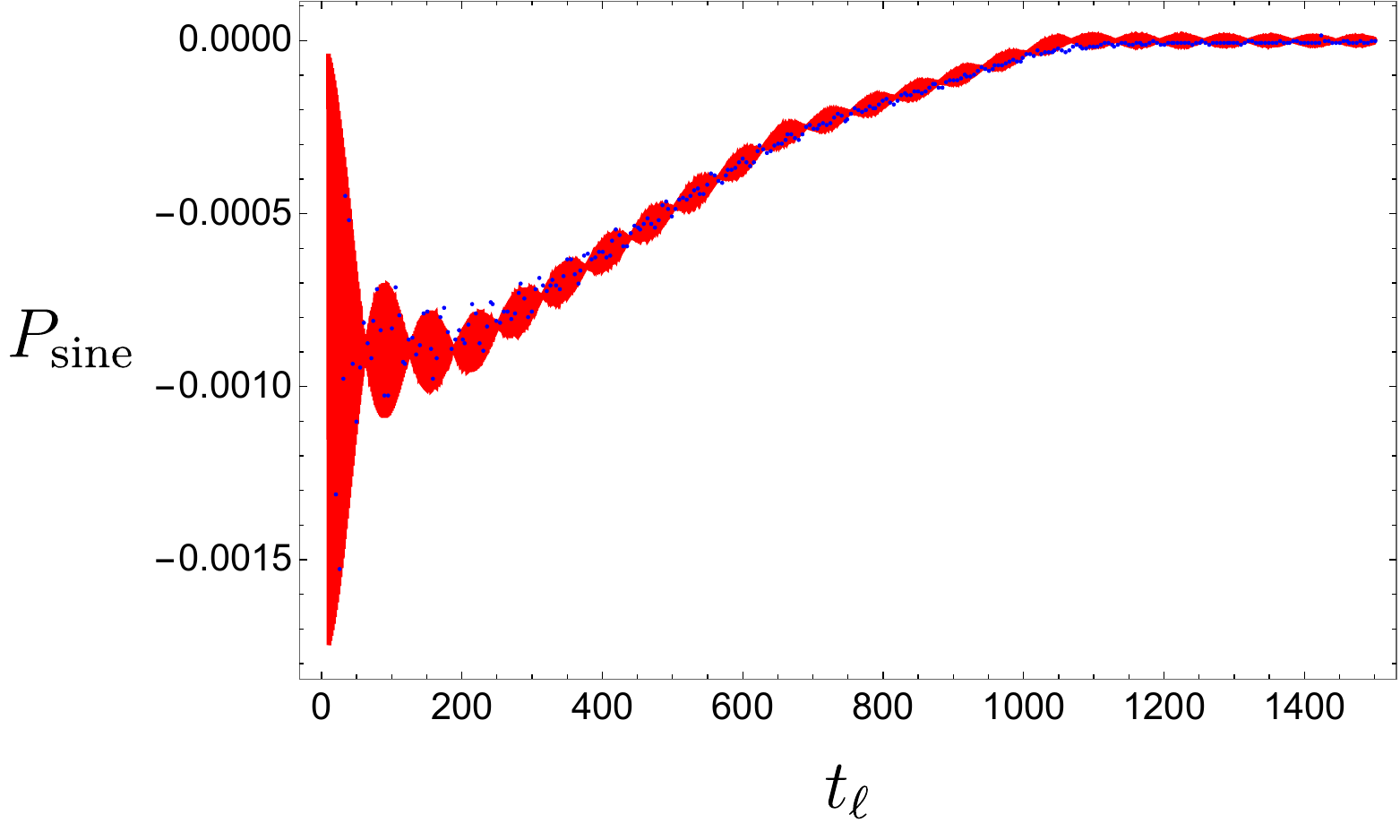}
    	\includegraphics[width=2.95in]{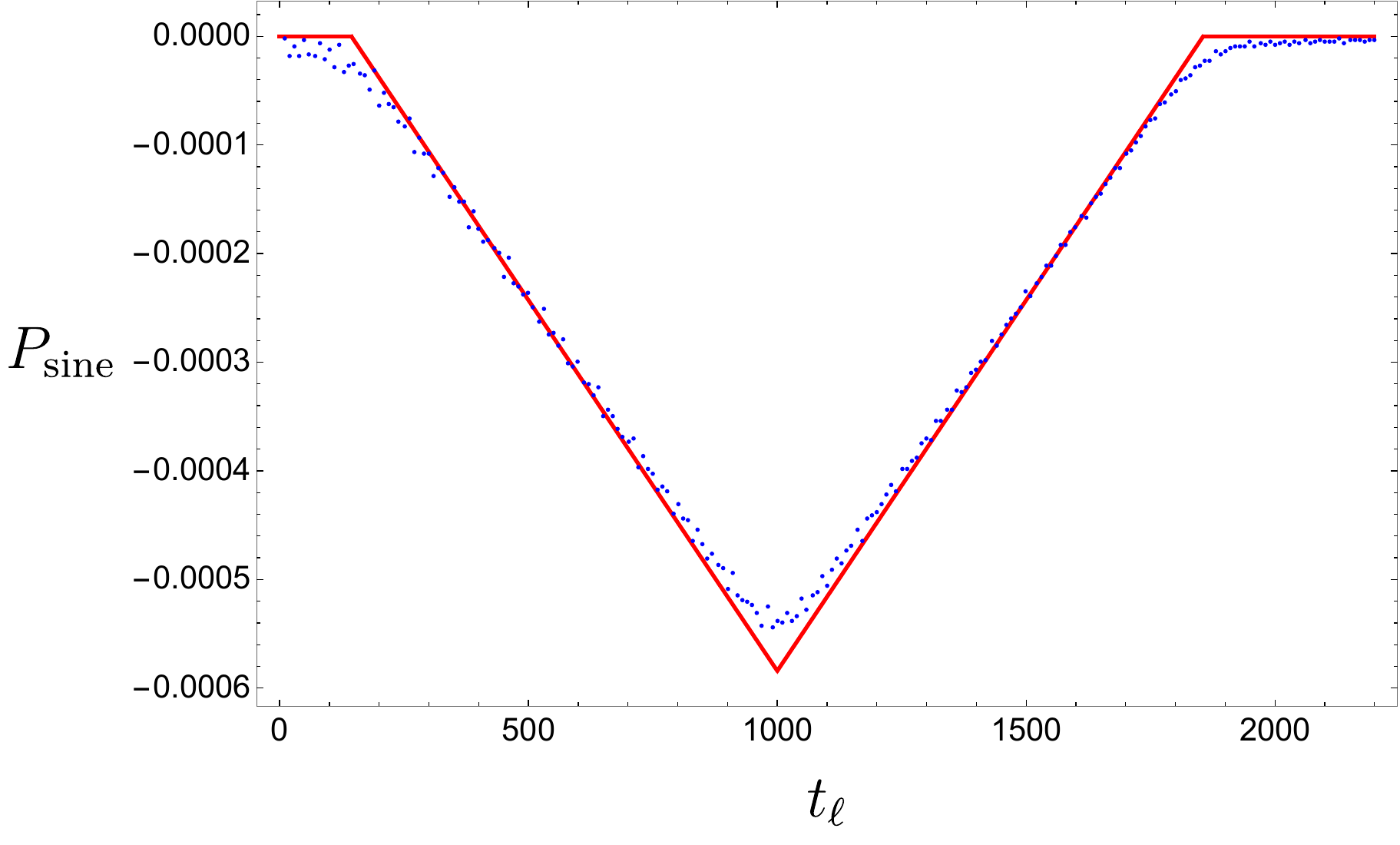}
	\caption{The (negative) probability contribution $P_{\mathrm{sine}}(\ell, t)$ originating from the sine kernel. The blue dots represent numerical results from its definition given in eq.~\eqref{eq:Psine}, while the analytical approximation is shown as the red curve. As in previous figures, we use parameter values $S_0=3$, $E_0=1$, and $\Delta E=\frac{1}{20}$. The left and right plots correspond to $t=200<T_{\mt{H}}$ and $t=1000>T_{\mt{H}}$, respectively.}
	\label{fig:Psine}
\end{figure}  

Finally, we consider the contribution from the sine-squared term, namely
\begin{equation}\label{eq:Psine}
\begin{split}
 P_{\rm{sine}}(\ell,t) &\equiv -\frac{1}{Z}\int_{E_0 -\frac{\Delta E}{2}}^{E_0+\frac{\Delta E}{2}} dE_1dE_2\,e^{-i(E_1-E_2)t} 
    \psi_{E_1}(\ell)  \psi_{E_2}(\ell) \frac{\sin^2(\pi e^{S_0}D_{\text{Disk}}(E_1)(E_1-E_2))}{\pi^2(E_1-E_2)^2} \,. 
\end{split}    
\end{equation}
It is straightforward to see that in the classical limit where $e^{S_0} \to \infty$, the term $P_{\rm{sine}}$ cancels out precisely with $P_{\rm{delta}}(\ell,t)$, since the sine-squared term reduces to a delta function:\footnote{This follows from the identity:
\begin{equation}
  \lim_{N\to \infty}   \frac{\left(\sin(N x) \right)^2}{Nx^2}   \sim \pi \delta (x) \,.
\end{equation}
which holds in the integral sense:
\begin{equation}
    \lim_{N \to \infty} \int_{-a}^{b}  \left( \frac{1}{N} \frac{\sin^2 \( N x\)}{x^2}  \right) x^n d x = 
   \begin{cases}
  \pi  \,,& n =0 \\
 0 \,,& n >0 
\end{cases}
 \end{equation}
for any $a, b >0$.} Substituting the approximate wavefunction \eqref{eq:appwave}, we derive the leading contribution:
\begin{equation}
 \begin{split}
   & P_{\rm{sine}}(\ell,t)  \approx 
   \frac{-4}{\pi \sqrt{E_0} T_{\mt{H}}^2 \Delta E}
    \int_{E_0-\Delta E/2}^{E_0+\Delta E/2}  dE_1dE_2 \,e^{-i(E_1-E_2)t}  \frac{\sin^2( T_{\mt{H}}(E_1-E_2)/2)}{(E_1-E_2)^2} \\
     & \times \cos \left(  2( E_0 t_{\ell} -\sqrt{E_0}) -\frac{\pi}{4} + t_{\ell} (E_1 -E_0)  \right) \cos\left( 2( E_0 t_{\ell} -\sqrt{E_0}) -\frac{\pi}{4} + t_{\ell} (E_2 -E_0)\right) \,.
 \end{split}
\end{equation}
By defining new variables
\begin{equation}\label{eq:variables}
\bar{E}=\frac{E_1+E_2}{2} \,,\qquad  E_{12}=E_1-E_2 \,,
\end{equation}
we rewrite the integral as
\begin{equation}
\begin{split}
\int_{E_0-\Delta E/2}^{E_0+\Delta E/2}dE_1dE_2 \rightarrow &\int_{E_0}^{E_0+\Delta E/2}d\bar{E}\int_{-\Delta E + 2(\bar{E}- E_0)}^{\Delta E - 2(\bar{E}- E_0)}dE_{12} + \int^{E_0}_{E_0-\Delta E/2}d\bar{E}\int_{-\Delta E + 2(E_0 - \bar{E})}^{\Delta E - 2(E_0-\bar{E})}dE_{12}  \,. 
\end{split}
\end{equation}
The first integral
\begin{equation}
\mathcal{I}_{12} =  \int dE_{12}~e^{-iE_{12}t}  \frac{\sin^2( T_{\mt{H}}E_{12}/2)}{(E_{12})^2}  \cos \left(  A + \frac{1}{2} t_{\ell} E_{12}   \right) \cos \left(  A - \frac{1}{2} t_{\ell} E_{12}   \right)  \,, 
\end{equation}
with identifying 
\begin{equation}
 A = 2 (E_0 t_{\ell} - \sqrt{E_0}) -\frac{\pi}{4}
 + t_{\ell} (\bar{E} - E_0 ) \,, 
 \end{equation} 
can be evaluated explicitly using the sine integral function $\mathrm{Si}(z)$. However, since we are interested in the late-time behavior, we will further assume
\begin{equation}
 t \sim T_{\mt{H}} \sim e^{S_0} \gg \frac{1}{\Delta E} \,, 
 \end{equation}
Under these conditions, we find that the explicit integral $\mathcal{I}_{12}$ is dominated by\footnote{We only use the series expansion of the sine integral function 
\begin{equation}
 \mathrm{Si}(z) \approx \frac{\pi }{2} -\frac{\cos (z)}{z}-\frac{\sin (z)}{z^2}+ \mathcal{O}\left(\frac{1}{z^3}\right) \,,
\end{equation}
with assuming $\left(t-t_{\ell} + T_{\mt{H}}, t-t_{\ell} - T_{\mt{H}}, t-T_{\mt{H}}, t- t_{\ell}, t\right) \times \Delta E \gg 1$.It is important to note that these approximations cannot be valid in the regime like $t-t_{\ell} + T_{\mt{H}} \sim 0$. This invalidation illustrates the sharp transitions in the linear regime of the approximate expression as shown in eq.~\eqref{eq:Ptotal}. See \eg figure \ref{fig:Psine} for example.}
\begin{equation}
\begin{split}
\mathcal{I}_{12} &\approx  \frac{\pi}{16} \bigg(  |t-t_{\ell} - T_{\mt{H}}| 
  + |t-t_{\ell} + T_{\mt{H}}| + |t+t_{\ell} - T_{\mt{H}}| -  (t+t_{\ell} - T_{\mt{H}}) 
\\
 &\qquad -2|t -t_{\ell}|  + 2 \left( |t- T_{\mt{H}}| - t +T_{\mt{H}}  
\right)\cos (2A) \bigg)  \\
&\qquad + \frac{\sin ( \Delta E t) \sin ^2\left(T_{\mt{H}}\Delta E /2\right) (\cos (2 A)+\cos ( \Delta E t_{\ell}))}{(\Delta E)^2 t} +\mathcal{O}\left(\frac{1}{t^2}\right) \,,
\end{split}
\end{equation}
where we keep the subleading correction in the late-time limit $t 
\to \infty$ in the last line. At the leading order, we can express it in terms of the geodesic length $t_{\ell}$ and rewrite it for various regimes as 
\begin{equation}
  \mathcal{I}_{12} \approx    
   \begin{cases}
  \left\{\begin{array}{lr}
        0\,, & \qquad \qquad (t_{\ell}  >  t + T_{\mt{H}})\\
         \frac{\pi}{8}  \left( T_{\mt{H}}  + t  - t_{\ell}  \right)\,, & ( t<t_{\ell}  <  t + T_{\mt{H}} )\\
       \frac{\pi}{8}  \left(    T_{\mt{H}} + t_{\ell}  -  t \right)\,, &  (t - T_{\mt{H}} <t_{\ell}  <  t )\\
       0\,, & \qquad \qquad  (t_{\ell}   < t - T_{\mt{H}} )\\
        \end{array}\right\}  \qquad \text{with} \qquad t > T_{\mt{H}} \,, \\
        \\
      \left\{\begin{array}{lr}
        \frac{\pi}{4}   (T_{\mt{H}} -t)  \cos (2 A)\,, & \qquad \qquad (t_{\ell}  >  t + T_{\mt{H}})\\
         
        \frac{\pi}{8}  \left(t+T_{\mt{H}}-t_{\ell} + 2 (T_{\mt{H}}-t)\cos (2 A) \right) \,, & ( \max [t, T_{\mt{H}}-t] < t_{\ell}  <  t + T_{\mt{H}} )\\
          \frac{\pi}{8}   \left(T_{\mt{H}}+t_{\ell} -t + 2(T_{\mt{H}}-t)\cos (2 A) \right) \,, &  ( T_{\mt{H}} -t <t_{\ell}  <  t )\\
    \frac{\pi}{4}   \left(T_{\mt{H}}-t_{\ell} + (T_{\mt{H}}-t)\cos (2 A) \right)  \,, &  (t <t_{\ell}  <   T_{\mt{H}} -t ) \\
       \frac{\pi}{4}   \left(T_{\mt{H}}-t + (T_{\mt{H}}-t)\cos (2 A) \right)\,, & \qquad \qquad  (t_{\ell}   < \min [t, T_{\mt{H}}-t] )\\
        \end{array}\right\}  \,\text{with} \quad t <  T_{\mt{H}}\,. \\
        \end{cases}
\end{equation}
To obtain the leading probability contribution from the sine kernel, we explicitly evaluate the second integral. For instance, we find that
\begin{equation}
P_{\rm{sine}}(\ell,t)   \approx  -\frac{1}{2\sqrt{E_0}T_{\mt{H}}^2} \left( T_{\mt{H}}  - |t_{\ell} - t|  \right) \,,
\end{equation}
when $ 0<t- T_{\mt{H}}<t_{\ell} <  t + T_{\mt{H}}$, or
\begin{equation}
P_{\rm{sine}}(\ell,t)   \approx  -\frac{\left( T_{\mt{H}}  - t  \right)}{\sqrt{E_0}T_{\mt{H}}^2}  \frac{\sin \left(4 \left( E_0 t_{\ell}-\sqrt{E_0}\right)\right) \sin (t_{\ell} \Delta E )}{t_{\ell} \Delta E } \,,
\end{equation}
for $ t_{\ell}  >  t + T_{\mt{H}} > 2 t$. By combining both contributions from the delta term, we express the quantum corrections in a compact form:
\begin{equation}\label{eq:Pquantum}
    \begin{split}
    P_{\rm{quantum}}(\ell ,t)  &\equiv P_{\rm{delta}} +P_{\rm{sine}}  
     \approx
    \frac{1}{2T_{\mt{H}}^2\sqrt{E_0}} 
     \Big(\min\big[T_{\mt{H}},~|t_{\ell}-t|\big]+\min\big[T_{\mt{H}},~t_{\ell}+t\big]
     \\
     &\qquad +2\left(T_{\mt{H}} -\theta(T_{\mt{H}}-t)\left(T_{\mt{H}}-t\right)\right)
    \frac{\sin \left(4 \left( E_0 t_{\ell}-\sqrt{E_0}\right)\right) \sin (t_{\ell} \Delta E )}{t_{\ell} \Delta E } 
    \Big) \,. 
    \end{split}
\end{equation}


\begin{figure}[t]
	\centering
	\includegraphics[width=2.95in]{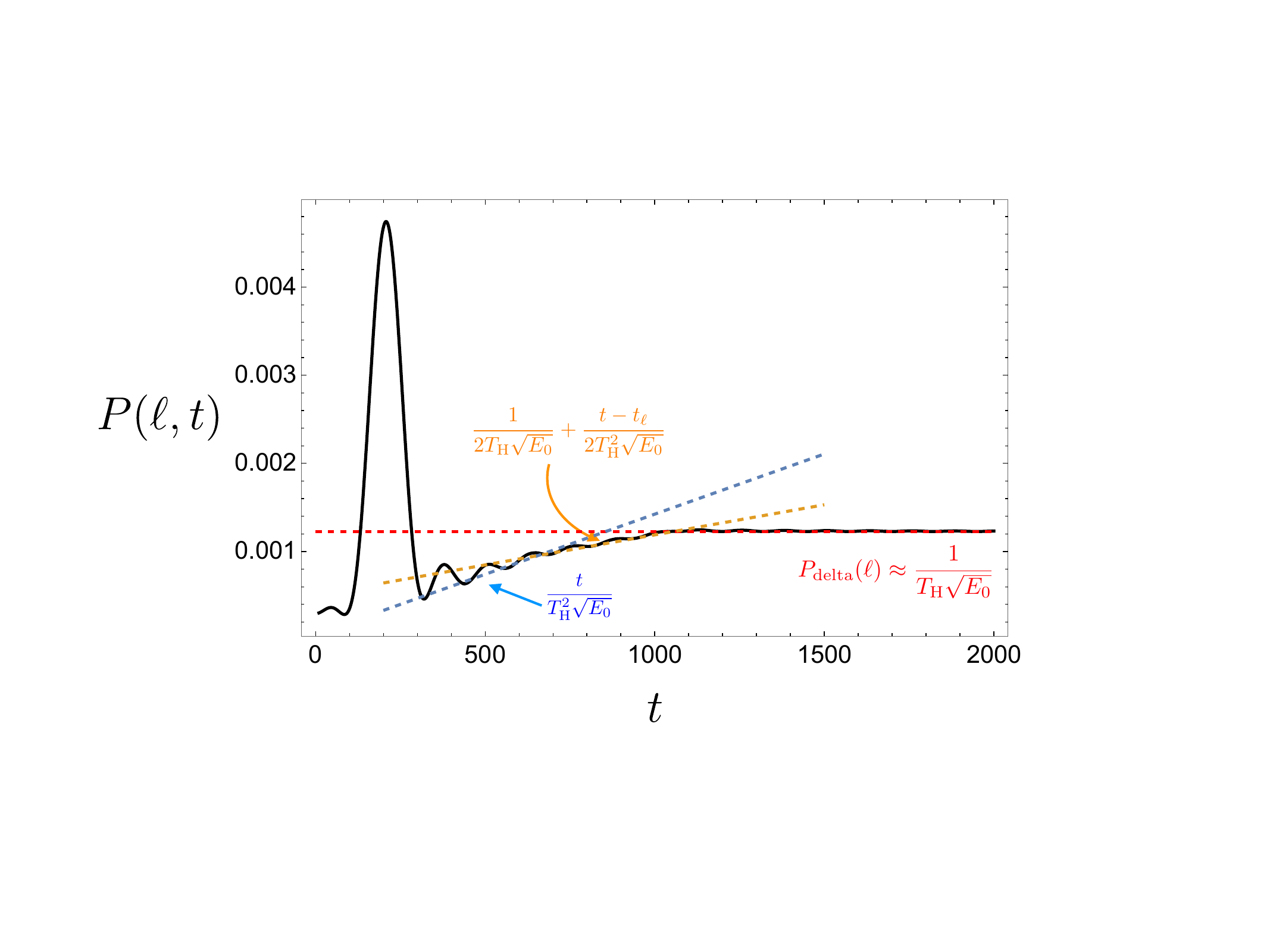}
    	\includegraphics[width=2.95in]{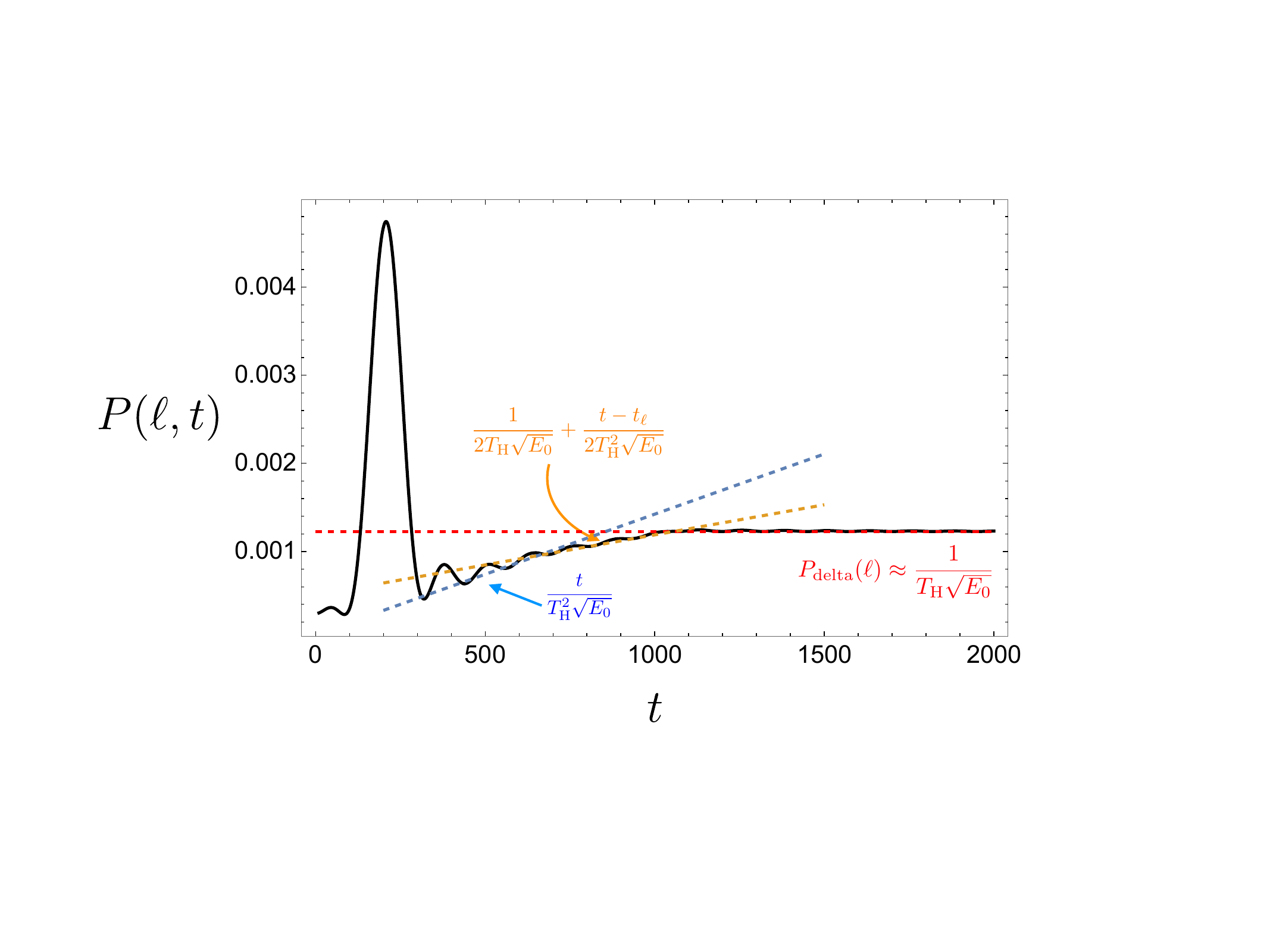}
	\caption{Total probability $P(\ell, t)$ in terms of the geodesic length $t_\ell$. Similar to other figures, we choose $S_0=3, E_0=1$, and $ \Delta E=\frac{1}{20}$ for this plot. The left/right plot corresponds to choosing $t=200<T_{\mt{H}}$ and $t=1000>T_{\mt{H}}$, respectively. The gray line represents $t_\ell=t$. The final plateau is described by the constant $\frac{1}{T_{\mt{H}}\sqrt{E_0}}$, as shown in eq.~\eqref{eq:finaltl}.}
	\label{fig:Ptotal01}
\end{figure}  

\begin{figure}[t]
	\centering
    \includegraphics[width=2.95in]{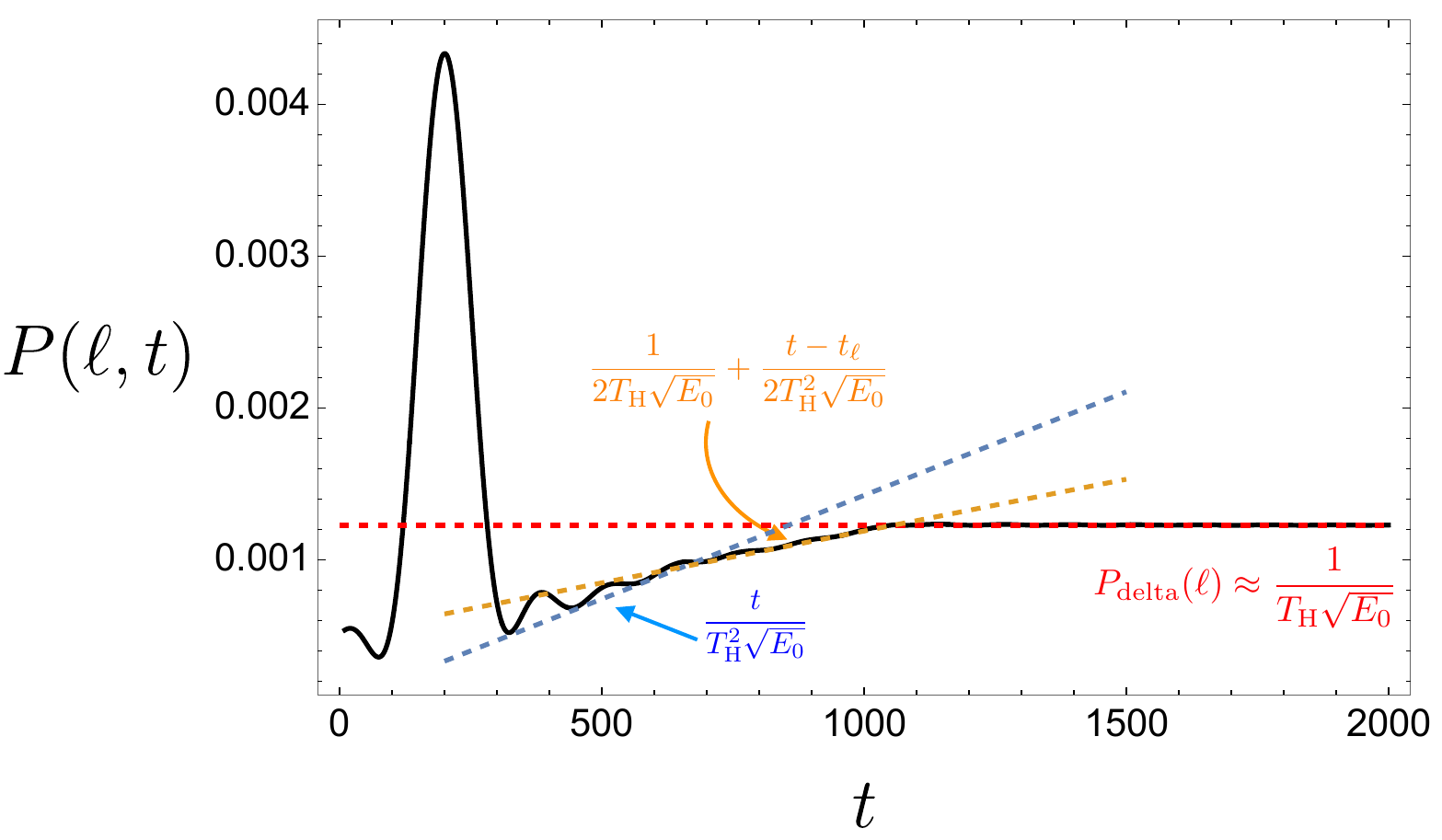}
    \includegraphics[width=2.95in]{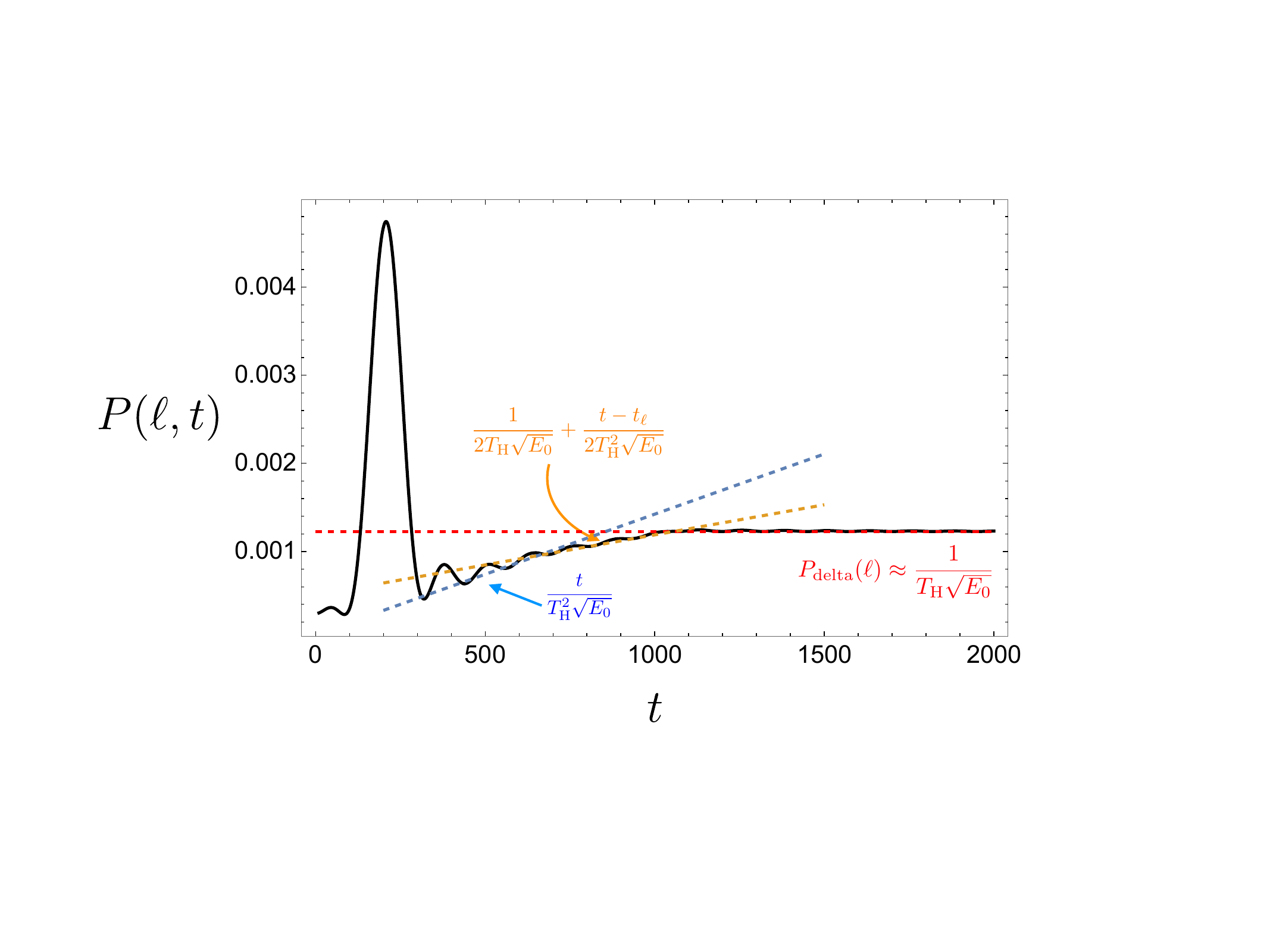}
	\caption{The time dependence of the total probability $P(\ell, t)$ with a fixed value of $t_\ell$. The left/right plot corresponds to $t_\ell < T_{\mt{H}}$ and $t_\ell > T_{\mt{H}}$, respectively.}
	\label{fig:Ptotal02}
\end{figure}  

Putting everything together, we conclude that the total probability distribution between the fixed-length state\footnote{We note that a similar probability distribution (transition probability) has been studied in \cite{Erdmenger:2023wjg} for Krylov states $| O_n \rangle$, which form the orthogonal Krylov basis. The authors show numerically that the time evolution of the transition probability for chaotic systems has the universal rise-slope-ramp-plateau structure similar to $P(\ell, t)$ for fixed-length states.} and the TFD state is approximately given by
\begin{equation}\label{eq:Ptotal}
    \begin{split}
     P(\ell,t)  
     \approx \frac{1}{\pi\sqrt{E_0}\Delta E}
    \frac{\sin^2\left((t_{\ell}-t)\frac{\Delta E}{2}\right)}{\big(t_{\ell}-t\big)^2} 
   &+ \frac{\Big(\min\big[T_{\mt{H}},~|t_{\ell}-t|\big]+\min\big[T_{\mt{H}},~t_{\ell}+t\big] \Big)}{2\sqrt{E_0}T_{\mt{H}}^2} 
      \\
     &+\mathrm{Oscillating \, \,terms}
    \end{split}
\end{equation}
where the oscillating terms are explicitly derived in eq.~\eqref{eq:Pclassical} and eq.~\eqref{eq:Pquantum}. The first and last oscillating terms in $P(\ell, t)$ depend on the width of the energy window but are suppressed by the geodesic length $t_\ell$. The behavior of the total probability distribution $P(\ell, t)$, as derived in eq.~\eqref{eq:Ptotal}, is illustrated in figures \ref{fig:Ptotal01} and \ref{fig:Ptotal02}. Notably, the distribution exhibits similar dependencies on both $t_\ell$ and $t$, with the time evolution characterized by significantly less oscillatory behavior. This is because the last term in $P_{\rm{delta}}(\ell, t)$ in eq.~\eqref{eq:Ptotal} remains fixed as a constant for a given fixed-length state. Ignoring these oscillatory corrections, $P_{\rm{delta}}(\ell, t)$ approximately exhibits symmetry under the exchange $t_\ell \longleftrightarrow t$. 

To focus the discussion, let us consider the time dependence. The total distribution initially exhibits oscillatory behavior around $t = t_\ell$, dominated by the contributions from the classical term. As $t$ (or equivalently $t_\ell$) increases, the oscillations gradually decay. Quantum corrections eventually become comparable to the classical contribution at a time scale given by\footnote{This transition time differs from that for the SFF in JT gravity or GUE because the classical term decays as $\frac{1}{(t_\ell - t)^2}$ rather than $\frac{1}{(t_\ell - t)^3}$. See eq.~\eqref{eq:canZZ} in Appendix.}
\begin{equation}
 |t_\ell - t| \quad \sim \quad \left(  \frac{T_{\mt{H}}^2}{\min (T_{\mt{H}}, t_\ell)} \right)^{1/2} \quad \sim \quad  e^{S_0/2}\,.
\end{equation}

After this time scale, the distribution transitions to a linear regime, approximately described by
\begin{equation}
  \text{Linear ramp regime:} \qquad   P(\ell,t)
  \approx 
    \frac{1}{2T_{\mt{H}} \sqrt{E_0}} +   \frac{t-t_\ell}{2T_{\mt{H}}^2 \sqrt{E_0}}   \,, 
\end{equation}
as shown in figure \ref{fig:Ptotal02}. Another distinct linear regime emerges for $t_\ell < T_{\mt{H}}$, where the distribution is approximated by
\begin{equation}
     P(\ell,t) \approx \frac{t_\ell}{2T_{\mt{H}}^2 \sqrt{E_0}} +   \frac{t-t_\ell}{2T_{\mt{H}}^2 \sqrt{E_0}}  = \frac{t}{T_{\mt{H}}^2 \sqrt{E_0}} \,, 
\end{equation}
for the regime $t_\ell < T_{\mt{H}}$. 
However, the total probability cannot increase linearly indefinitely because the finite dimension of the Hilbert space imposes an upper bound fixed by the constant $S_0$. As a result, after the linear ramp phase, the total probability stabilizes and approaches a plateau. In other words, we conclude that in the late-time limit, the total probability $P(\ell,t)$ is dominated by a time-independent constant:
\begin{equation}
 \lim_{t \gg  t_{\ell} + T_{\mt{H}}} P(\ell,t)    \approx   P_{\rm{delta}}(\ell) \approx  \frac{1}{T_{\mt{H}} \sqrt{E_0}}  \left(1 +
    \frac{\sin\left(4( E_0 t_{\ell} -\sqrt{E_0}) \right)  \sin(t_{\ell}\Delta E)}{t_{\ell}\Delta E}
    \right)  \,, 
\end{equation}
which predominantly originates from the delta term $P_{\rm{delta}}(\ell, t)$. Similarly, for a fixed-length state with a large geodesic length, the distribution approaches a similar finite constant, \ie 
\begin{equation}\label{eq:finaltl}
  \lim_{t_{\ell} \gg T_{\mt{H}} +t} P(\ell,t)    \approx \frac{1}{T_{\mt{H}}\sqrt{E_0}} \,.  
\end{equation}
In summary, the squared overlap $P(\ell,t)$ between TFD states and fixed-length states exhibits a universal peak-ramp-plateau structure. 

The meticulous reader may recognize that the time evolution of the probability $P(\ell, t)$ after the peak at $t = t_\ell$ universally follows a slope-ramp-plateau structure. This behavior closely resembles the time evolution of the spectral form factor in JT gravity, or more generally, the universality class of random matrix theory described by the Gaussian Unitary Ensemble (GUE). This resemblance is not coincidental. In the next section, we will demonstrate that the underlying mechanism governing the universal slope-ramp-plateau structure of the SFF is the same as that driving the probability distribution $P(\ell, t)$ for fixed-length states $|\ell\rangle$.

\section{Overlap between Time-Shifted TFD States}\label{section:timeshiftstate}

In the previous section, we focused on states $| \ell \rangle$ with a fixed geodesic length. However, the choice of basis states is arbitrary. We now turn our attention to the TFD state but with a fixed relative shift time. If we prepare the Hartle-Hawking state at $t=0$, \ie $|\mathrm{TFD} (0)\rangle$, as defined in eq.~\eqref{eq:TFD}, it evolves with the left and right boundary times $t_{\mt{L}}, t_{\mt{R}}$ under the time evolution operator $e^{- \frac{i}{2}( H_{\mt{L}} t_{\mt{L}} +  H_{\mt{R}} t_{\mt{R}} )}$. 

Since the TFD state is invariant under evolution with the Hamiltonian $H_{\mt{R}}- H_{\mt{L}}$, it is an eigenvector of $H_{\mt{R}}- H_{\mt{L}}$ with eigenvalue zero. Equivalently, we can evolve it only with $H_{\mt{R}}+ H_{\mt{L}}$. Without loss of generality, we label the time-evolved TFD state by the shift time $\delta$ as follows:
\begin{equation}
   |\delta\rangle \equiv | \mathrm{TFD}(\delta) \rangle  = e^{- i\frac{\delta}{2} ( H_{\mt{L}}  +  H_{\mt{R}}) }| \mathrm{TFD}(0) \rangle  =\frac{e^{S_0}}{\sqrt{Z}}\int_{E_0-\Delta E/2}^{E_0+\Delta E/2}dE~D_{\text{Disk}}(E)e^{-iE \delta}|E\rangle \,, 
\end{equation}
where the eigenvalue is given by $\delta =t_{\mt{L}}+t_{\mt{R}}$. Similar to the fixed-length state $|\ell\rangle$, we may consider $ |\delta\rangle$ as an eigenstate of the particular time-shifted operator $\hat{\delta}$. In analogy to the fixed-length state \eqref{eq:lengthstate}, we rewrite the time-evolved TFD state as
\begin{equation}\label{eq:deltastate}
 |\delta\rangle   = e^{S_0} \int_{E_0 - \Delta E/2}^{E_0 + \Delta E/2} dE \, D_{\rm{Disk}}(E) \phi_{E}(\delta) |E\rangle \,, 
\end{equation}
where we identify the corresponding wavefunction for the $\delta$ eigenstates as
\begin{equation}
    \phi_{\delta}(E,\delta)= \langle E |\delta\rangle =  \frac{e^{-iE \delta}}{\sqrt{Z}} \,.
\end{equation}  
The wavefunction is purely a phase factor with an overall normalization factor of $1/\sqrt{Z}$, ensuring the normalization $\langle \delta | \delta \rangle = 1$. However, different from fixed-length states, the time-evolved TFD states only approximately satisfy the completeness relation due to
\begin{equation}\label{eq:diagonal}
\begin{split}
    \int_{-\infty}^{\infty}d\delta~|\delta\rangle\langle \delta| &= 2\pi e^{S_0} \int_{E_0-\Delta E/2}^{E_0+\Delta E/2}dE~ D_{\rm{Disk}}(E) \, \frac{ e^{S_0} D_{\rm{Disk}}(E)}{Z} \, |E\rangle\langle E| \\
    &\approx \frac{2\pi}{\Delta E} \, \hat{\mathbb{I}} \,.
\end{split}
\end{equation}

We begin with a TFD state at a particular time $t$ and examine its overlap with the time-shifted TFD state $|\delta \rangle$, namely,
\begin{equation}
\begin{split}
    \langle \delta |\text{TFD}(t)\rangle  &=  \langle \mathrm{TFD} (0) | \, e^{i\frac{\delta}{2} ( H_{\mt{L}}  +  H_{\mt{R}}) }  \, |\mathrm{TFD}(t)\rangle  \\
    &=  \langle \mathrm{TFD} (0) | \, e^{-i\frac{(t-\delta)}{2} ( H_{\mt{L}}  +  H_{\mt{R}}) }  \, |\mathrm{TFD}(0)\rangle  \,, 
\end{split}
\end{equation}
which is nothing but the return amplitude (or the survival amplitude) of the TFD state. Expressing this in terms of the (analytically continued) partition function, the return amplitude simplifies to
\begin{equation}
 \langle \delta |\text{TFD}(t)\rangle = \frac{Z(0+i(t -\delta))}{Z(0)} \,,
\end{equation}
with a vanishing inverse temperature $\beta=0$. Considering the overlap squared distributions between time-shifted TFD states, we find that the return probability is equivalent to the spectral form factor \cite{mehta2004random}, \ie
\begin{equation}
    \begin{split}
        P^{\mt{TFD}}(\delta , t):=\langle \text{TFD}(t)|\delta\rangle\langle \delta|\text{TFD}(t)\rangle =\frac{ |Z(0+i(t -\delta))|^2}{Z(0)^2}  = \mathrm{SFF} \left(|t-\delta|\right)  \,,
    \end{split}
\end{equation}
which depends only on the relative time shift $|t -\delta|$. Similar to the definition \eqref{eq:Plt} for the fixed-length state, we can express the probability in terms of the spectral two-point function, namely,
\begin{equation}\label{eq:Pdeltat}
    \begin{split}
       P^{\mt{TFD}}(\delta , t)
    &= \frac{e^{2S_0}}{Z}\int dE_1\int dE_2~e^{-i(E_1-E_2)t}
    \phi^\ast_{E_1}(\delta)  \phi_{E_2}(\delta)
    \langle D(E_1)D(E_2) \rangle \\
    &= \frac{e^{2S_0}}{Z^2}\int dE_1\int dE_2~e^{-i(E_1-E_2)(t-\delta)}    \langle D(E_1)D(E_2) \rangle \,,
\end{split}
\end{equation}
where the integral is defined within the energy window $[E_0 - \frac{\Delta E}{2}, E_0 + \frac{\Delta E}{2}]$. Alternatively, one may take the limit $E_0 = \frac{\Delta E}{2} \to \infty$ to recover the canonical ensemble, in which case the SFF reduces to the Fourier transform of the connected two-point spectral correlation function $\langle D(E_1)D(E_2) \rangle$.

\begin{figure}[t]
	\centering
\includegraphics[width=2.95in]{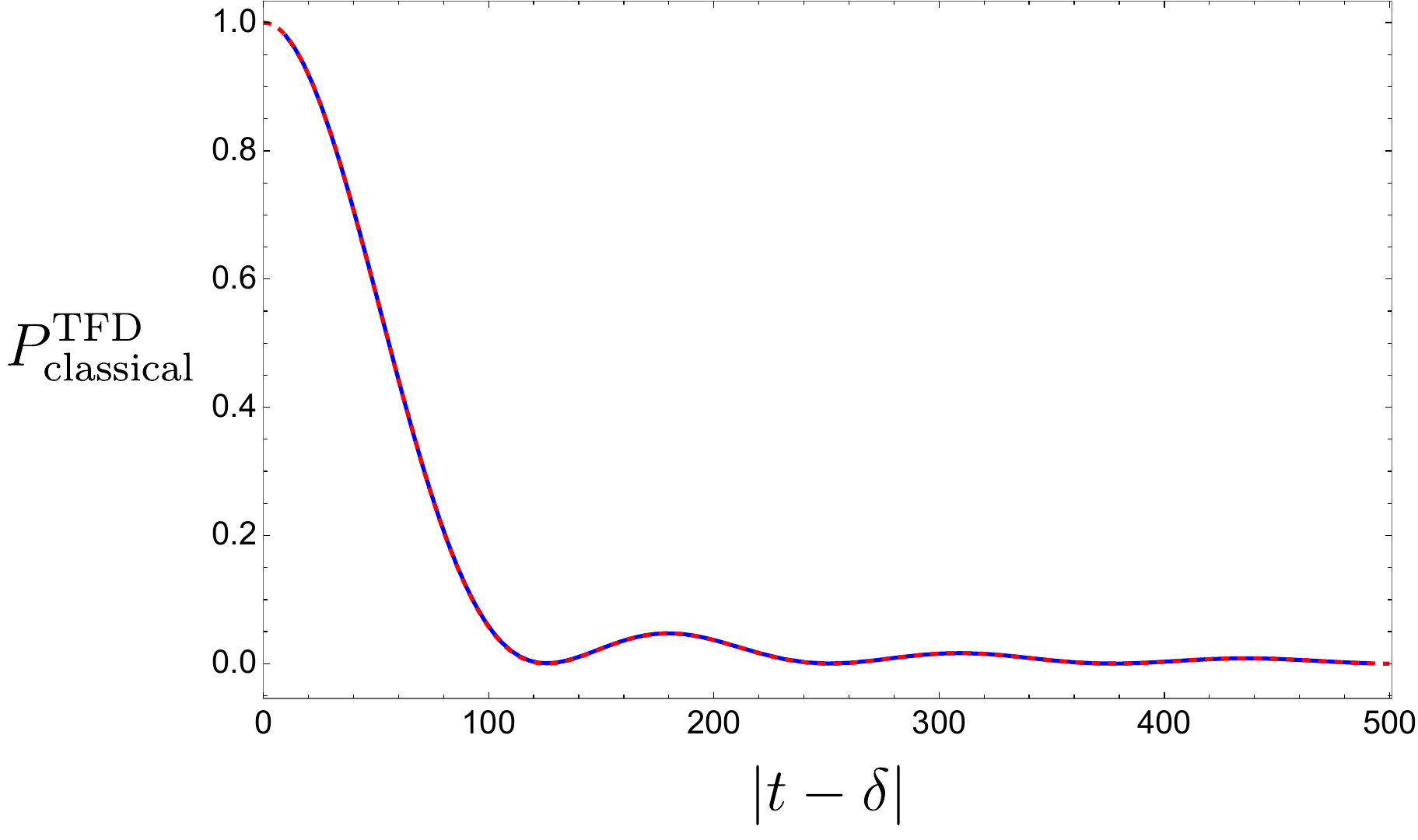}
 \includegraphics[width=2.9in]{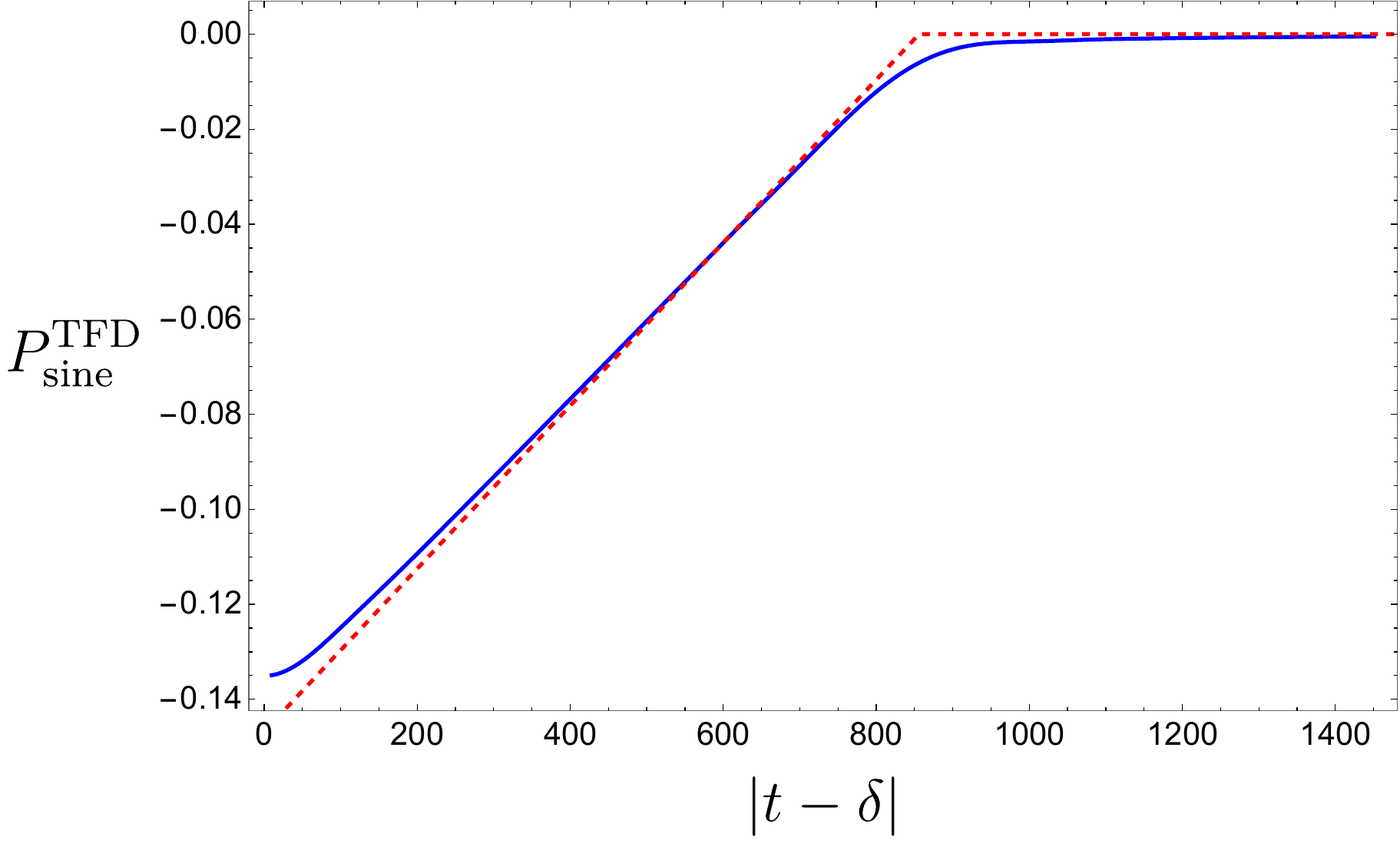}
	\caption{The classical term $P_{\rm{classical}}^{\mt{TFD}}(\delta,t)$ and the contribution $P_{\rm{sine}}^{\mt{TFD}}(\delta,t)$ from the sine kernel. The red dashed curves represent the approximate expressions derived in eq.~\eqref{eq:PTFDclassical} and eq.~\eqref{eq:PTFDsine}. The numerical results are obtained with the following parameter choices: $S_0=3, E_0=1, \Delta E=1/20$ ($T_{\mt{H}}\approx 856$).}
	\label{fig:PTFD01}
\end{figure}  

\begin{figure}[t]
	\centering
\includegraphics[width=4in]{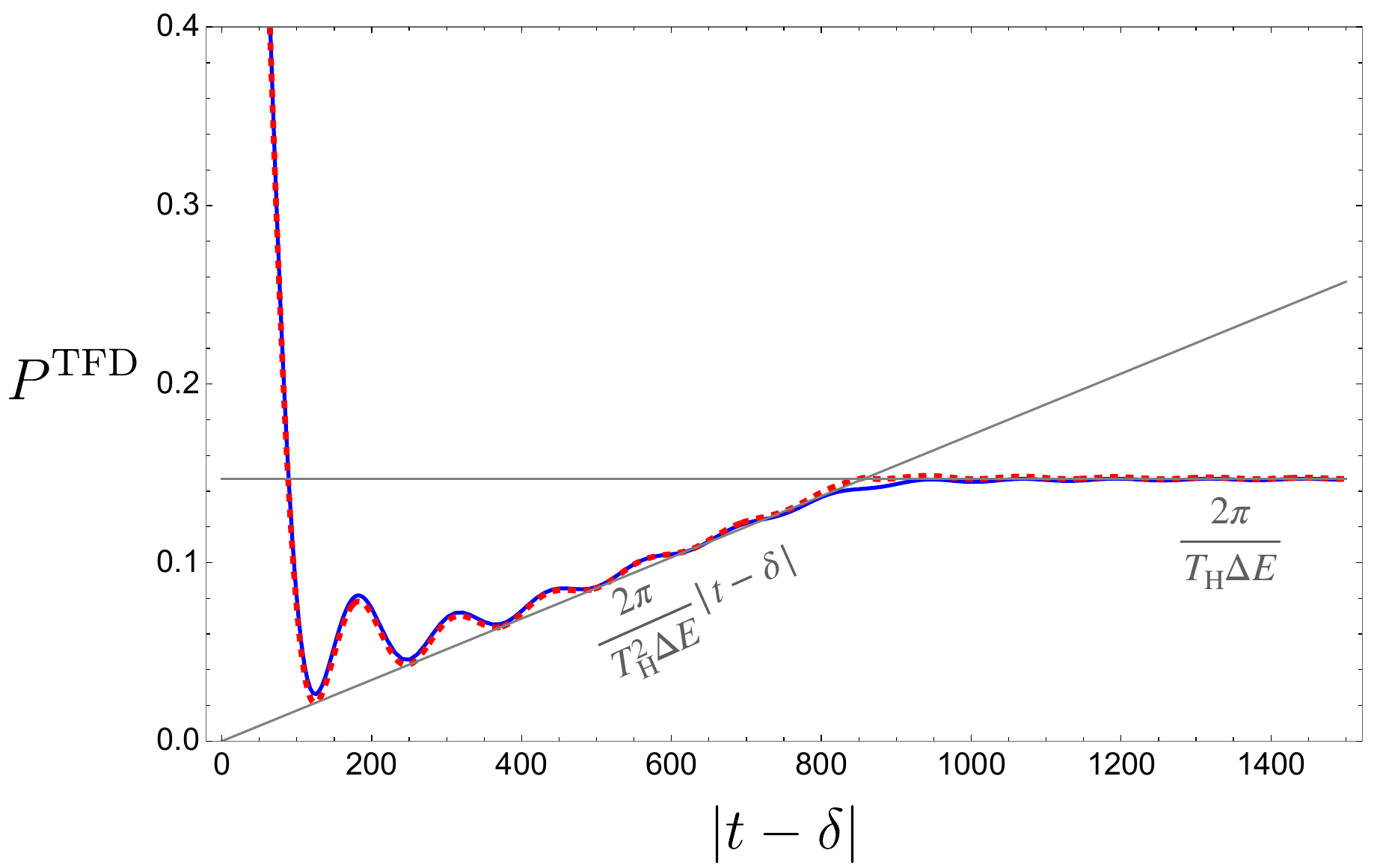}
	\caption{The transition probability between time-evolved TFD states, $ P^{\mt{TFD}}(\delta,t)$, \ie the spectral form factor $\mathrm{SFF}(|t-\delta|)$. The blue curve and red dashed curve denote the numerical integral and analytical approximation derived in eq.~\eqref{eq:PTFD}, respectively. The linear ramp and final plateau are indicated by gray lines. The numerical results use the same parameter values as in other figures, \ie $S_0=3, E_0=1$, and $\Delta E=1/20$ with $T_{\mt{H}} \approx 856$.}
	\label{fig:PTFD}
\end{figure}

Using the approximate spectral correlation \eqref{eq:sine} with the sine kernel, it is straightforward to evaluate the contribution for each term. The classical contribution corresponds to the factorized two-point function and is defined by 
\begin{equation}\label{eq:PTFDclassical}
    \begin{split}
     P^{\mt{TFD}}_{\rm{classical}}(\delta , t)  &= \frac{e^{2S_0}}{Z^2}\int dE_1\int dE_2~e^{-i(E_1-E_2)(t-\delta)}     D_{\rm{Disk}}(E_1) \cdot D_{\rm{Disk}}(E_2) \\ 
     &\approx 
   \left( \frac{2\sin\big((\delta- t)\frac{\Delta E}{2}\big)}{\left(\delta- t\right)\Delta E} \right)^2  + \mathcal{O}\left( \frac{1}{(t-\delta)^4}\right) \,,
    \end{split}
\end{equation}
where we assume the late-time regime $t-\delta \gg 1$ to obtain the final expression. The second contribution originating from the delta term obviously reduces to a constant, \ie 
\begin{equation}
    \begin{split}
   P^{\mt{TFD}}_{\rm{delta}}(\delta , t) \approx \frac{1}{e^{S_0}D_{\rm{disk}}(E_0)\Delta E} =   \frac{2\pi}{T_{\mt{H}}\Delta E} \,,
    \end{split}
\end{equation}
which corresponds to the final plateau, as shown in figure \ref{fig:PTFD}. The contribution from the sine kernel is defined in terms of 
\begin{equation}
    \begin{split}
     P_{\rm{sine}}^{\mt{TFD}}(\delta,t)
     &=-\frac{1}{Z^2}\int dE_1\int dE_2~e^{-i(E_1-E_2)(t-\delta)}    \frac{\sin^2(\pi e^{S_0}D_{\text{Disk}}(E_1)(E_1-E_2))}{\pi^2(E_1-E_2)^2} \,. 
     \end{split}
\end{equation}
Introducing the same variables \eqref{eq:variables} as before, we find that the leading late-time contribution is 
\begin{equation}\label{eq:PTFDsine}
    P_{\rm{sine}}^{\mt{TFD}}(\delta,t)   \approx 
    \frac{2\pi}{T_{\mt{H}}^2 \Delta E}  
    \times 
    \begin{cases}
    0 \,, \qquad\qquad \quad \quad \,\,  T_{\mt{H}} < |t-\delta| \\
     |t-\delta| - T_{\mt{H}} \,, \qquad T_{\mt{H}} > |t-\delta| \\
    \end{cases}\,,
\end{equation}
where we neglect subleading terms of order $\frac{1}{\Delta E^2 ( T_{\mt{H}}^2 - |t-\delta|^2)}$. 

As a summary, we conclude that the probability distribution between time-evolved TFD states, \ie the spectral form factor, is approximately given by
\begin{equation}\label{eq:PTFD}
    \begin{split}
    P^{\mt{TFD}}(\delta,t)  = \mathrm{SFF} \left(|t-\delta|\right)    \approx 
    \left( \frac{2\sin\big((\delta- t)\frac{\Delta E}{2}\big)}{\left(\delta- t\right)\Delta E} \right)^2 +
    \frac{2\pi}{T_{\mt{H}}^2 \Delta E}  \left( \min \big[T_{\mt{H}}\,, |\delta-t|  \big]  \right) \,. 
    \end{split}
\end{equation}
As shown in figure \ref{fig:PTFD}, this explicitly produces the well-known slope-dip-ramp-plateau behavior of the spectral form factor in the Gaussian Unitary Ensemble (GUE) of random matrix theory. The linear ramp is described by 
\begin{equation}
  \text{Linear ramp regime:} \qquad    P^{\mt{TFD}}(\delta,t)
  \approx 
    \frac{2\pi}{T_{\mt{H}}^2 \Delta E} |t-\delta|   \,.
\end{equation}
At late times, it reaches the final plateau:
\begin{equation}\label{eq:PTFDplateau}
\text{Plateau:} \qquad  \lim_{ |t-\delta| \gg T_{\mt{H}} } P^{\mt{TFD}} = \frac{2\pi}{ T_{\mt{H}} \Delta E} = \frac{1}{e^{S_0}D(E_0) \Delta E}  \approx \frac{1}{Z}\,,
\end{equation}
which is represented by the gray line in figure \ref{fig:PTFD}. It is associated with the partition function in terms of 
\begin{equation}
  \lim_{ |t-\delta| \gg T_{\mt{H}} } P^{\mt{TFD}} =   \lim_{\beta \to 0} \left(  \frac{Z(2\beta)}{Z(\beta)^2}\right) = \frac{1}{Z}  \,,
\end{equation}
by taking vanishing inverse temperature. The transition time between the linear ramp and late-time plateau at the leading order is determined by the Heisenberg time, \ie
\begin{equation}
|t-\delta| \approx T_{\mt{H}} =2\pi e^{S_0}D(E_0)  \,. 
\end{equation}

\section{Generating Functions of Complexity and Spectrum Probes}\label{section:probe}

In previous sections, we explored two distinct overlaps with the TFD states, namely $P(\ell, t)$ for the fixed-length states $|\ell \rangle$ and $P^{\mt{TFD}}(\delta, t)$ for the time-shifted TFD states. Although the fixed-length and time-shifted TFD states differ significantly, the probability distributions $P(\ell, t)$ and $P^{\mt{TFD}}(\delta, t)$ exhibit a similar structure. In this section, we employ these distributions to define the generating functions associated with the geodesic length and time shift. More interestingly, by taking the $\alpha \to 0$ limit, we can illustrate the time evolution of quantum complexity for a chaotic system from the universal time evolution of the generating functions of complexity.

\subsection{Time evolution of the expectation value of the length operator}\label{sec:lengthvalue}
\subsubsection{Expectation values from the probability distribution}

A primary application of the probability distribution, \ie the squared overlap, is that it provides a direct method for calculating the expectation value of specific operators associated with the corresponding eigenstates. Considering the fixed-length states $|\ell \rangle$, it is natural to define the geodesic length operator $\hat{\ell}$ as
\footnote{This construction yields the expected eigenvalue equation $\hat{\ell}|\ell \rangle = \int_{-\infty}^{\infty} \ell' \, |\ell'\rangle\langle \ell' | \ell \rangle \, d\ell = \ell |\ell \rangle$ in the canonical ensemble, where the energy spectrum is given by $E \in (0, +\infty)$ and the fixed-length states are classically orthogonal, \ie $\langle \ell | \ell'\rangle = \delta(\ell - \ell')$.}
\begin{equation}\label{eq:loperator}
\hat{\ell} = \int_{-\infty}^{\infty} \ell\, |\ell\rangle\langle \ell | \, d\ell \,.
\end{equation}
We are interested in the expectation values corresponding to the TFD state $|\text{TFD}(t)\rangle$. At the classical level, the expectation value of the length operator is defined as
\begin{equation}
\begin{split}
 \langle \hat{\ell} \rangle_{\text{classical}} &\equiv \langle \text{TFD}(t)| \hat{\ell} |\text{TFD}(t)\rangle \\
 &= \int_{-\infty}^{\infty} \Bigl( \langle \text{TFD}(t)| \ell'\rangle \langle \ell'|\text{TFD}(t)\rangle \, \ell' \Bigr) d\ell' \\
 &= \int_{-\infty}^{\infty} P_{\text{classical}}(\ell',t) \, \ell' \, d\ell' \\
 &= (2 \sqrt{E}_0 )^2 \int_{-\infty}^{\infty} P_{\text{classical}} t_\ell \, d t_\ell - \log (4 E_0) \,,
\end{split}
\end{equation}
where we have used the normalization $\int_{-\infty}^{\infty} P_{\text{classical}}(\ell, t) \, d\ell = 1$ to obtain the last line. More explicitly, by employing the approximate probability distribution, one can perform the implicit integral and obtain
\begin{equation}
(2 \sqrt{E}_0 )^2 \int^{t_\ell} P_{\text{classical}} t_\ell' \, d t_\ell' \quad \overset{t_\ell \to \infty}{\sim} \quad \sqrt{E_0} t + \frac{2\sqrt{E}_0}{\pi \Delta E} \log \left( \frac{t_\ell \Delta E}{2} \right) \,.
\end{equation}
However, a straightforward evaluation of the integral reveals that $\langle \hat{\ell} \rangle_{\text{classical}}$ diverges due to a logarithmic divergence arising from the region $\ell \sim \infty$.

At the quantum level, the expectation value of the geodesic length operator is defined in terms of the probability density $P(\ell, t)$ as
\begin{equation}
\langle \hat{\ell} \rangle = \langle \text{TFD}(t)| \hat{\ell} |\text{TFD}(t)\rangle = \int_{-\infty}^{\infty} P(\ell,t) \, \ell \, d\ell \,.
\end{equation}
The divergence of the expectation value $\langle \hat{\ell} \rangle$ is due to the presence of infinite fixed-length states $|\ell \rangle$ with arbitrarily large $\ell$. From the perspective of the quantum probability $P_{\mathrm{quantum}}$, the inclusion of infinite fix-length states renders the probability distribution non-normalizable. Indeed, using the probability distribution derived before, we can find
\begin{equation}
\begin{split}
 \int_{-\infty}^{\infty} P(\ell,t) \, d\ell &= \int_{-\infty}^{\infty} \Bigl( P_{\mathrm{classical}} + P_{\mathrm{quantum}} \Bigr) d\ell \\
 &= 1 + \int_{-\infty}^{\infty} P_{\mathrm{quantum}} \, d\ell \to \infty \,,
\end{split}
\end{equation}
where the quantum contribution diverges due to infinite contributions from the plateau regime characterized by
\begin{equation}
\lim_{t_{\ell} \gg T_{\mt{H}} + t} P(\ell,t) \approx \frac{1}{T_{\mt{H}}\sqrt{E_0}} \,.
\end{equation}
This divergence is also reflected in the matrix elements of the geodesic length operator, \eg
\begin{equation}
\begin{split}
\langle E_i | \, \hat{\ell} \, | E_j \rangle = \int_{-\infty}^{\infty} \ell \, \langle E_i | \ell' \rangle \langle \ell' | E_j \rangle \, d\ell' \quad \longrightarrow \infty \,.
\end{split}
\end{equation}
Furthermore, we emphasize that the divergent component is time-independent because of 
\begin{equation}
\begin{split}
 \frac{d}{dt}\int_{-\infty}^{\infty} P(\ell,t) \, d\ell = 0 \,.
\end{split}
\end{equation}
It follows from the explicit evaluation of the integral in eq.~\eqref{eq:integrals02} and the identity
\begin{equation}
\partial_t \Bigl( \delta(E_i - E_j)e^{-i(E_i-E_j)t} \Bigr) = 0 \,.
\end{equation}

Clearly, the divergent sum of probabilities contradicts the finiteness of the total dimension of the Hilbert space. All these divergences can be traced back to the fact that the infinite set of fixed-length states $|\ell\rangle$, when used as a basis for a finite Hilbert space, is over-complete. Specifically,
\begin{equation}\label{eq:suml02}
\begin{split}
 \int_{-\infty}^{\infty} d\ell \, |\ell\rangle \langle \ell| &= e^{2S_0} \int dE_1 \int dE_2~ \langle D(E_1)D(E_2) \rangle \, |E_1\rangle\langle E_2| \int_{-\infty}^{\infty} d\ell \, \psi_{E_1}(\ell)\psi_{E_2}(\ell) \\
 &= e^{S_0} \int dE ~ \frac{\langle D(E_1)D(E_2) \rangle}{D_{\mathrm{Disk}}(E)} \, |E\rangle\langle E| \\
 &\ne \hat{\mathbbm{1}} \,.
\end{split}
\end{equation}
This issue has been investigated and resolved in \cite{Miyaji:2024ity} by constructing a discrete spectrum for the eigenvalues $\ell$. See also \cite{Banerjee:2024fmh} for a recent discussion on the discrete energy spectrum and the factorization puzzle.

Naively, one might expect that a regularization procedure can be implemented to define a regulated expectation value for the length operator $\hat{\ell}$. For instance, a simple approach is to introduce a {\it cut-off} on the spectrum of the geodesic length, \ie by considering $\ell \in (-\ell_{\mathrm{cut}}, \ell_{\mathrm{cut}})$. Taking the expectation value at $t=0$ as a regulator:
\begin{equation}
\langle \hat{\ell} \rangle_{\mathrm{reg}} := \langle \hat{\ell} \rangle - \langle \hat{\ell} \rangle|_{t=0}\,,
\end{equation}
one can remove any {\it time-independent divergence}. This regularization method has been explicitly used in \cite{Iliesiu:2021ari}. In light of the fact that the divergent contribution originates from the regime $\ell \sim \infty$, one may also regularize the length expectation value by introducing a damping factor $e^{-\alpha \ell}$. After taking the limit $\alpha \to 0$, one obtains the regularized length expectation 
\begin{equation}
\begin{split}
\langle \hat{\ell} \rangle_{\alpha \to 0} 
 &= \lim_{\alpha \to 0} \left( - \frac{d}{d \alpha}
\int_{-\infty}^{\infty} P(\ell',t) e^{-\alpha \ell'} \, d\ell' \right) \\
&=\lim_{\alpha \to 0} \left( \int_{-\infty}^{\infty} P(\ell',t) e^{-\alpha \ell'} \ell' \, d\ell' \right) \,.
\end{split}
\end{equation}
Note that the exponential factor $e^{-\alpha \ell}$ does not introduce any new divergence from the negative length region because the wavefunction (or the probability density) is doubly exponentially suppressed for
\begin{equation}
( \psi_E (\ell))^2 \overset{\ell \ll -1}{\quad \sim \quad} e^{-4 e^{-\ell/2} + \ell } \,.
\end{equation}
As a result, it effectively overcomes the exponential factor $e^{-\alpha \ell}$ for any positive $\alpha$.

While infinitely many regularization methods can be devised, due to the ill-defined nature of the length operator $\hat{\ell}$ associated with a continuous spectrum (as shown in eq.~\eqref{eq:loperator}), our view is that the regulated expectation value is not physically meaningful—even though certain divergences can be removed via regularization. A simple indication of this is that the regulated length depends on the choice of regularization scheme. For example, one has\footnote{The $\alpha$-divergence appearing in the canonical ensemble as $\alpha \to 0$ is formulated as a $\frac{1}{\alpha^2}$ term. The additional divergence $\log (\alpha t)$ originating from the classical part is characteristic of the microcanonical ensemble with $\beta = 0$.}
\begin{equation}\label{eq:alphatozero}
\langle \hat{\ell} \rangle_{\mathrm{reg}} \ne \langle \hat{\ell} \rangle_{\alpha \to 0} = \lim_{\alpha \to 0} \left( \#\log \left(\frac{1}{\alpha \, t}\right) + \frac{\#}{\alpha^2} + \text{time dependent part} \right) \,.
\end{equation}
Moreover, the limit $\alpha \to 0$ is ambiguous because it does not commute with the late-time limit, \ie 
\begin{equation}
\lim_{\alpha \to 0} \lim_{t \to \infty} \int_{-\ell}^{\ell} P(\ell',t) e^{-\alpha \ell'}\, d\ell' \ne \lim_{t \to \infty} \lim_{\alpha \to 0} \int_{-\ell}^{\ell} P(\ell',t) e^{-\alpha \ell'}\, d\ell' \,,
\end{equation}
due to the presence of an $\alpha \log (t)$ term. In this sense, the regularization procedure merely shifts the $t_\ell$-divergence from $t_\ell \to \infty$ to a corresponding $\alpha$-divergence as $\alpha \to 0$.

Rather than attempting to regularize the expectation value of the ill-defined length operator $\hat{\ell}$, we focus on the well-defined {\it generating function}\footnote{Note that the parameter $\alpha$ carries the dimensions of mass/energy. It is similar to conformal dimension of the scalar operator showing in the geodesic approximation \eqref{eq:geodesic}.} for the geodesic length, \viz 
\begin{equation}\label{eq:definegenerating}
\begin{split}
\langle e^{-\alpha \ell} \rangle 
:= \langle \widehat{e^{-\alpha \ell}} \rangle := \int_{-\infty}^{\infty} P(\ell,t) e^{-\alpha \ell} \, d\ell \,.
\end{split}
\end{equation}
Indeed, the generating function encapsulates all the physical information about the expectation value of the length operator and its time evolution. This generating function corresponds to the expectation value of a particular operator defined by
\begin{equation}
\widehat{e^{-\alpha \ell}} = \int_{-\infty}^{\infty} e^{-\alpha \ell}\, |\ell\rangle\langle \ell| \, d\ell \,.
\end{equation}
Note that it is distinct from the exponential of the length operator, namely 
\begin{equation}
e^{-\alpha \hat{\ell}} := \hat{\mathbbm{1}} - \alpha \hat{\ell} + \frac{\alpha^2}{2} \hat{\ell}^2 + \cdots \,,
\end{equation}
because the orthogonality condition, \ie $\langle \ell | \ell' \rangle = \delta(\ell - \ell')$, is lost after including quantum corrections. We refer to $\langle e^{-\alpha \ell} \rangle$ as the generating function.
This is because the (regularized) length expectation value and its time evolution can be derived from the generating function. This will be illustrated in the next subsection \ref{sec:expl}.

To close this subsection, we now make explicit connections to previous studies \cite{Yang:2018gdb,Saad:2019pqd,Iliesiu:2021ari}. First, note that the matrix element associated with the length operator $\hat{\ell}$ defined in eq.~\eqref{eq:loperator} takes the form of
\begin{equation}
\begin{split}
\langle E_i | \widehat{ e^{-\alpha \ell } }  | E_j \rangle = \int_{-\infty}^{\infty} e^{-\alpha \ell' } \langle E_i | \ell' \rangle \langle \ell' | E_j \rangle \, d\ell' \,,
\end{split}
\end{equation}
for any two energy eigenstates $|E_i \rangle$. By recalling the definition of the wavefunction $\psi_E(\ell)$ in eq.~\eqref{eq:definepsiEl} for the fixed-length states and by performing the integrals involving modified Bessel functions, one obtains \cite{Yang:2018gdb,Saad:2019pqd,Iliesiu:2021ari}
\begin{equation}\label{eq:E1expalE2}
\begin{split}
\langle E_i |\widehat{ e^{-\alpha \ell } } | E_j \rangle &\equiv \int_{-\infty}^{\infty} e^{-\alpha \ell}\psi_{E_i}(\ell)\psi_{E_j}(\ell) \, d\ell \\
&\quad = \frac{2}{e^{S_0}} \frac{ |\Gamma\left(\alpha + i (\sqrt{E_i} +\sqrt{E_j})\right)|^2 \, |\Gamma\left(\alpha + i (\sqrt{E_i}-\sqrt{E_j})\right)|^2 }{ \Gamma (2\alpha)} \,.
\end{split}
\end{equation}
Correspondingly, the regularized matrix element associated with the length operator $\hat{\ell}$ is derived as
\begin{equation}\label{eq:E1lE2}
\begin{split}
\left( \langle E_i | \, \hat{\ell} \, | E_j \rangle \right)\Big|_{\alpha \to 0} 
&= - \lim_{\alpha \to 0} \partial_\alpha \left( \langle E_i |\widehat{ e^{-\alpha \ell } } | E_j \rangle \right) \\
&\quad = \frac{-8 \pi^2 e^{-S_0}}{(E_i-E_j) \left(\cosh \left(2 \pi \sqrt{E_i}\right)-\cosh \left(2 \pi \sqrt{E_j}\right)\right)} \,.
\end{split}
\end{equation}
It is important to emphasize that the order of limits is crucial: the $\alpha \to 0$ limit must be taken at the end of the calculation. To explicitly reveal the divergence similar to that in eq.~\eqref{eq:alphato0}, we can consider the diagonal matrix element:
\begin{equation}\label{eq:ElE}
\lim_{\alpha \to 0} \langle E |\widehat{ e^{-\alpha \ell } } | E \rangle = \frac{1}{2\pi e^{S_0} D_{\rm Disk}(E) \sqrt{E}} \frac{1}{\alpha} + \mathcal{O}(\alpha^0) \,,
\end{equation}
which originates from the contact term $\delta (E_i - E_j)$ in the connected spectral correlation function. 
Summing the contributions over the energy spectrum yields the same result as $\langle \hat{\ell} \rangle_{\alpha \to 0}$ with the identical time-independent divergence at $\alpha \to 0$. This divergence arises because the infinite-long plateau of the probability $P(\ell, t)$ \eqref{eq:finaltl} for the large length is dominated by the contribution from the contact term $\delta (E_i -E_j)$ in the connected spectral correlation function.     

\subsubsection{From Linear Growth to the Late-Time Plateau}\label{sec:ldot}

\begin{figure}[t]
	\centering
	\includegraphics[width=3.5in]{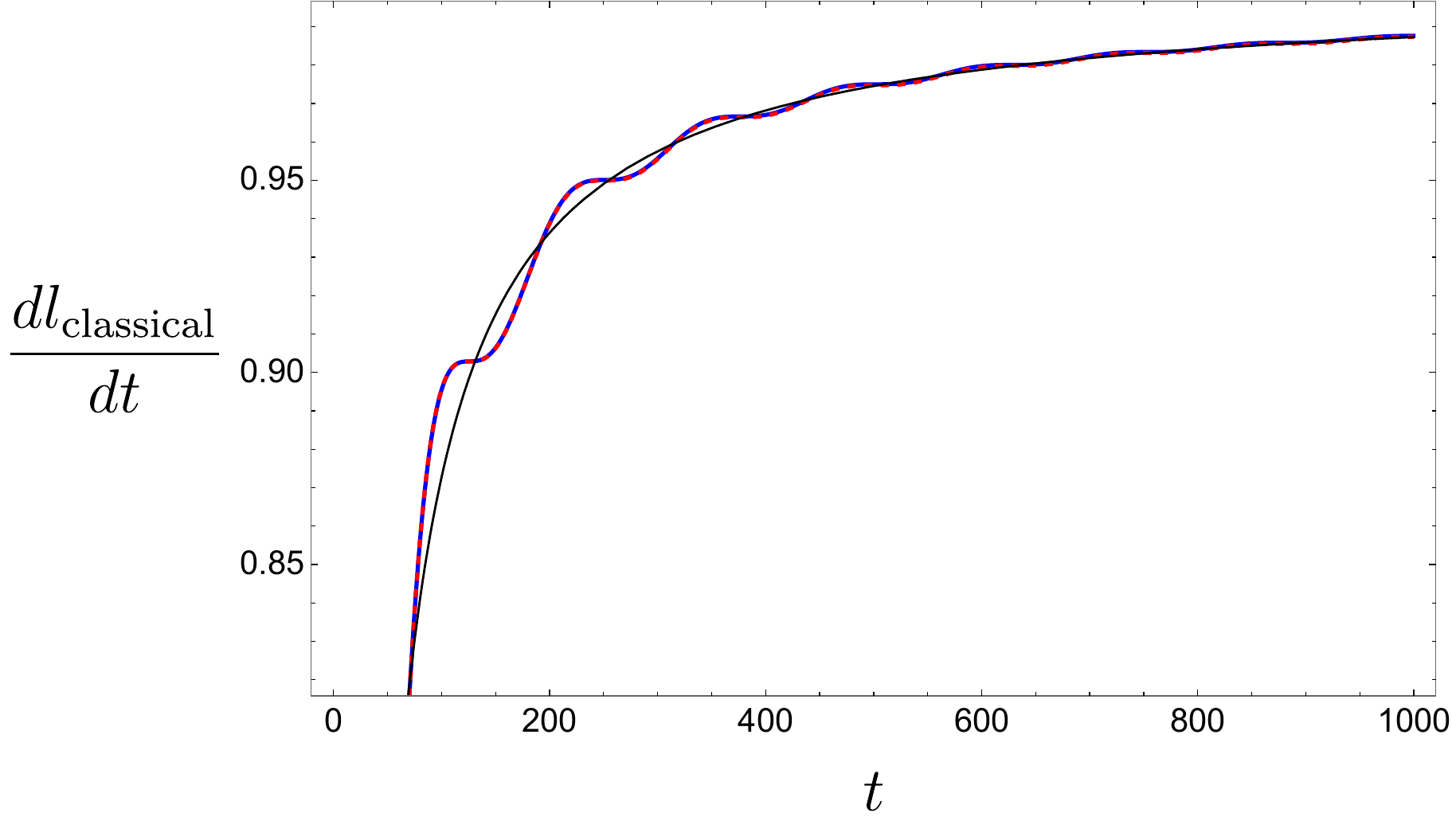}
	\caption{Time growth rate of the classical geodesic length (rescaled by $\frac{1}{2\sqrt{E_0}}$). The blue and red curves represent the numerical result obtained from the definition in eq.~\eqref{eq:dldtc} and the analytical approximation derived in eq.~\eqref{eq:dldtcapp}, respectively. The thin black curve represents the leading approximation $1 - \frac{2}{\pi t \Delta E}$.}
	\label{fig:dldtclassical}
\end{figure}

Before introducing the generating function $\langle e^{-\alpha \ell} \rangle$, we first demonstrate why the length, or equivalently the volume of a black hole interior, cannot increase linearly forever, \viz it saturates to a plateau after the Heisenberg time $T_{\mt{H}} \sim e^{S_0}$. Instead of evaluating the expectation value of the ill-defined geodesic length operator, it is instructive to consider its time derivative, namely 
\begin{equation}
\frac{d \langle \hat{\ell} \rangle}{dt} = \frac{d}{dt} \int_{-\infty}^{\infty} P(\ell,t) \, \ell \, d\ell = \int_{-\infty}^{\infty} \partial_t P(\ell,t) \, \ell \, d\ell \,,
\end{equation}
which is free from divergence issues since the contributions from the infinite-length region do not affect the time evolution. More importantly, we would like to highlight that all information about the time evolution has been encoded in the squared overlap $P(\ell,t)$. 

First, we examine the time evolution of the geodesic length in the classical limit. In this case, one can use the classical probability distribution derived in eq.~\eqref{eq:Pclassical} to obtain the time derivative as follows:
\begin{equation}\label{eq:dldtc}
\frac{d \langle \hat{\ell} \rangle}{dt}\bigg|_{\mathrm{classical}} = (2\sqrt{E_0})^2 \int_{-\infty}^{\infty} \left( t_\ell \, \partial_t P_{\text{classical}} \right) dt_\ell \,.
\end{equation}
Here, we have changed the length variable $\ell$ to $t_\ell = \frac{\ell + \log(4E_0)}{2\sqrt{E_0}}$ and used
\begin{equation}\label{eq:sumdPdt}
\int_{-\infty}^{\infty} \partial_t P_{\mathrm{classical}} \, d\ell = 0 \,,
\end{equation}
which follows from the completeness relation of the states $|\ell\rangle$ at the classical level. One expects that the leading contribution from eq.~\eqref{eq:dldtc} exhibits a simple linear growth, corresponding to the growth of the geodesic length in a classical AdS$_2$ black hole spacetime, \ie  
\begin{equation}
\ell_{\text{classical}} \sim 2\sqrt{E_0}\, t \,.
\end{equation}
By employing the approximate classical probability $P_{\mathrm{classical}}$ derived in eq.~\eqref{eq:Pclassical}, one can explicitly perform the integral to obtain
\begin{equation}\label{eq:dldtcapp}
\begin{split}
\frac{d \langle \hat{\ell} \rangle}{dt}\bigg|_{\mathrm{classical}} &\approx (2\sqrt{E_0})^2 \int_{0}^{\infty} \left( t_\ell \, \partial_t P_{\text{classical}} \right) dt_\ell \\
&= 2\sqrt{E_0} \left( \frac{2}{\pi}\,\mathrm{Si}(t \Delta E) + \frac{2 (\cos(\Delta E t)-1)}{\pi t \Delta E} \right) \,,
\end{split}
\end{equation}
as illustrated in figure \ref{fig:dldtclassical}. Similarly, we have neglected the contributions from the negative $t_\ell$ regime\footnote{The lower limit of the $t_\ell$ integration is chosen to be zero for clarity. Choosing any value of order one would not affect the final result, given the doubly exponential suppression of the wavefunction and probability in that region, as discussed in eq.~\eqref{eq:psinegative}.}. Furthermore, by focusing on the regime $\Delta E\, t \gg 1$ and neglecting short-time effects, the time derivative simplifies to
\begin{equation}
\frac{d \langle \hat{\ell} \rangle}{dt}\bigg|_{\mathrm{classical}} \approx 2\sqrt{E_0} \left( 1 - \frac{2}{\pi t \Delta E} + \cdots \right) \,.
\end{equation}
As expected,we find that the time derivative of the length expectation is approximately reduced to a constant $2\sqrt{E_0}$. It implies a linear growth of the geodesic length after early times, \ie 
\begin{equation}\label{eq:applclassical}
\langle \hat{\ell} \rangle\big|_{\mathrm{classical}} \sim 2\sqrt{E_0}\, t - \frac{4\sqrt{E_0}}{\pi \Delta E}\log(t) + \text{constant} \,,
\end{equation}
where the divergent constant part has been omitted. An equivalent result can be also obtained by first evaluating the regularized quantity
\begin{equation}
-\partial_t \partial_\alpha \int_{0}^{\infty} \left( e^{-\alpha \ell}\, P_{\text{classical}} \right)d\ell = \int_{0}^{\infty} \left( e^{-\alpha \ell}\, \ell\, \dot{P}_{\text{classical}} \right)d\ell \,,
\end{equation}
and then taking the following limit
\begin{equation}
\lim_{\alpha \to 0} \int_{0}^{\infty} \left( e^{-\alpha \ell}\, \ell\, \dot{P}_{\text{classical}} \right)d\ell = \frac{d \langle \hat{\ell} \rangle}{dt}\bigg|_{\mathrm{classical}} \,.
\end{equation}

\begin{figure}[t]
	\centering
	\includegraphics[width=5.9in]{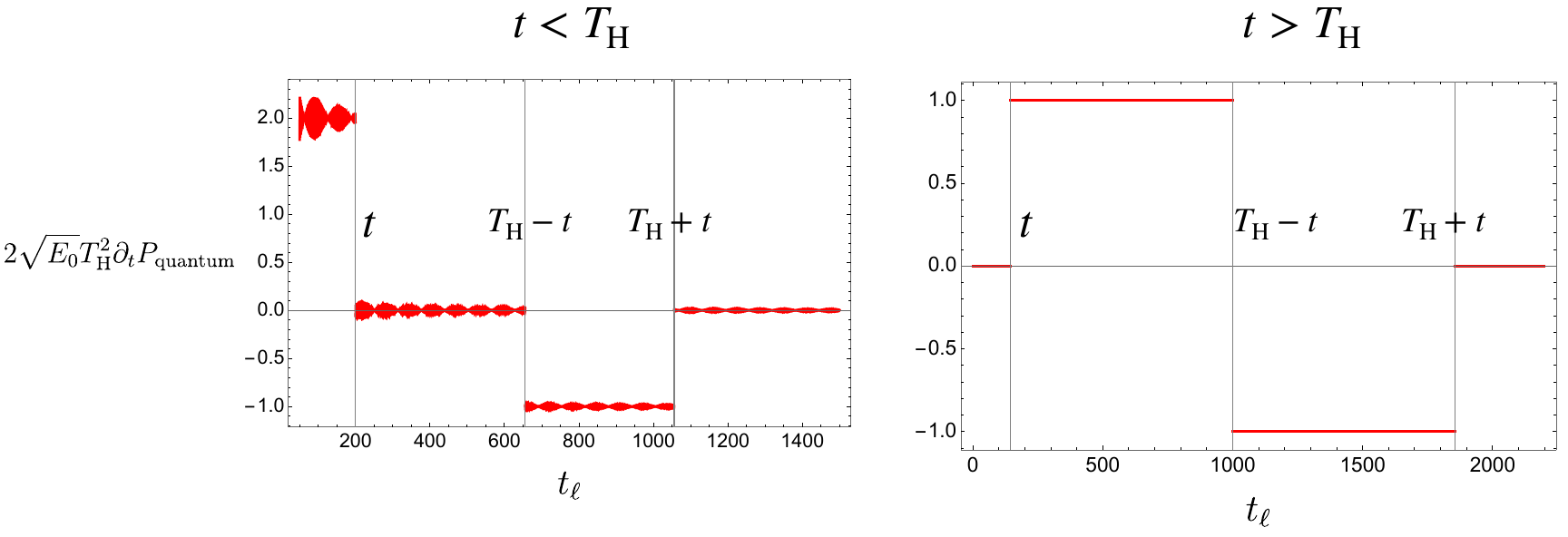}
	\caption{Time derivative of the quantum probability $P_{\rm quantum}$ for $t< T_{\mt{H}}$ and $t> T_{\mt{H}}$, respectively. The parameters for both numerical plots are the same as those used in the other figures.}
	\label{fig:dPdtquantum}
\end{figure}

\begin{figure}[t]
	\centering
	\includegraphics[width=5.9in]{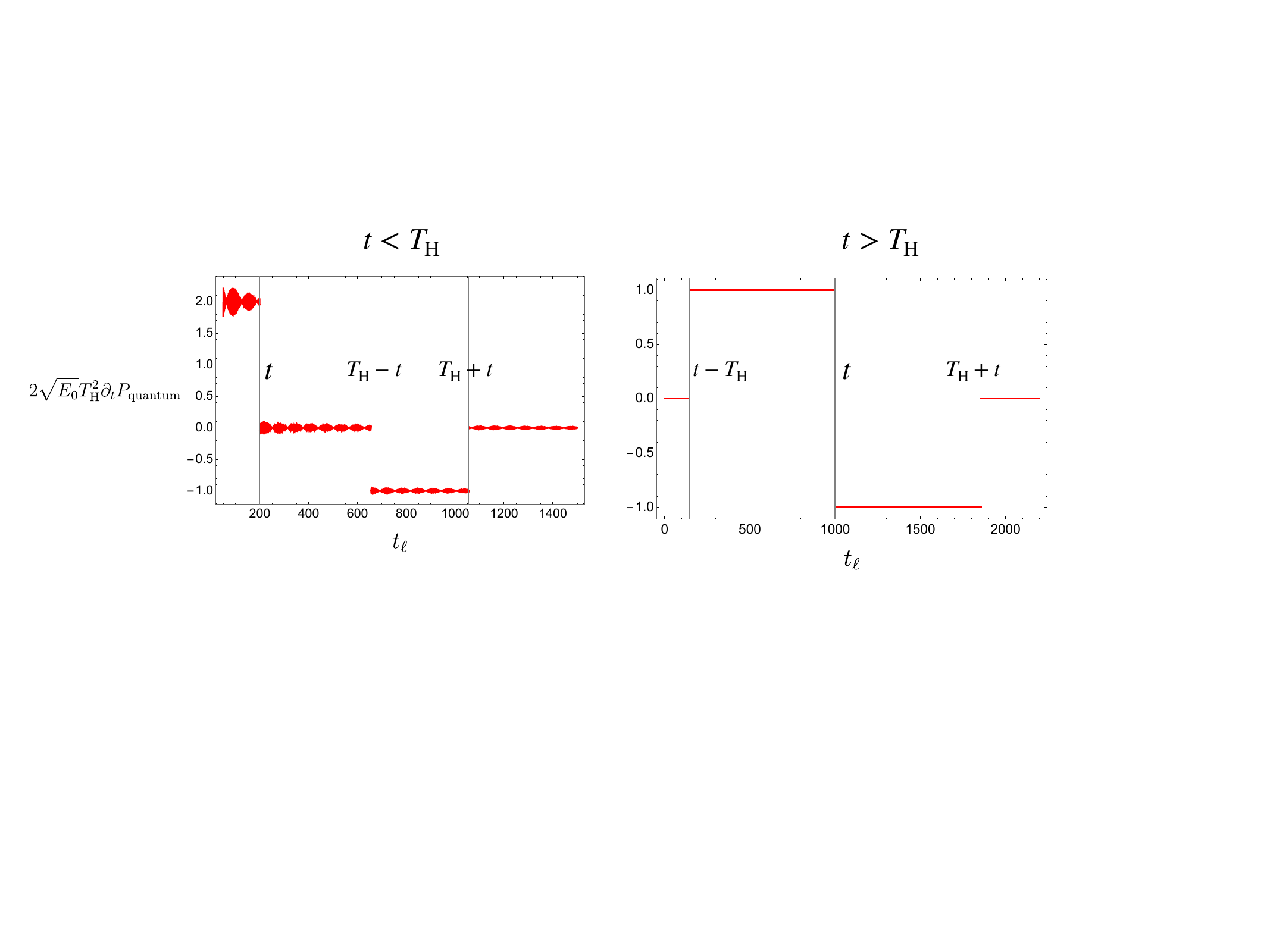}
	\caption{Time evolution of the time derivative of the length expectation value and its integration, as derived in eqs.~\eqref{eq:dLdtquantum} and \eqref{eq:Lvalues}, respectively. The constant term in the integration has been omitted.}
	\label{fig:Lvalues}
\end{figure}

However, we expect that the wormhole size of AdS black hole spacetime should not grow forever due to the finiteness of the Hilbert space dimension. In other words, after including the quantum contributions, \ie $P_{\mathrm{quantum}}$, the linear growth of the geodesic length should be exactly canceled at late times. To demonstrate this cancellation explicitly, we evaluate the time derivative by including the quantum part:
\begin{equation}
\frac{d \langle \hat{\ell} \rangle}{dt}\bigg|_{\mathrm{quantum}} = \int_{-\infty}^{\infty} \left( \partial_t P_{\mathrm{quantum}} \right) \, \ell \, d\ell = (2\sqrt{E_0})^2 \int_{-\infty}^{\infty} \left( \partial_t P_{\mathrm{quantum}} \right) \, t_\ell \, dt_\ell \,,
\end{equation}
where we have used a similar relation as eq.~\eqref{eq:sumdPdt} for the quantum contribution part.
By employing the approximate probability $P_{\rm quantum}$ derived in eq.~\eqref{eq:Pquantum}, one can find that its time derivative, $\partial_t P_{\rm quantum}$, exhibits an extremely simple structure, as illustrated in figure~\ref{fig:dPdtquantum}. The non-vanishing contributions of $\partial_t P_{\rm quantum}$ can be expressed as
\begin{equation}
 (2\sqrt{E_0} T^2_{\mt{H}}) \partial_t P_{\rm quantum} \approx    
   \begin{cases}
  \left\{\begin{array}{lr}
        2\,, & \quad    t_{\ell}  <  \left[ 0, t \right]  \\
        -1\,, &  \quad  t_\ell \in \left[ T_{\mt{H}}-t, T_{\mt{H}}+t \right] \\
        \end{array}\right\}  \quad \text{with} \quad t < T_{\mt{H}} \,, \\[1em]
      \left\{\begin{array}{lr}
       +1\,, & \qquad   t_\ell \in \left[t- T_{\mt{H}}, t \right] \\      
       -1\,, & \qquad   t_\ell \in \left[t, t+T_{\mt{H}} \right] \\
        \end{array}\right\}  \quad \text{with} \quad t >  T_{\mt{H}} \,. \\
   \end{cases}
\end{equation}
Integration over $t_\ell$ then yields
\begin{equation}\label{eq:dLdtquantum}
 \frac{d \langle \hat{\ell} \rangle}{dt}\bigg|_{\mathrm{quantum}}  
 \approx 
 \begin{cases}
  2\sqrt{E_0} \left( -1 + \left( 1 - \frac{t}{T_{\mt{H}}} \right)^2 \right) + \mathcal{O}\left(\frac{1}{T_{\mt{H}}^2}\right) \,, \quad (t < T_{\mt{H}})  \\[1em]
  - 2\sqrt{E_0} + \mathcal{O}\left(\frac{1}{T_{\mt{H}}^2}\right) \,, \quad (t > T_{\mt{H}}) \,.
 \end{cases}
\end{equation}
Consequently, the linear growth of the length expectation value arising from the classical contribution is precisely canceled by the leading quantum corrections. In other words, the expectation value of the length operator saturates after the Heisenberg time $T_{\mt{H}}$, reaching a plateau. By performing a time integration, the regularized length expectation value can be derived as
\begin{equation}\label{eq:Lvalues}
 \langle \hat{\ell}\rangle_{\rm reg} \approx \text{Constant} + \int_0^t \left( \frac{d \langle \hat{\ell} \rangle}{dt} \right) dt \approx 
\begin{cases}
  2\sqrt{E_0}\, t \left( 1 - \frac{t}{T_{\mt{H}}} + \frac{t^2}{3T^2_{\mt{H}}} \right) \,, \quad (t < T_{\mt{H}})  \\[1em]
  \frac{2\sqrt{E_0}}{3}\, T_{\mt{H}} \,, \quad (t > T_{\mt{H}}) \,,
 \end{cases}
\end{equation}
where the constant corresponding to the length expectation value at $t=0$, \ie $\langle \hat{\ell}\rangle|_{t=0}$, is associated with the regularization of $\hat{\ell}$. It is important to note that the logarithmic term $-\frac{4\sqrt{E_0}}{\pi \Delta E}\log(t)$ arising from the classical contribution has been ignored here, as it becomes significant only on a timescale of order $e^{e^{S_0}}$, which far exceeds the regime that can be probed by this calculation (see section \ref{section:comment} for further discussion). For completeness, figure~\ref{fig:Lvalues} displays the time evolution of both the time derivative $\frac{d \langle \hat{\ell} \rangle}{dt}$ and its integration, \ie the regularized expectation value of the geodesic length operator $\langle \hat{\ell}\rangle_{\rm reg}$.
 
\subsection{Generating Function for the Length Expectation}\label{sec:expl}

\begin{figure}[t]
	\centering
	\includegraphics[width=4in]{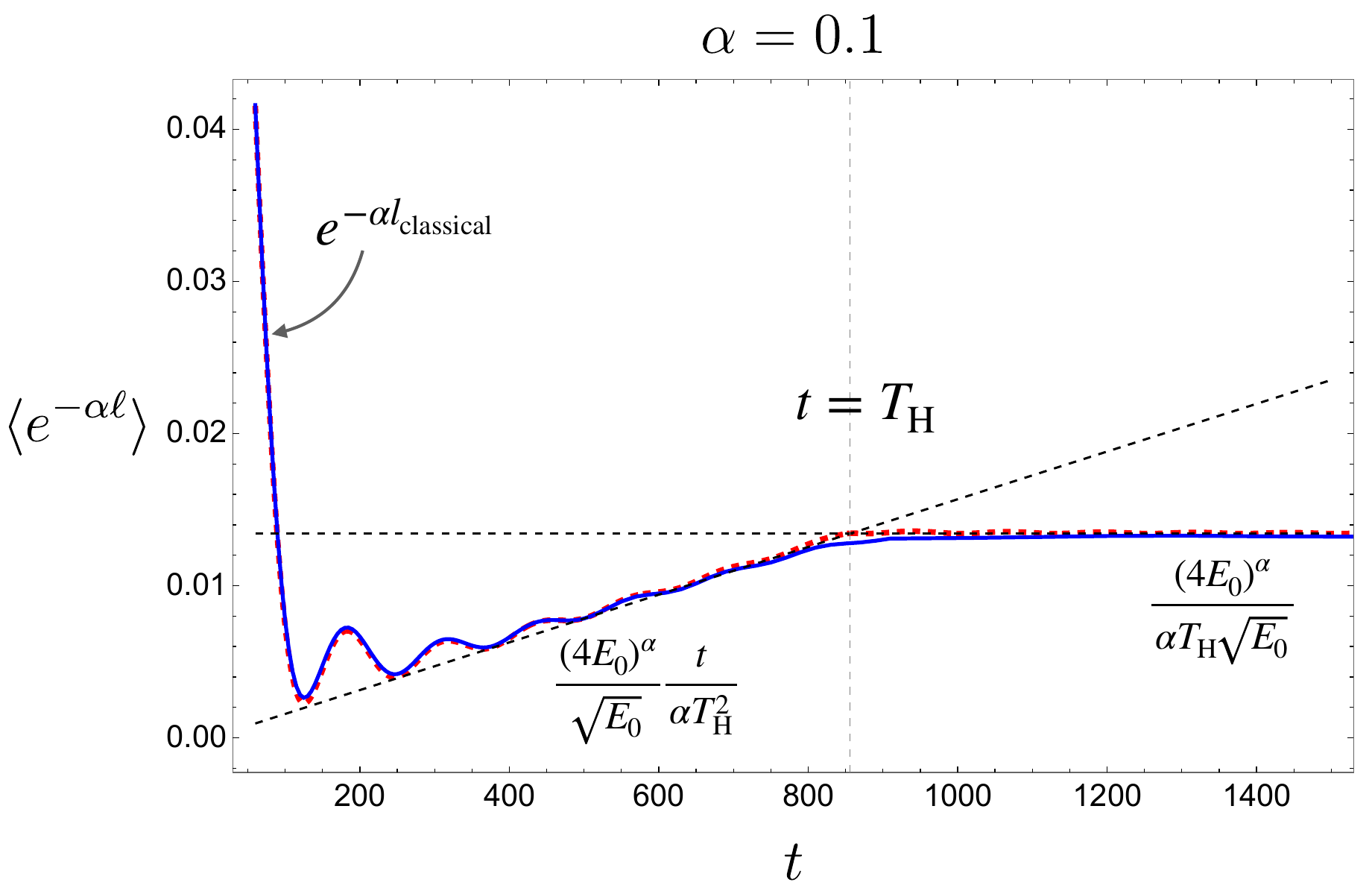}
	\caption{Characteristic time evolution of the generating function $\langle e^{-\alpha \ell } \rangle$. This function with a generic $\alpha$ exhibits a universal slope-ramp-plateau structure analogous to the spectral form factor. The blue and red curves represent, respectively, the numerical results obtained from the precise wavefunction and the corresponding analytical approximation. The plot is generated by using the same parameters as in the previous figures with $T_{\mt{H}} \approx 856$ and $\alpha = 1/10$.}
	\label{fig:expal}
\end{figure}

One interesting result we would like to highlight is that the time evolution of such as the geodesic length operator $\hat{\ell}$ is universal and is fully encoded in its generating function $\langle e^{-\alpha \ell} \rangle$. Moreover, this generating function bears a striking resemblance to the spectral form factor, which is characterized by the well-known slope-ramp-plateau structure (see figure \ref{fig:PTFD} and figure \ref{fig:expal}). This behavior captures the chaotic nature of the system, distinguishing it from integrable systems.

In this section, we first illustrate the universal time evolution of the generating function for geodesic lengths, namely
\begin{equation}
\begin{split}
 \langle e^{-\alpha \ell } \rangle   
=  \int_{-\infty}^{\infty} P(\ell,t)  e^{-\alpha \ell } \, d \ell  
= \langle e^{-\alpha \ell } \rangle \big|_{\rm classical}  
+ \langle e^{-\alpha \ell } \rangle \big|_{\rm quantum} \,.
\end{split}
\end{equation}
This quantity is well defined for any finite $\alpha$ and was introduced in the previous section as a potential regularization for the expectation value of the ill-defined geodesic length operator. We refer to it as the generating function because it gives rise to the expectation value of the length operator via
\begin{equation}\label{eq:fromexpaltol}
\lim_{\alpha \to 0} \left( - \partial_\alpha \langle e^{-\alpha \ell } \rangle \right) = \text{Divergent Constant} + \langle \ell \rangle_{\rm reg} \,.
\end{equation}
In the regime where $\alpha T_{\mt{H}} \ll 1$, the slope-ramp-plateau behavior of $\langle e^{-\alpha \ell} \rangle$ simplifies because the linear ramp part effectively disappears. The inverse of this simplified time evolution of the generating function then yields the time evolution of the regularized length expectation value—that is, it has a long-time linear growth and then transits to a late-time plateau. More explicitly, taking the limit $\alpha \to 0$ of the derivative of the generating function reproduces the time evolution of $\langle \ell \rangle_{\rm reg}$, as discussed in subsection \ref{sec:ldot}, by evaluating the finite expectation value of its time derivative.

\subsubsection{Universal time evolution of the generating function}

\begin{figure}[t]
	\centering
	\includegraphics[width=5.9in]{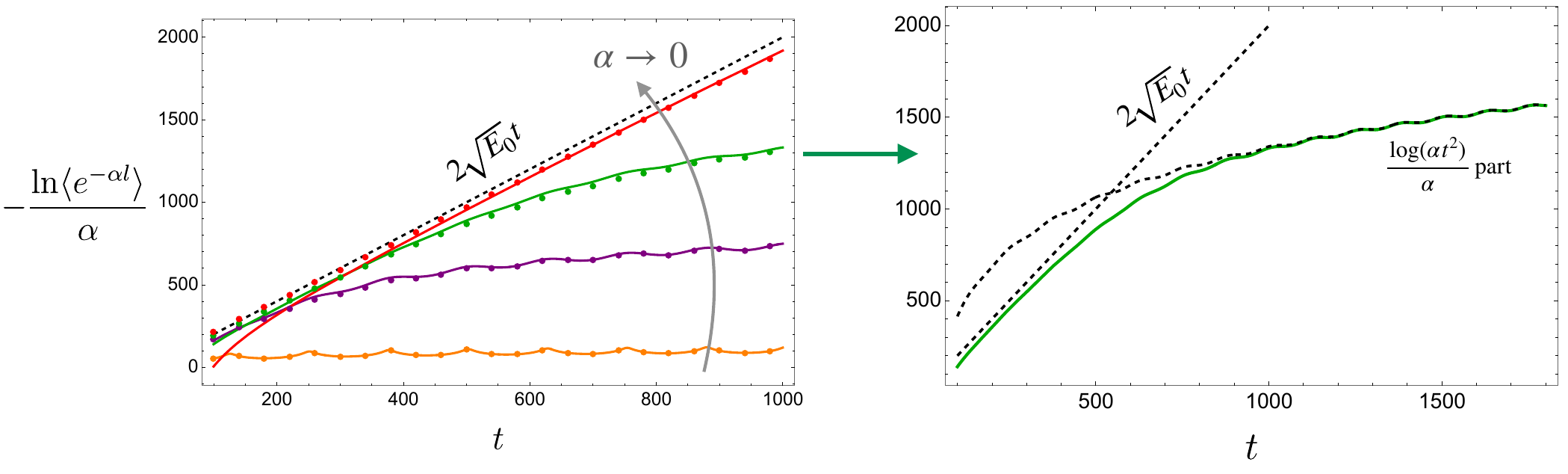}
	\caption{Classical contribution of the generating function, \ie $\langle e^{-\alpha \ell } \rangle \big|_{\rm classical}$. \textit{Left:} The orange, purple, green, and red curves represent the approximate results for fixed values $\alpha=\frac{1}{10}$, $\frac{1}{100}$, $\frac{1}{200}$, and $\frac{1}{500}$, respectively. The corresponding colored dots denote the numerical results obtained from the integral using the wavefunction $\psi_E(\ell)$. \textit{Right:} Time evolution of $\langle e^{-\alpha \ell } \rangle \big|_{\rm classical}$ for $\alpha=\frac{1}{200}$. In this panel, the dashed curves indicate the contributions from the linear term and the logarithmic term, which dominate in the regimes $\alpha t \sim 1$ and $\alpha t \gg 1$, respectively.}
	\label{fig:ExpaLClassical}
\end{figure}

Similarly to previous sections, we now analyze the classical and quantum contributions to the generating function separately. By substituting the approximate classical probability derived in eq.~\eqref{eq:Pclassical} and neglecting the doubly exponentially suppressed contributions from negative $t_\ell$, one obtains the following approximate expression:
\begin{equation}
\begin{split}
\langle e^{-\alpha \ell } \rangle \big|_{\rm classical}  \approx  &\frac{1}{2\pi \Delta E} \left[  e^{-\alpha ( 2\sqrt{E_0}t - \log (4E_0) ) } \left( 
2\bar{\mathcal{E}}\mathrm{Ei}(\bar{\mathcal{E}}t)-\mathcal{E}_-\mathrm{Ei}(\mathcal{E}_-t)-\mathcal{E}_+\mathrm{Ei}(\mathcal{E}_+t)
\right) 
\right.\\
&\quad\left.  
 +e^{\alpha ( 2\sqrt{E_0}t + \log (4E_0) ) } \left( 
-2\bar{\mathcal{E}}\mathrm{E}_1(\bar{\mathcal{E}}t)+\mathcal{E}_-\mathrm{E}_1(\mathcal{E}_-t)+\mathcal{E}_+\mathrm{E}_1(\mathcal{E}_+t)
\right) 
\right] \,,
\end{split}
\end{equation}
where we have introduced the variables $
\mathcal{E}_\pm = 2\alpha\sqrt{E_0}\pm i\Delta E \quad \text{and} \quad \bar{\mathcal{E}} = \frac{\mathcal{E}_+ + \mathcal{E}_-}{2} = 2\alpha\sqrt{E_0}$
to simplify the expression, and have dropped the leading oscillatory terms present in $P_{\mathrm{classical}}$ (see eq.~\eqref{eq:Pclassical}). Here, $\mathrm{Ei}(z)$ and $\mathrm{E}_1(z)$ denote the exponential integral functions\footnote{The definitions of these exponential integral functions are given by
\begin{equation}
   \begin{split}
    \mathrm{Ei}(z) &= \int_{-z}^\infty \frac{e^{-t}}{t}\, dt \,, \qquad     
    \mathrm{E}_n (z)= - \int_1^\infty \frac{e^{-zt}}{t^n}\, dt \,.
   \end{split}
\end{equation}
For a real argument $x$, one has the relation $\mathrm{E}_1 (x) = -\mathrm{Ei}(-x)$; however, this relation does not extend to complex arguments due to the branch cut on the complex $z$-plane running from $(-\infty, 0)$.}. Notice that the time-reflection symmetry $t \leftrightarrow -t$ is manifest in the above expression. As depicted in figure~\ref{fig:ExpaLClassical}, the classical contribution to the generating function decays monotonically.

Neglecting the early-time regime with $\alpha t \sim 0$, the classical generating function is further approximated by
\begin{equation}\label{eq:expalclassicalapp}
\langle e^{-\alpha \ell } \rangle \big|_{\rm classical}  \approx  e^{-\alpha \ell_{\rm classical}} +  \frac{(4E_0)^\alpha}{ \pi \Delta E \sqrt{E_0}} \left(  1  - \frac{4 E_0 \cos (\Delta E t)}{ 4E_0 + (\Delta E)^2/\alpha^2}  \right) \frac{1}{\alpha t^2} + \mathcal{O}\left( \frac{1}{t^4}\right) \,. 
\end{equation}
The first term represents the exponential decay with the decay rate governed by the classical geodesic length, namely 
\begin{equation}
\ell_{\rm classical}  = 2\sqrt{E_0}\, t - \log(4 E_0) \,.
\end{equation}
It is important to note that this term dominates in the regime where $\alpha t \sim 1$, whereas the second (polynomial) term in eq.~\eqref{eq:expalclassicalapp} becomes dominant in the large-time regime with $\alpha t \gg 1$. Although this polynomial term deviates from the classical exponential decay, its effect diminishes as $\alpha t\to 0$, implying that the linear regime characterized by $e^{-\alpha \ell_{\rm classical}}$ extends over longer times, as demonstrated in figure~\ref{fig:ExpaLClassical}. Of course, we expect that by taking the limit $\alpha\to 0$, the generating function $\langle e^{-\alpha \ell} \rangle \big|_{\rm classical}$ reproduces the linear growth of the classical geodesic length. With $t$ fixed (so that $\alpha t \to 0$), the approximate expression reduces to
\begin{equation}\label{eq:expalclassical}
\langle e^{-\alpha \ell } \rangle \big|_{\rm classical}  \overset{\alpha \to 0}{\approx} 1 - \alpha \left(  \ell_{\rm classical} - \frac{4\sqrt{E_0}}{\pi \Delta E} \left( \log (\alpha t)  + \gamma + \log \sqrt{4E_0}  \right) \right) + \mathcal{O}(\alpha^2) \,.
\end{equation}
This result is consistent with the expression for $\langle \hat{\ell} \rangle \big|_{\mathrm{classical}}$ in eq.~\eqref{eq:applclassical}, which was derived from the finite time derivative. Finally, we emphasize that the limit $\alpha \to 0$ is subtle because the time $t$ may also tend to infinity. In order to recover the classical result from the $\alpha\to 0$ limit, one must ensure that $\alpha t \ll 1$ even in the late-time limit $t\to \infty$.


The infinite decay of the generating function $\langle e^{-\alpha \ell } \rangle$ at the classical level is analogous to the decay observed in the two-point correlation function of generic operators in an eternal AdS black hole background. This eternal decay arises because infalling particles monotonically fall into the black hole. In the context of a chaotic quantum mechanical theory, this decay regime can be interpreted as an indicator of the thermalization process. However, the quantum correlator for a finite Hilbert space would not decay forever otherwise, the information would be lost. In a similar fashion, one naturally expects that the generating function $\langle e^{-\alpha \ell } \rangle$ will eventually cease to decay and, instead, begin to increase after a characteristic time scale associated with the Heisenberg time $T_{\mt{H}}$. To show this behavior explicitly, we must introduce the quantum corrections to the generating function at late times by including the quantum probability contribution $P_{\rm quantum}(\ell, t)$, \viz
\begin{equation}
\begin{split}
 \langle e^{-\alpha \ell } \rangle \big|_{\rm quantum}  
=  \int_{-\infty}^{\infty} P_{\rm quantum}(\ell,t) \, e^{-\alpha \ell } d\ell  
\approx  2\sqrt{E_0}\,(4E_0)^\alpha \int_{0}^{\infty} P_{\rm quantum} e^{-\tilde{\alpha} t_\ell } \, dt_\ell,
\end{split}
\end{equation}
where we have introduced the rescaled parameter $\tilde{\alpha} = 2\sqrt{E_0}\,\alpha$.

\begin{figure}[t]
	\centering
	\includegraphics[width=5.5in]{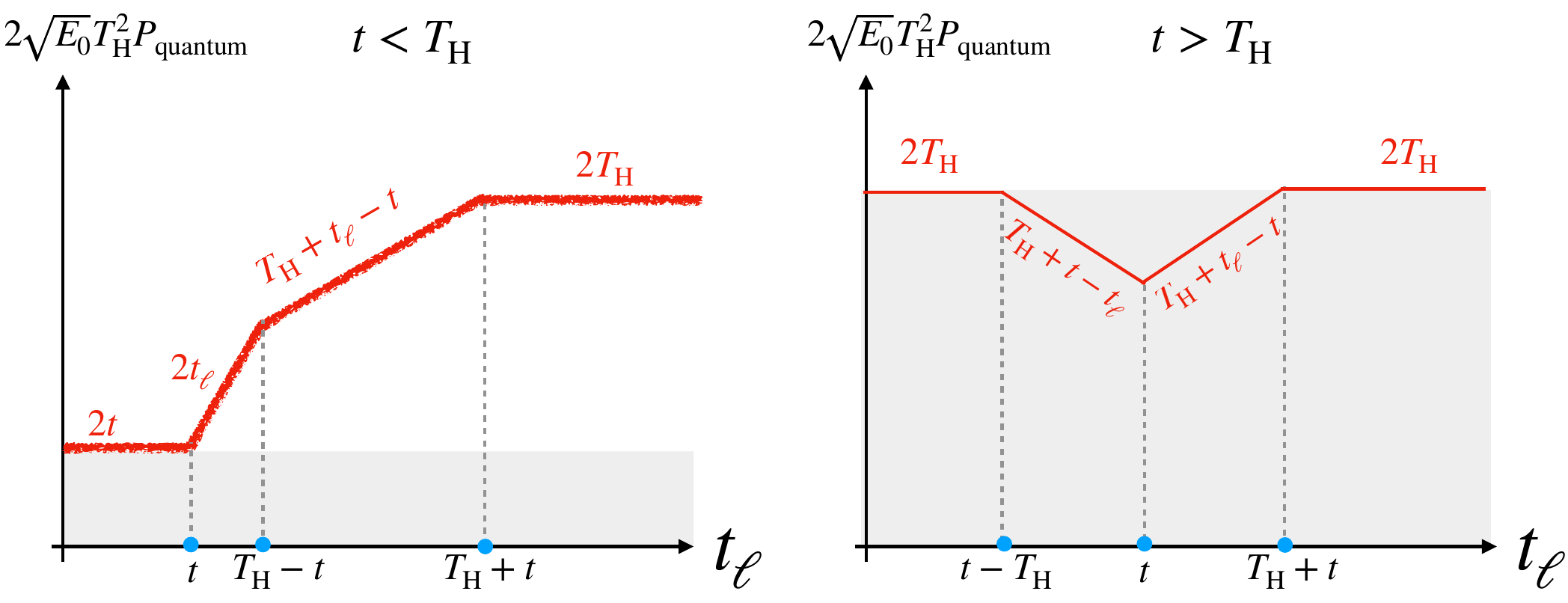}
	\caption{A schematic diagram of the quantum probability distribution $P_{\rm quantum}(\ell, t)$ as a function of the rescaled geodesic length $t_\ell$, as approximated in eq.~\eqref{eq:Pquantum}. The gray regions in the two plots result in the linear ramp and late-time plateau of the generating function $\langle e^{-\alpha \ell } \rangle$, respectively.}
	\label{fig:Pquantum}
\end{figure}

We first focus on the time regime $t < T_{\mt{H}}$, where the approximate expression for $P_{\rm quantum}(\ell, t)$ has been derived in eq.~\eqref{eq:Pquantum} (see also figure~\ref{fig:Pquantum}). Substituting this expression into the integral yields
\begin{equation}\label{eq:expalquantum01}
\begin{split}
 \langle e^{-\alpha \ell } \rangle \big|_{\rm quanutm}  \approx  \frac{(4E_0)^\alpha}{ T_{\mt{H}}^2} &\left[   \frac{2t}{\tilde{\alpha}}  - \frac{1}{\tilde{\alpha}^2} 
 \left(  e^{-\talpha ( T_{\mt{H}} -t)} +e^{-\talpha ( T_{\mt{H}} + t)}  - 2 e^{-\talpha t} \right) \right.\\
 &\quad  \left. -
 \frac{2t \left(\talpha  \sin \left(4 \sqrt{E_0}\right)-4 E_0 \cos \left(4 \sqrt{E_0}\right)\right)}{\talpha ^2+16 E_0^2} 
 \right] \,.
\end{split}
\end{equation}
Here, the last term originates from the oscillatory component in $P_{\rm quantum}(\ell, t)$. For a generic parameter $\alpha$,\footnote{To ensure the validity of the approximate wavefunction, we restrict our analysis to the regime $\alpha/E_0 \lesssim 1$. This constraint is necessary because a large $\alpha$ introduces an exponential suppression factor $e^{-\alpha \ell}$, leading to dominant contributions from the region $t_{\ell} \sim 0$, where the approximation in eq.~\eqref{eq:psiDiskapp} is no longer reliable.} it is evident that the quantum contribution to the generating function is dominated by a linear growth term, \ie 
\begin{equation}
\text{linear ramp:} \quad  \langle e^{-\alpha \ell } \rangle \big|_{\rm quantum}  \approx \frac{(4E_0)^\alpha}{ \sqrt{E_0}} \frac{t}{\alpha \, T_{\mt{H}}^2} + \cdots \,, \qquad \text{for} \quad t < T_{\mt{H}} \,.
\end{equation}
Although this linear quantum correction is of order $1/T_{\mt{H}}^2$, it can compete with and even dominate over the decaying classical contribution $\langle e^{-\alpha \ell } \rangle \big|_{\rm classical}$. This competition gives rise to the linear ramp regime of the generating function, as depicted in figure~\ref{fig:expal}. Notably, the origin of the linear ramp is traced to the small-$t_\ell$ region (\eg $t_\ell \in (0,t)$) where $P_{\rm quantum} \approx \frac{t}{\sqrt{E_0}T_{\mt{H}}^2}$. The linear growth of the generating function eventually ceases once the time grows beyond the Heisenberg time $T_{\mt{H}}$. For $t > T_{\mt{H}}$, using the branch of $P_{\rm quantum}$ corresponding to $t_\ell > T_{\mt{H}}$ (see figure~\ref{fig:Pquantum}), the integral can be expressed as
\begin{equation}\label{eq:expalquantum02}
\begin{split}
 \langle e^{-\alpha \ell } \rangle \big|_{\rm quanutm}  \approx  \frac{(4E_0)^\alpha}{ T_{\mt{H}}^2} &\left[   \frac{2T_{\mt{H}}}{\tilde{\alpha}}   - \frac{e^{-\talpha (t+T_{\mt{H}}) } ( e^{\talpha T_{\mt{H}}} -1)^2}{\talpha^2}   \right. \\
 &\qquad -\left. \frac{2T_{\mt{H}} \left(\talpha  \sin \left(4 \sqrt{E_0}\right)-4 E_0 \cos \left(4 \sqrt{E_0}\right)\right)}{\talpha ^2+16 E_0^2} \right] \,.
\end{split}
\end{equation}
In the late-time limit, this expression approaches a constant, thereby establishing the plateau in the generating function:
\begin{equation}\label{eq:expplateau}
\text{plateau:} \quad \lim_{\alpha t \to \infty} \langle e^{-\alpha \ell } \rangle \approx \frac{(4E_0)^\alpha}{ \sqrt{E_0}} \frac{1}{\alpha \, T_{\mt{H}}} \,.
\end{equation}

Combining the contributions from the classical part eq.~\eqref{eq:expalclassicalapp} and the two piece‐wise quantum corrections (eqs.~\eqref{eq:expalquantum01} and \eqref{eq:expalquantum02}), we obtain our approximate result for the generating function. For a generic value of $\alpha$\footnote{To obtain these approximate expressions, we assume that $\alpha$ is neither too large ($\alpha/E_0 \gg 1$) nor too small ($\alpha T_{\mt{H}} \ll 1$).}, the generating function exhibits the well‐known slope–ramp–plateau structure, which can be explicitly expressed as
\begin{equation}\label{eq:expalphalsum}
 \langle e^{-\alpha \ell }  \rangle  \approx 
\begin{cases}
e^{-\alpha \ell_{\rm classical}} + \mathcal{O}(\frac{1}{\alpha t^2 \Delta E})\,, \qquad \text{slop at early times} \quad \alpha t \sim 1  \,,\\[1em]
\frac{(4E_0)^\alpha}{ \sqrt{E_0}} \frac{t}{\alpha \, T_{\mt{H}}^2} + \cdots \,,\qquad \text{ramp at the middle stage} \quad  t <   T_{\mt{H}}  \,,\\[1em]
\frac{(4E_0)^\alpha}{ \sqrt{E_0}} \frac{1}{\alpha \, T_{\mt{H}}} + \mathcal{O}(e^{-\alpha (t-T_{\mt{H}})})\,,\qquad \text{plateau at late times} \quad  t >   T_{\mt{H}}  \,,\\
\end{cases}
\end{equation}
Here, the transition between the ramp and the plateau occurs at approximately $t \approx T_{\mt{H}}$. This approximation \eqref{eq:expalphalsum} thus explicitly captures the time evolution of $\langle e^{-\alpha \ell }  \rangle$, as illustrated in figure~\ref{fig:expal}.

As indicated by its name, the generating function $\langle e^{-\alpha \ell} \rangle$ also contains the information about the time evolution of the length expectation value $\langle \hat{\ell} \rangle$. To wit, 
\begin{equation}
 \lim_{\alpha \to 0} \langle e^{-\alpha \ell} \rangle  \sim  \Biggl( 1 + \frac{(4E_0)^\alpha}{ \sqrt{E_0}}\,\frac{1}{\alpha \, T_{\mt{H}}} \Biggr)- \alpha\, \langle \hat{\ell} \rangle + \mathcal{O}(\alpha^2) \,,
\end{equation}
where the first (divergent) term is derived from the integral $\int (P_{\rm classical}+P_{\rm quantum})\,d\ell$. Neglecting all time‐independent divergences (\ie the $\log (\alpha)$ and $\frac{1}{\alpha^2}$ terms arising from the over‐complete basis $|\ell \rangle$), we may formally write
\begin{equation}
 \langle \hat{\ell} \rangle 
  \sim \lim_{\alpha \to 0} \Bigl(- \partial_\alpha \langle  e^{-\alpha \ell} \rangle \Bigr)
  = \langle \hat{\ell} \rangle_{\alpha \to 0} \,.
\end{equation}
A similar calculation has been explicitly investigated in \cite{Iliesiu:2021ari} within the canonical ensemble.

\begin{figure}[t]
	\centering
\includegraphics[width=5in]{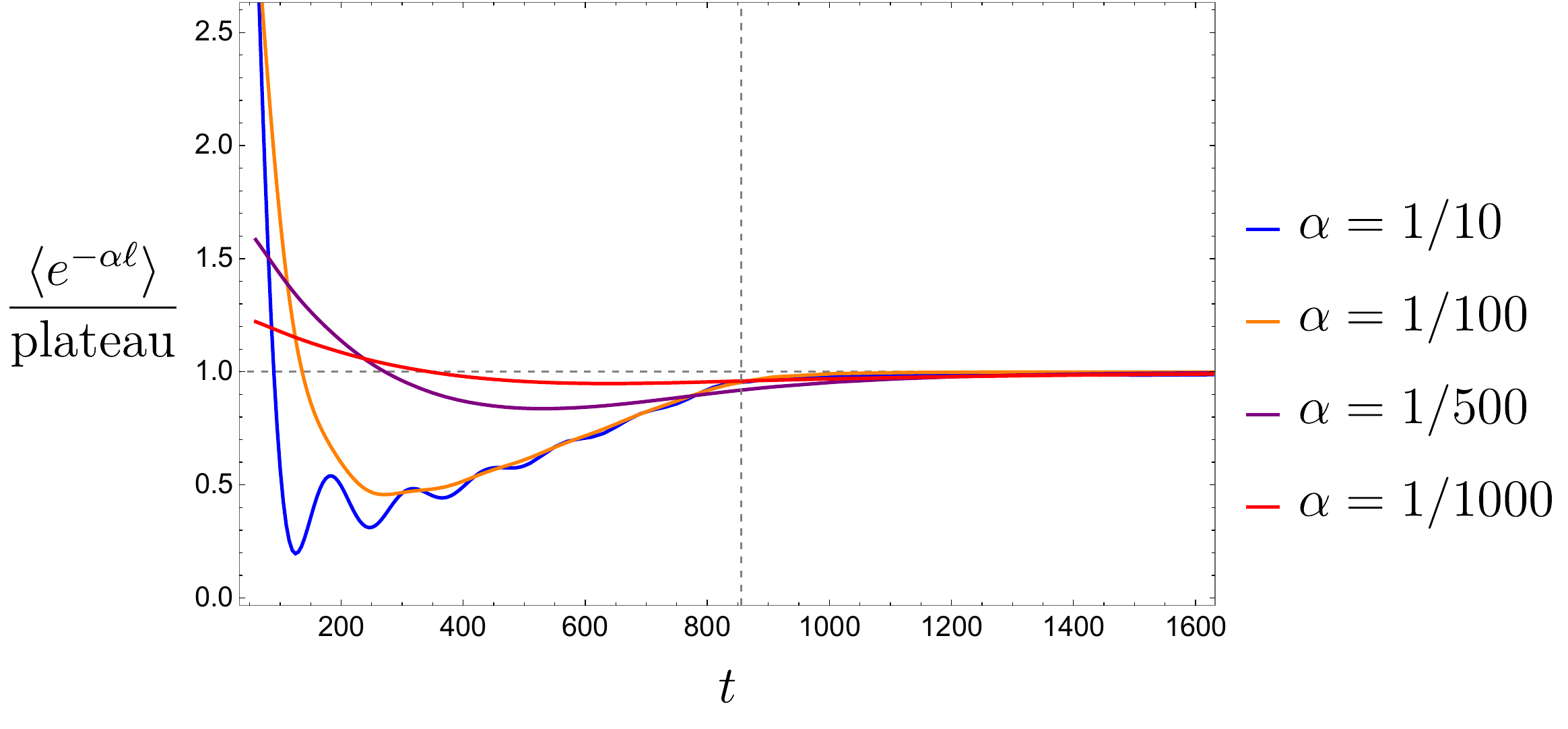}
	\caption{The time evolution of the generating function $\langle e^{-\alpha \ell } \rangle$ (rescaled by its value \eqref{eq:expplateau} on the plateau) with decreasing $\alpha$. The numerical plot is generated by using the precise wavefunction $\psi_E(\ell)$ and the same parameters as the others.}
	\label{fig:expal02}
\end{figure}

The crucial point is that the linear growth and the late-time plateau of the length expectation value $\langle \hat{\ell} \rangle$ are directly traced back to the slope and plateau in the generating function whose time evolution is typically similar to that of the spectral form factor. 
The key fact is that the decrease of $\alpha$ results in the absence of the liner ramp part. Recall that, for $t < T_{\mt{H}}$, the quantum correction given by eq.~\eqref{eq:expalquantum01} involves an exponential suppression that becomes significant when $\alpha t < \alpha T_{\mt{H}} \ll 1$. In other words, the linear term is not dominant in this regime. Instead, the ramp is replaced by
\begin{equation}\label{eq:expalquantum03}
\begin{split}
 \lim_{\alpha T_{\mt{H}} \ll 1} \langle e^{-\alpha \ell } \rangle \big|_{\rm quantum}  
 \approx  (4E_0)^\alpha \Biggl(  \frac{1}{\sqrt{E_0}\,\alpha\, T_{\mt{H}} }  - 1 +  \mathcal{O} (\alpha T_{\mt{H}}) \Biggr) \,,
\end{split}
\end{equation}
which indicates the absence of a linear ramp in this parameter region. For comparison, figure~\ref{fig:expal02} displays the numerical results for $\langle e^{-\alpha \ell } \rangle$ as $\alpha$ decreases.

In summary, by combining the classical and quantum contributions, the generating function $\langle e^{-\alpha \ell} \rangle$ reproduces the expected slope–ramp–plateau structure. Moreover, in the limit $\alpha \to 0$ (with the constraint $\alpha t \ll 1$ even at late times) this generating function gives rise to the transition of $\langle \hat{\ell} \rangle$ from a regime of linear growth to a late-time plateau.


\subsubsection{Spectral representation of generating functions}
Similar to the spectral form factor, the expectation value of the generating function $ \langle e^{-\alpha \ell}\rangle$ also provides a novel probe to the spectrum. Especially its characteristic behavior also distinguishes chaotic and non-chaotic quantum systems. To explicitly show that the generating function is a probe of the spectrum, let us recall its definition \eqref{eq:definegenerating} and recast it in terms of 
\begin{equation}
    \begin{split}
   \langle e^{-\alpha \ell}\rangle  &= \int_{-\infty}^{\infty} d\ell~P(\ell,t)e^{-\alpha \ell} =  \langle \widehat{e^{-\alpha \ell}}  \rangle = \langle  \text{TFD}(t) | \widehat{e^{-\alpha \ell}} | \text{TFD}(t) \rangle \\
   &=  \frac{e^{2S_0}}{Z}\int dE_1\,dE_2~e^{-i(E_1-E_2)t}   \langle  E_2 | \widehat{e^{-\alpha \ell}} | E_1 \rangle   \langle D(E_1)D(E_2) \rangle \,,
    \end{split}
\end{equation}
where we expand the TFD states in the energy basis $|E_i \rangle$. Substituting the matrix element \eqref{eq:E1expalE2} derived in JT gravity \cite{Yang:2018gdb,Saad:2019pqd,Iliesiu:2021ari}, we can obtain an alternative representation of the generating function $ \langle e^{-\alpha \ell}\rangle$, \ie 
\begin{equation}
    \begin{split}
    \frac{2e^{S_0}}{Z}\int dE_1 dE_2~e^{-i(E_1-E_2)t}
    \langle D(E_1)D(E_2) \rangle \frac{ |\Gamma\left(\alpha + i (\sqrt{E_1} +\sqrt{E_2})  \right)|^2 |\Gamma\left(\alpha + i (\sqrt{E_1}-\sqrt{E_2})  \right)|^2 }{ \Gamma (2\alpha)}.
    \end{split}
\end{equation}
The same expression has been defined and calculated before in \cite{Iliesiu:2021ari}. Similarly, we can rewrite the expectation value of the length operator derived from the generating function as 
\begin{equation}
    \begin{split}
    \langle \hat{\ell} \rangle_{\alpha \to 0}
    &=\frac{e^{2S_0}}{Z}\int dE_1\,dE_2~\langle D(E_1)D(E_2) \rangle e^{-i(E_1-E_2)t}
   \left( \langle E_1 | \, \hat{\ell} \, | E_2 \rangle \right)\big|_{\alpha \to 0}  \,,
    \end{split}
\end{equation}
where the matrix element $ \left( \langle E_1 | \, \hat{\ell} \, | E_2 \rangle \right)\big|_{\alpha \to 0} $ is formulated as eq.~\eqref{eq:E1lE2}, \ie 
\begin{equation}
\begin{split}
\left( \langle E_2 | \, \hat{\ell} \, | E_1 \rangle \right)\big|_{\alpha \to 0}  
&=  \frac{8 \pi ^2 e^{-S_0}}{(E_2-E_1) \left(\cosh \left(2 \pi  \sqrt{E_1}\right)-\cosh \left(2 \pi  \sqrt{E_2}\right)\right)}  \,. 
\end{split}
\end{equation}
Obviously, the explicit calculations of expectation values associated with the fixed-length state require information about the wavefunction $\psi_E(\ell)$. Without using the approximate density of state at the disk level, we can replace the energy integral with the discrete sum over the spectrum, namely 
\begin{equation}
    \int{dE_1\,dE_2}\,e^{2S_0}\langle D(E_1)D(E_2)\rangle\quad \sim \quad\sum_{E_1,E_2} \,. 
\end{equation}
a prescription that can also be applied to more generic quantum systems. As a result, the discrete spectral representation of the generating function is recast as 
\begin{equation}\label{eq:ealdiscrete}
  \langle e^{-\alpha \ell}\rangle
    =\frac{2e^{S_0}}{Z}\sum_{E_1,E_2} e^{-i(E_1-E_2)t}\,\frac{ |\Gamma\left(\alpha + i (\sqrt{E_1} +\sqrt{E_2})  \right)|^2 |\Gamma\left(\alpha + i (\sqrt{E_1}-\sqrt{E_2})  \right)|^2 }{ 2\,\Gamma (2\alpha)} .
\end{equation}
Similarly, taking $\alpha \to 0$ limit of its derivative leads to the (non-regularized) expectation value of the geodesic length operator:  
\begin{equation}\label{eq:ldiscrete}
 \langle \hat{\ell} \rangle_{\alpha \to 0} =  \frac{8\pi^2 e^{S_0}}{Z} \sum_{E_1,E_2}   e^{-i (E_1- E_2)t}  \frac{1}{(E_2-E_1) \left(\cosh \left(2 \pi  \sqrt{E_1}\right)-\cosh \left(2 \pi  \sqrt{E_2}\right)\right)}  \,.
\end{equation}
These discrete expressions coincide with those obtained in JT gravity after using the approximate density of states provided in eq.~\eqref{eq:Ddisk}. Naturally, they are specifically associated with the geodesic length operator $\hat{\ell}$ or the fixed-length states. In light of the complexity=volume conjecture, one may reinterpret expressions \eqref{eq:ldiscrete} and \eqref{eq:ealdiscrete} as representing the holographic complexity and its generating function, respectively. From the perspective of the dual boundary theory, these quantities correspond to the quantum complexity of the boundary system. However, it has been pointed out in \cite{Belin:2021bga,Belin:2022xmt,Jorstad:2023kmq} that there exist infinitely many gravitational observables —such as the maximal volume—that can serve as candidates for holographic complexity, \ie complexity=anything conjecture. In the same spirit, we can also find infinite spectral representations for complexity (and the corresponding generating functions), which present the universal linear growth before the Heisenberg time and transit to a late-time plateau. The reason behind the universal time evolution of quantum complexity is related to the universality of complexity functional for small energy separation $E_1 \sim E_2$, \ie the same pole structure \cite{toappear}.

\subsubsection{Spectral complexity}
To show this idea explicitly, let us consider the simplest example: the spectral complexity in microcanonical ensemble, \ie 
\begin{equation}\label{eq:SC}
    \mathrm{SC}(t)\sim\sum_{E_1\neq E_2}\frac{1-\cos[(E_1-E_2)t]}{(E_1-E_2)^2} \,,
\end{equation}
which is proposed in \cite{Iliesiu:2021ari} and motivated by the fact that it has a similar time evolution to the maximal volume (\ie geodesic length) in JT gravity. Although the spectral complexity \eqref{eq:SC} and geodesic length \eqref{eq:ldiscrete} have distinct expressions, it is obvious that they are quite similar in the region $E_1 \sim E_2$, which is the universal part. To see this universality, we can approximate the geodesic length in the microcanonical ensemble with $E_i \in \left[ E_0 - \frac{\Delta E}{2}, E_0 + \frac{\Delta E}{2} \right]$. Using the two variables $E_{12}= E_1 -E_2$ and $ \bar{E}= (E_1 +E_2)/2$, the approximate expression can be derived as 
\begin{equation}
 \langle \hat{\ell} \rangle_{\alpha \to 0} \approx \frac{32\pi^3}{\sinh^2(2\pi \sqrt{E_0})\Delta E}  \sum_{E_1,E_2}   \frac{\cos \left( (E_1 -E_2)t \right)}{(E_2-E_1)}  \left( \frac{\sqrt{E_0} +\mathcal{O}(\Delta E)}{E_1 -E_2}  + \mathcal{O}(E_1 - E_2) \right)   \,,
\end{equation}
where we zoom into the region with $E_{12} \ll 1$ and have used $ \Delta E/E_0 \ll 1$. The first term dominates while the second term is oscillating but subleading. Ignoring the divergent constant from the diagonal part \eqref{eq:ElE} and a non-relevant factor, we can find that the dominating contribution of the regularized length expectation value is 
\begin{equation}
 \langle \hat{\ell} \rangle_{\rm reg} \sim \sum_{E_1 \ne E_2}   -\frac{\cos \left( (E_1 -E_2)t \right)}{(E_1-E_2)^2}  + \cdots \,. 
\end{equation}
It is obviously equivalent to the spectral complexity up to a time-independent constant. It is thus natural to expect that the universal part of the generating function $\langle e^{-\alpha \ell}\rangle$ \eqref{eq:ealdiscrete} corresponds to the generating function of spectral complexity, denoted as $G_{\mt{SC}} (\alpha ,t)$. For simplicity, let us focus on the small $\alpha$ region by expanding the spectral representation \eqref{eq:ealdiscrete} around $E_{12}\ll 1$ and $\alpha\ll 1$. This approximation yields 
\begin{equation}
 \langle e^{-\alpha \ell}\rangle \approx \frac{8\pi e^{S_0}}{Z} \sum_{E_1, E_2}  e^{-(E_1 -E_2)t} \left( \frac{ \alpha \sqrt{\bar{E}} \csch \left( 2\pi \sqrt{\bar{E}}\right)  + \mathcal{O} ( \alpha^2)  }{(E_1-E_2)^2 + 4 \bar{E} \alpha^2}  + \mathcal{O}(E_1 -E_2)  \right)\,,
\end{equation}
where the leading term reproduces the slope-ramp-plateau structure. Taking this lesson, we can thus define the generating function of the spectral complexity $\rm{SC}(t)$ in terms of  \footnote{The canonical version can be given by adding a thermal factor $e^{- \frac{\beta}{2}(E_1 +E_2)}$. See eq.~\eqref{eq:GSCcan}. 
}
\begin{equation}\label{eq:GSC}
G_{\mt{SC}}(\alpha,t) = 
  \sum_{E_1, E_2} \frac{ \alpha  \cos \left( (E_1 -E_2)t \right) }{(E_1-E_2)^2 + 2(E_1 +E_2) \alpha^2}  \,.  
\end{equation}
which also applies to more generic quantum systems. Performing the $\alpha \to 0$ limit of the derivative of the generating function gives rise to the spectral complexity, namely  
\begin{equation} \label{eq:CSCreg}
    - \lim_{\alpha \to 0} \left( \frac{d}{d\alpha} G_{\mt{SC}}(\alpha,t)  \right)    =  \mathrm{SC}(t) +  \text{Divergent Constant}  \,. 
\end{equation}
It is worth noting that the choice of the generating function $G_{\mt{SC}}(\alpha,t)$ for the spectral complexity is not unique. Our definition \eqref{eq:GSC} is chosen to make the connection to the generating function $ \langle e^{-\alpha \ell}\rangle $ more explicit. As a result, we can find that the generating function $G_{\mt{SC}}(\alpha,t)$ for a generic $\alpha$ presents the slope-ramp-plateau structure for chaotic systems just like the spectral form factor as well as the generating function $ \langle e^{-\alpha \ell}\rangle $ for geodesic length in JT gravity.

\begin{figure}[t] 
    \centering
   	\includegraphics[width=2.95in]{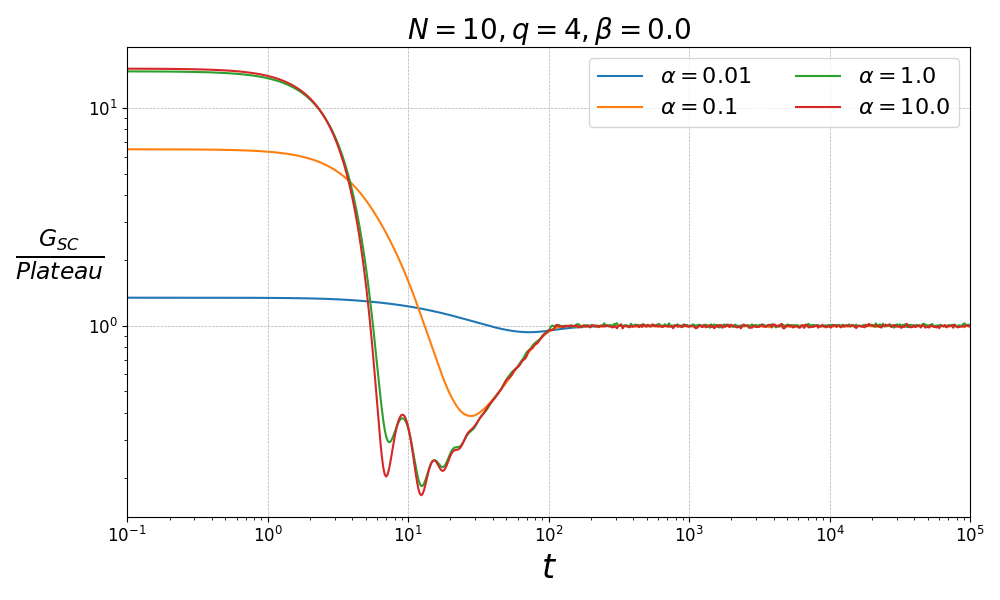}
   	 \includegraphics[width=2.95in]{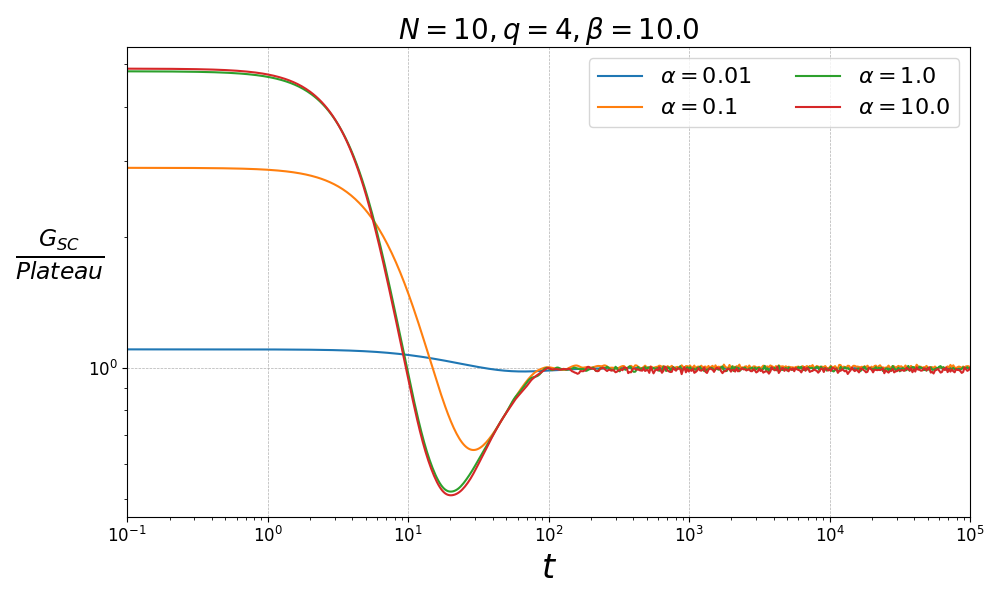}
\caption{The generating function $G_{\mt{SC}}(\alpha,t)$ (defined in eq.~\eqref{eq:GSCcan}) of the spectral complexity $\text{SC}(t)$ for different parameters $\alpha$. The numerical results are obtained by using the SYK model of $\beta=0$ and $\beta=10$ with $N=10,\,q=4$ and taking 10000 samples. Note that $N=10$ corresponds to GUE. In each sample, the spectrum is shifted by $1$ so that the center of the spectrum becomes $1$, and all eigenvalues are positive with a high probability. For a generic $\alpha$, it shows the slope-ramp-plateau structure as the spectral form factor. With decreasing $\alpha$, the linear ramp region disappears, resulting in a long-time linear growth and the transition to a late-time plateau of spectral complexity.}
\label{fig:SYKspecs}
\end{figure}

\begin{figure}[t]
\centering
    \includegraphics[width=4in]{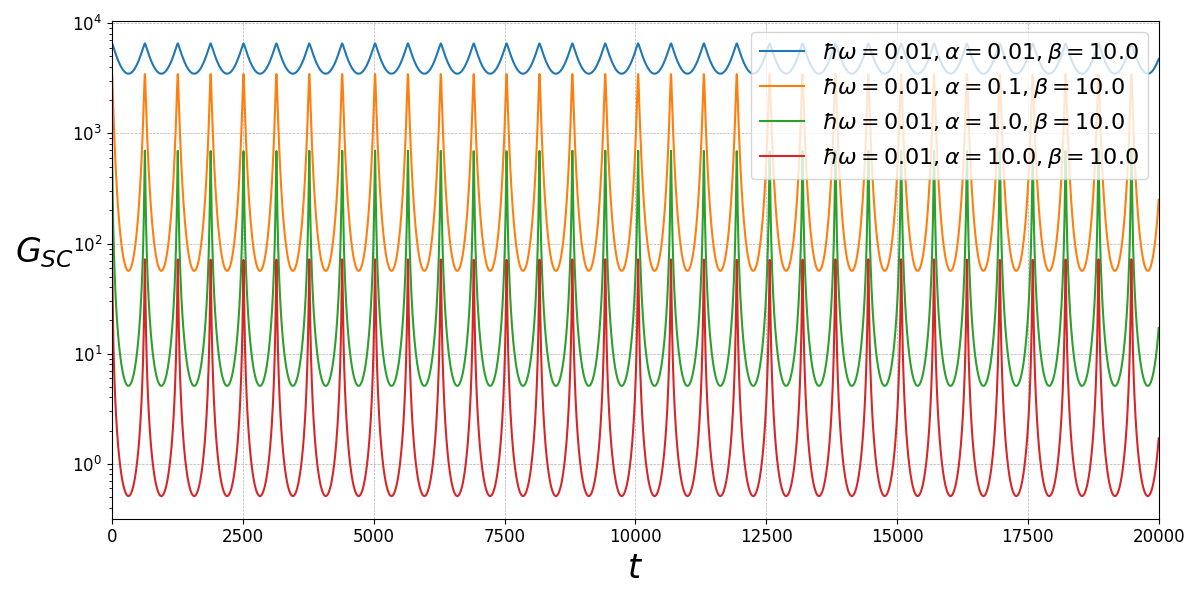}
\caption{The generating function $G_{\mt {SC}}$ \eqref{eq:GSCcan} of the spectral complexity for a harmonic oscillator with a frequency $\hbar\omega=0.01$. We choose the canonical ensemble with the inverse temperature $\beta =10$. It shows a recurrence whose periodicity is determined by the level spacing, which is nothing but $\hbar\omega$ in harmonic oscillators. In contrast to a chaotic system, such as the SYK model, universal time-evolution features, \ie the slope-ramp-plateau structure, are absent in this integrable model.}
\label{fig:oscillatorspec}
\end{figure}

To close this section, we finally remark that both quantum complexity and the corresponding generating function can diagnose the chaotic system and integrable systems. As an example, let us consider the SYK model \cite{Sachdev_1993,KitaevTalks}  whose Hamiltonian with $N$ Majorana fermions and  $q$-body interactions is defined by \footnote{We follow the conventions in \cite{Maldacena:2016hyu}.} 
\begin{equation}
    H=\frac{i^{\frac{q}{2}}}{q!}\sum_{i_1,\ldots,i_q}^N J_{i_1\ldots i_q}\psi_{i_1}\cdots\psi_{i_q} \,.
\end{equation} 
where $\psi$ represents a Majorana fermion and indices $i_q$ run from $1$ to $N$. The random coupling constant denoted as $J$ is drown from a Gaussian distribution with zero mean and variance given by $\sigma^2=\frac{(q-1)!}{N^{q-1}}J^2$. We simply set $J=1$ and focus on $N=10,\,q=4$ case, which corresponds to the Gaussian unitary ensemble as that in JT gravity. In particular, we focus on the generating function $G_{\rm{SC}}$ of the spectral complexity in the canonical ensemble, namely
\begin{equation}\label{eq:GSCcan}
G_{\mt{SC}}(\alpha,t) = 
  \sum_{E_1, E_2} \frac{ \alpha  \, e^{-\frac{\beta}{2} (E_1 +E_2)}e^{- i(E_1 -E_2)t} }{(E_1-E_2)^2 + 2(E_1 +E_2) \alpha^2}   \,.  
\end{equation}
The numerical results for $G_{\mt{SC}}(\alpha,t)$ in SYK model and harmonic oscillator are shown in figure \ref{fig:SYKspecs} and \ref{fig:oscillatorspec}, respectively. In the SYK model, we chose the canonical ensemble with $\beta=0$ and $\beta=10$. For a generic value of $\alpha$ parameter, its time evolution is similar to the spectral form factor or the generating function $ \langle e^{-\alpha \ell}\rangle $. With decreasing the value of $\alpha$, the linear ramp region will disappear. As a result, it produces the linear-plateau structure for the spectral complexity $ \mathrm{SC}(t)$. Comparing the two plots with $\beta=0$ and finite temperature with $\beta \sim \mathcal{O}(1)$, we can see that the main difference is that the subleading oscillating terms are suppressed in the finite temperature case.

We can expect that this pattern of time evolution is absent in integrable systems. As a comparison, we can find that the parallel results for a quantum harmonic oscillator is highly oscillatory\footnote{One can also evaluate a simpler generating function 
\begin{equation}
G_{\mt{SC}}(\alpha,t) = 
  \sum_{E_1, E_2} \frac{ \alpha  \, e^{-\frac{\beta}{2} (E_1 +E_2)}e^{- i(E_1 -E_2)t} }{(E_1-E_2)^2 + \alpha^2}   \,, 
\end{equation}
which will be introduced in the next subsection. An analytical result for a harmonic oscillator can be derived by using the hypergeometric function $\, _2F_1(a,b;c;z)$.}, with the periodicity determined by the level spacing. Summing over the discrete spectrum $E_n= \frac{1}{2} +n$, we can also analytically derive the spectral complexity, \ie
\begin{equation}
\begin{split}
  - \lim_{\alpha \to 0} \left( \frac{d}{d\alpha} G_{\mt{SC}}(\alpha,t)  \right) = \frac{\tanh ^{-1}\left(e^{-\frac{\beta}{2}}\right)}{2 \alpha^2}  +  \frac{  e^{\beta /2}}{1-e^{\beta }}
\left(\mathrm{Li}_2\left(e^{-i t-\frac{\beta }{2}}\right)+\mathrm{Li}_2\left(e^{i t-\frac{\beta }{2}}\right)\right),
\end{split}
\end{equation}
where the first term is the divergent constant at the limit $\alpha \to 0$ and the second term denotes the spectral complexity (up to a constant)\footnote{After taking a vanishing inverse temperature limit $\beta \to 0$, we can use the inversion formula of the polylogarithm to derive a simpler expression for the spectral complexity, as follows 
\begin{equation}
\mathrm{SC}(t) = \text{Constant} -\frac{1}{\beta}\left(  \frac{1}{2} \left( \arccos(\cos  t )\right)^2-\pi  \arccos(\cos t) +\frac{\pi ^2}{3}   \right) \,,
\end{equation}
which explicitly oscillates with the time evolution. 
}. This oscillatory spectral complexity for a harmonic oscillator is obviously distinguishable from that of a chaotic system. To demonstrate the oscillatory nature of this spectral complexity, it is helpful to calculate the second time derivative, namely  
\begin{equation}
\frac{d^2 \mathrm{SC}(t)}{dt^2} =  \frac{1}{1-e^{\beta }}\frac{e^{\beta /2} \cos (t)-1}{ \cos (t)-\cosh \left(\frac{\beta }{2}\right)} \,.
\end{equation}

\subsection{Generating function for the time shift}\label{sec:timeshift}
Similar to the fixed-length states $| \ell \rangle$, we have also investigated the non-perturbatively overlaps between time-shifted TFD states $|\delta \rangle$ in section \ref{section:timeshiftstate}. The squared overlap $P^{\mt{TFD}}(\delta, t)$ is nothing but the spectral form factor $\mathrm{SFF}(|t- \delta|)$. Correspondingly, we can also define a time-shift operator $\hat{\delta}$ in terms of 
\begin{equation}\label{eq:deltaoperator}
\hat{\delta}    = \int_{-\infty}^{\infty} \,
\delta \, |\delta \rangle\langle \delta |  \, d \delta  
\,. 
\end{equation}
In this subsection, we focus on studying the expectation value of the time-shift operator and its relevant generating functions.

\subsubsection{Expectation value of the time-shift operator}
Considering a TFD state located at the boundary time $t$, the time-shift operator plays the role of measuring the boundary time via its expectation value, \ie  
\begin{equation}\label{eq:deltavalue}
\langle \hat{\delta}   \rangle \equiv    \langle {\text{TFD}}(t)| \hat{\delta} |\text{TFD}(t)\rangle = \int_{-\infty}^{\infty} P^{\mt{TFD}}(\delta,t) \, \delta \, d \delta \,, \\
\end{equation}
where the non-perturbative probability $P^{\mt{TFD}}$ has been approximately derived in eq.~\eqref{eq:PTFD}.

\begin{figure}[t]
	\centering
\includegraphics[width=3in]{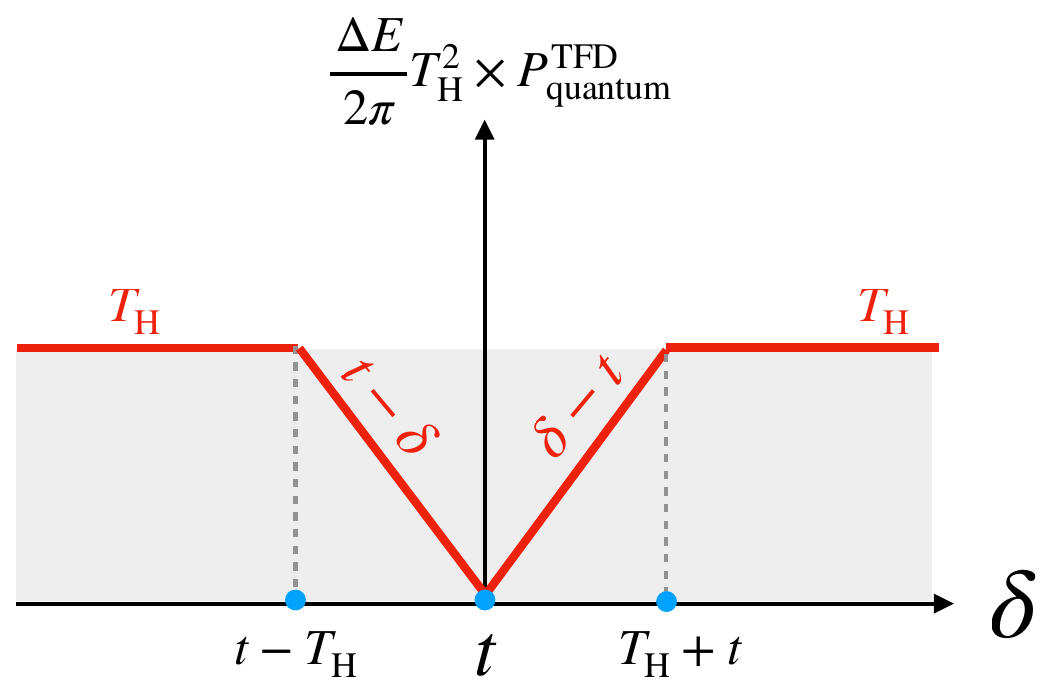}
	\caption{The characteristic plot for the quantum correction parts of the squared overlap $P^{\mt{TFD}}(\delta, t)$ as a function of the time-shift value $\delta$. The gray region represents the contribution of the plateau, which corresponds to the origin of the divergence.}
	\label{fig:PTFDquantum}
\end{figure}

Before we move to the detailed calculations. We would like to highlight two basic properties of the squared overlap $P^{\mt{TFD}}(\delta,t)$ associated with a infinite and continuous $\delta$-spectrum, namely 
\begin{itemize}
  \item It  is only a function of relative time shift $\delta -t$ by definition \eqref{eq:Pdeltat};
  \item It has a $Z_2$ symmetry between $\delta -t$ and $t-\delta$ due to the spectral symmetry between $E_i$ and $E_j$. 
\end{itemize}
As a result, we should always find that the overlap squared $P^{\mt{TFD}}(\delta,t)$ only depends on the absolute value of the parameter $|t-\delta|$, \ie 
\begin{equation}
P^{\mt{TFD}}(\delta,t) = P^{\mt{TFD}}(|t-\delta|) \,. 
\end{equation}
Using this symmetry and introducing a new variable $\tilde{\delta}= \delta -t$, it is easy to find that the expectation value can be recast as 
\begin{equation}\label{eq:detlavalue}
\begin{split}
\langle \hat{\delta}   \rangle  
&= \int_{-\infty}^{\infty} P^{\mt{TFD}}(|\tilde{\delta}|) \,  (\tilde{\delta}+t)\, d \tilde{\delta} \\
&= t \times \int_{-\infty}^{\infty} P^{\mt{TFD}}(|\tilde{\delta}|) \, d \tilde{\delta}  \,,\\
\end{split}
\end{equation}
which is precisely proportional to the boundary time\footnote{We stress that the above analysis with changing the time shift variable $\delta$ to $\tilde{\delta}$ is very sensitive to the fact that the time-shifted spectrum $\tilde{\delta}$ remains the same, since the original spectrum of $\delta$ is infinite.} of TFD state $\mathrm{TFD}(t)$ as one can generally expect. However, the first caveat is that the time expectation value associated with infinite time-shift states $|\delta \rangle$ is not well-defined because the basis expanded by $|\delta\rangle$ is over-complete, which is similar to the situation in the fixed-length states $|\ell \rangle$. More explicitly, one can notice that the summing over all time-shift states leads to a divergence in the total probability, namely
\begin{equation}
    \int_{-\infty}^{\infty} P^{\mt{TFD}}(|\tilde{\delta}|) \, d \tilde{\delta} = \int_{-\infty}^{\infty} \left( P^{\mt{TFD}}_{\rm classical} (|\tilde{\delta}|) + P^{\mt{TFD}}_{\rm quantum} (|\tilde{\delta}|) \right) \, d \tilde{\delta} \quad  \longrightarrow  \quad \infty \,, 
\end{equation}
because there is a infinite-long plateau for large $|t-\delta|$, as shown in eq.~\eqref{eq:PTFDplateau} and figure \ref{fig:PTFDquantum}. As we will show in the next subsection, we instead first evaluate the finite generating functions of the time-shift operator by introducing a control parameter $\alpha$.

Before moving to the generating functions, let us mention a surprising result, \ie the regularized expectation value of the time shift would always be vanishing at the leading order. There are two ways to show this conclusion. First of all, let us naively assume a cut-off given by $\delta = \delta_{\rm cut}$ to avoid the divergence. Using the explicit results derived in eq.~\eqref{eq:PTFD}, we can obtain 
\begin{equation}
\begin{split}
   \int_{-\delta_{\rm cut}}^{\delta_{\rm cut}} P^{\mt{TFD}}(|\tilde{\delta}|) \, d \tilde{\delta} &= \int_{-\delta_{\rm cut}}^{\delta_{\rm cut}}\left( P^{\mt{TFD}}_{\rm classical} (|\tilde{\delta}|) + P^{\mt{TFD}}_{\rm quantum} (|\tilde{\delta}|) \right) \, d \tilde{\delta}  \\
    & \approx \left(  \frac{2\pi}{\Delta E}  - \frac{4}{\delta_{\rm cut} \Delta E^2 }  + \mathcal{O}\left(\frac{1}{\delta_{\rm cut}^2} \right) \right)  + \left( \frac{4 \pi }{\Delta E } \delta_{\rm cut} -   \frac{2\pi}{\Delta E}   + \mathcal{O}\left(\frac{1}{\delta_{\rm cut}} \right) \right)\\
    & \approx  \frac{4 \pi }{\Delta E } \delta_{\rm cut} + 0    + \mathcal{O}\left(\frac{1}{\delta_{\rm cut}} \right) 
\end{split}
\end{equation}
After introducing a naive counterterm, the regularized sum of probability is thus vanishing, \ie 
\begin{equation}
\left(  \int_{-\infty}^{\infty} P^{\mt{TFD}}(|\tilde{\delta}|) \, d \tilde{\delta}  \right)_{\rm reg} \approx 0  + \mathcal{O}(\Delta E)\,.  
\end{equation}
where the correction we have ignored is at the order of $\Delta E$. Substituting the regularized result to eq.~\eqref{eq:detlavalue}, we thus find that the regularized expectation value also vanishes. On the second hand, we can find that this conclusion is not relevant to the regularization method. To show this, we note that the time derivative of the expectation value is still finite, which is not sensitive to the choice of regularization scheme. However, the explicit probability given in eq.~\eqref{eq:PTFD} yields 
\begin{equation}
\begin{split}
\frac{ d }{ d t}  \langle \hat{\delta}\rangle  
&=\int_{-\infty}^{\infty} \partial_t P_{\rm classical}^{\mt{TFD}}(\delta,t) \, \delta \, d \delta  + \int_{-\infty}^{\infty} \partial_t P_{\rm quantum}^{\mt{TFD}}(\delta,t) \, \delta \, d \delta \\
&\approx \frac{2\pi }{\Delta E}   +  \left( -  \frac{2\pi }{\Delta E} \right)   \\
&\approx 0  + \mathcal{O} (\Delta E) \,.
\end{split}
\end{equation}
Combining this result with the fact given in eq.~\eqref{eq:deltavalue}, the only consistent result for the regularized expectation value of the time-shift operator is that it vanishes:
\begin{equation}
\label{eq:totdelta}
    \langle \hat{\delta} \rangle_{\rm reg} = 0 + \mathcal{O}(\Delta E)\,.
\end{equation}
This surprising result implies that it is not possible to measure the boundary time of the TFD state using the time-shift operator. In what follows, we will evaluate the same time expectation value via the generating functions and arrive at the same conclusion. From the gravitational perspective, this vanishing can be attributed to a cancellation between the classical spacetime geometry and the non-perturbative contributions from Euclidean wormholes.

\subsubsection{Generating functions for the time-shift operator}\label{sec:expdelta}
\begin{figure}[t]
	\centering
\includegraphics[width=5in]{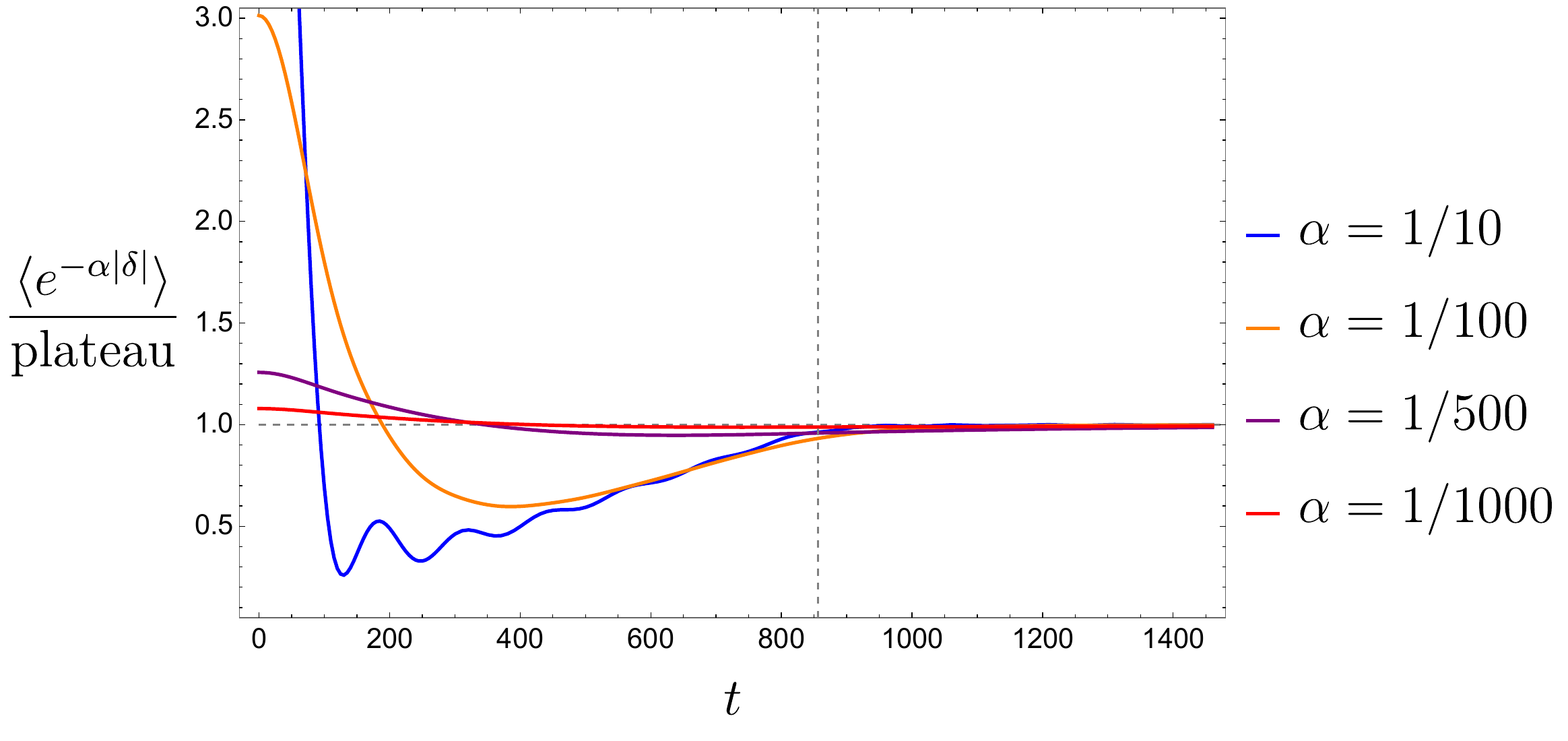}
	\caption{The time evolution of the generating function $\langle e^{-\alpha |\delta|} \rangle$ associated with the time-shift operator. With decreasing $\alpha$, the linear ramp region would gradually disappear. The numerical plot is generated by using the same parameters as the others.}
	\label{fig:expadelta}
\end{figure}  

As we have seen from the previous analysis about the fixed-length states, it is worth evaluating the generating functions that encapsulate the time evolution of the complexity or geodesic length in the $\alpha \to 0$ limit. Inspired by the definition of $\langle e^{-\alpha \ell} \rangle$ \eqref{eq:definegenerating}, we can define a new spectrum probe associated with the time-shift operator in terms of 
\begin{equation}\label{eq:definegeneratingdelta}
\begin{split}
 \langle e^{-\alpha |\delta|} \rangle   
:= \langle  \widehat{e^{-\alpha |\delta|} } \rangle := \int_{-\infty}^{\infty} P^{\mt{TFD}}(\delta,t) e^{-\alpha |\delta |} \, d \delta  \,. \\
\end{split}
\end{equation}
where we use the absolute value $|\delta|$ to avoid the exponential divergence from $\delta \to -\infty$. One can expect that this type of generating function presents a similar time evolution, \ie a slope-ramp-plateau structure, as shown in figure \ref{fig:expadelta}. With the decreasing of $\alpha$, one can similarly find that the linear ramp region is suppressed.

\begin{figure}[t]
	\centering
\includegraphics[width=4in]{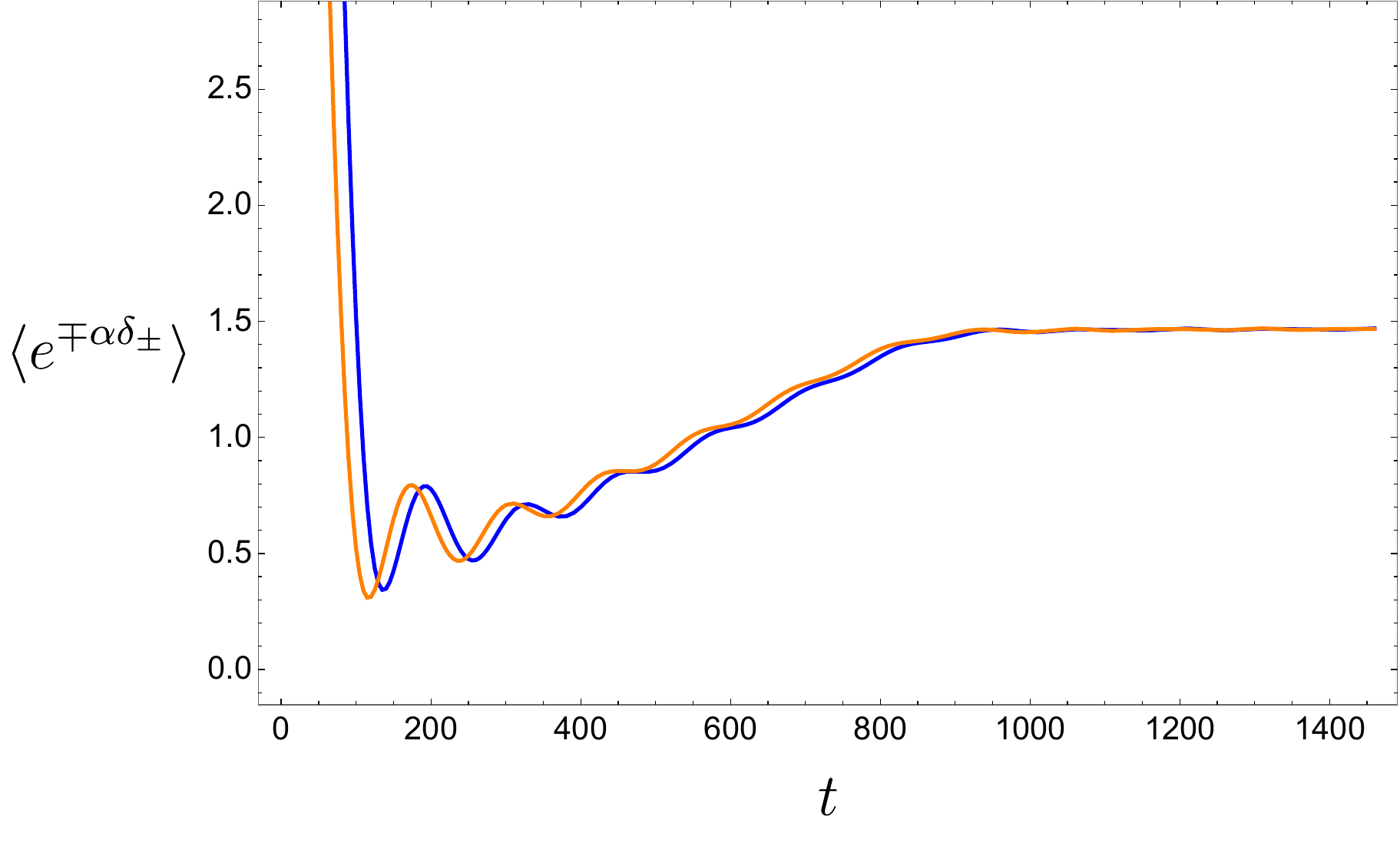}
	\caption{The time evolution of the two generating functions $\langle e^{-\alpha\delta_+} \rangle$ (blue) and $\langle e^{\alpha\delta_-} \rangle$ (orange). They present a similar slope-ramp-plateau structure.}
	\label{fig:expalphadeltapm}
\end{figure}

From the perspective of the expectation value of the time shift, it will be more useful to consider the generating functions for the positive and negative time shift, respectively. To wit, 
\begin{equation}\label{eq:gendeltapm}
    \langle e^{-\alpha\delta_+} \rangle :=\int_{0}^{\infty}\, d\delta\,e^{-\alpha \delta} P^{\mt{TFD}}(\delta,t)  \,, \qquad  \langle e^{\alpha\delta_-} \rangle :=\int^{0}_{-\infty}\, d\delta\,e^{+\alpha \delta} P^{\mt{TFD}}(\delta,t)  \,,
\end{equation}
which are associated with the following two distinct operators: 
\begin{equation}\label{eq:deltapmoperator}
\begin{split}
\hat{\delta}_{+}:=\int_{0}^{\infty}d\delta~\delta|\delta\rangle\langle \delta | \,, \qquad 
 \hat{\delta}_{-}:=\int_{-\infty}^{0}d\delta~\delta|\delta\rangle\langle \delta | \,.
\end{split}
\end{equation}
The generating function for the absolute value is thus given by the sum of these two, \ie $\langle e^{-\alpha |\delta|} \rangle = \langle e^{-\alpha \delta_+} \rangle +  \langle e^{\alpha \delta_-} \rangle$. Using these two corresponding generating functions, we can obtain a regularized time-shift value $\langle\delta\rangle_{\rm reg}$ by
\begin{equation}
\label{eq:definetotdelta}
    \begin{split}
        \langle \, \hat{\delta} \, \rangle_{\rm reg} :=  - \lim_{\alpha\to0} \frac{\partial}{\partial \alpha}  \left(  \langle e^{-\alpha\delta_+}\rangle - \langle e^{\alpha\delta_-}\rangle  \right)  :=  \langle \, \hat{\delta}_+ \, \rangle +\langle \, \hat{\delta}_- \, \rangle   \,.
    \end{split} 
\end{equation}
which is parallel to the regularized length value defined in eq.~\eqref{eq:fromexpaltol}. However, we highlight here that the above limit is free of divergence\footnote{The divergence remains in the absolute value case with 
\begin{equation}
   \langle \, |\hat{\delta}| \, \rangle :=  - \lim_{\alpha\to 0} \frac{\partial}{\partial \alpha}  \left(  \langle e^{-\alpha |\delta|}\rangle   \right)  :=\langle \, \hat{\delta}_+ \, \rangle   -  \langle \, \hat{\delta}_- \, \rangle \,.
\end{equation}
} due to the cancellation between $ \langle e^{-\alpha\delta_+}\rangle $ and $  \langle e^{\alpha\delta_-}\rangle $, \eg see figure \ref{fig:expalphadeltapm}. From the perspective of the expectation value of time shift, this cancellation is because $\langle \hat{\delta}_\pm  \rangle $ contains the same positive/negative divergence.

Using the approximate probability $P^{\mt{TFD}}$ derived in eq.~\eqref{eq:PTFD}, one can directly evaluate the $\delta$-integral to get the approximate expressions for different generating functions, which illustrate the numerical results shown in figure \ref{fig:expalphadeltapm}. Let us take $ \langle e^{-\alpha\delta_+} \rangle $ as an example. The classical part is approximately formulated as 
\begin{equation}
\langle e^{-\alpha\delta_+} \rangle \big|_{\rm classical} \approx  \frac{1}{(\Delta E)^2} \left[  -\frac{4\sin^2( \frac{\Delta E t}{2})}{t}+e^{-\alpha t}\left(2\alpha\mathrm{Ei}(\alpha t)-\alpha^{-}
\mathrm{Ei}(\alpha^{-}t)-\alpha^{+}\mathrm{Ei}(\alpha^{+}t)\right)  \right] \,, 
\end{equation}
with $\alpha^{\pm}=\alpha\pm i\Delta E$. In particular, let us focus on the regime with $\alpha t \gg 1$, which leads to 
\begin{equation}
\langle e^{-\alpha\delta_+} \rangle \big|_{\rm classical} \approx     \frac{2\pi }{\Delta E} e^{- \alpha t } +
\frac{2}{\Delta E^2 t^2}\left( \frac{1}{\alpha}
- \frac{\alpha  \cos (\Delta E t)+\Delta E \sin (\Delta E t)}{\alpha ^2+\Delta E^2}\right) + \mathcal{O}\left(\frac{1}{t^3}\right) \,. 
\end{equation}
It is evident that the classical generating function—acting as a classical correlation function—continues decaying with time, corresponding to the slope region of its characteristic behavior. By incorporating the contributions from the quantum corrections, denoted as $P^{\mt{TFD}}_{\rm quantum}$ (see figure~\ref{fig:PTFDquantum}), one can perform the $\delta$-integral in a straightforward manner to derive the quantum part of the generating function as
\begin{equation}
    \langle e^{-\alpha\delta_+} \rangle \big|_{\rm quantum} \approx  
    \dfrac{2\pi}{\Delta E}  \times 
    \begin{cases}
             \dfrac{\left(  \alpha t - 1+2e^{-\alpha t}- e^{-\alpha(t+T_{\mt{H}})}\right)}{\alpha^2 T_{\mt{H}}^2} \,, \qquad t<T_{\mt{H}} \\[1em]
              \dfrac{(\alpha T_{\mt{H}}-e^{-\alpha(t+T_{\mt{H}})}(e^{\alpha T_{\mt{H}}}-1)^2)}{\alpha^2 T_{\mt{H}}^2} \,,\qquad t>T_{\mt{H}} 
        \end{cases} \,.
\end{equation}
Obviously, the quantum corrections lead to a linear ramp for $t< T_{\mt{H}}$ and a final-time plateau after the Heisenberg time for a generic $\alpha$ case. As a summary, the generating function $ \langle e^{-\alpha\delta_+} \rangle$ presents the slope-ramp-plateau structure and it is described by 
\begin{equation}\label{eq:expadeltaphalsum}
  \langle e^{-\alpha\delta_+} \rangle \approx 
\frac{2\pi}{\Delta E} \times \begin{cases}
e^{-\alpha t} + \mathcal{O}(\frac{1}{\alpha t^2 \Delta E})\,, \qquad \text{slop at early times} \quad \alpha t \sim 1  \,,\\[1em]
\dfrac{t}{\alpha \, T_{\mt{H}}^2} + \mathcal{O} (e^{-\alpha t}) \,,\qquad \text{ramp at the middle stage} \quad  t <   T_{\mt{H}}  \,,\\[1em]
 \dfrac{1}{\alpha \, T_{\mt{H}}} + \mathcal{O}(e^{-\alpha (t-T_{\mt{H}})})\,,\qquad \text{plateau at late times} \quad  t >   T_{\mt{H}}  \,. \\
\end{cases}
\end{equation}
Given the similarities between the two probability distributions $P^{\mt{TFD}} (\delta,t)$ and $P(\ell,t)$, 
it is natural that the two generating functions, $\langle e^{-\alpha\delta_+} \rangle$ and $\langle e^{-\alpha \ell} \rangle$, bear a striking resemblance to each other, as shown in eq.~\eqref{eq:expalphalsum} and eq.~\eqref{eq:expadeltaphalsum}. The presence of distinct constant factors can be attributed to the normalization conditions applied to the states $|\ell\rangle$ and $|\delta\rangle$. Similar integral yields the approximate expression for the generating function for $\hat{\delta}_-$ (with $t>0$), \ie 
\begin{equation}\label{eq:expdelta02}
    \begin{split}
     \langle e^{\alpha\delta_-}\rangle \approx &\frac{1}{(\Delta E)^2}\left[ \frac{4\sin^2{ (\frac{\Delta E}{2}t})}{t} - e^{\alpha t} \left(2\alpha \mathrm{E}_1(\alpha t) 
     -\alpha^{-}\mathrm{E}_1(\alpha^{-}t)
     -\alpha^{+}\mathrm{E}_1(\alpha^{+}t)\right) \right] \\
     \qquad\qquad&+ \frac{2\pi}{\Delta E} \times 
     \begin{cases}
         \dfrac{1}{\alpha^2  T_{\mt{H}} ^2}\left(\alpha t +1-e^{\alpha(t- T_{\mt{H}})} \right) \,, \qquad t<  T_{\mt{H}} \\[1em]
         \dfrac{1}{\alpha  T_{\mt{H}}} \,, \qquad \qquad\qquad\qquad\qquad \quad \,\,  t> T_{\mt{H}}
     \end{cases} \,,
    \end{split}
\end{equation}
which exhibits a similar slope-ramp-plateau structure as its counterpart $ \langle e^{-\alpha\delta_+}\rangle$ shown in eq.~\eqref{eq:expadeltaphalsum}.

Equipped with these generating functions, one can subtract the time evolution by carefully taking the $\alpha\to 0$ limit for the time-shift operators $\hat{\delta}_\pm$. From a physical standpoint, one is primarily interested in the expectation value of the complete time-shift operator $\hat{\delta}$ defined in eq.~\eqref{eq:deltavalue}. As discussed earlier, in the regularization scheme the leading contribution to the regularized expectation value $\langle \hat{\delta} \rangle_{\rm reg}$ vanishes. In fact, by using the two approximate generating functions, $\langle e^{-\alpha\delta_+}\rangle$ and $\langle e^{\alpha\delta_-}\rangle$, one arrives at the same conclusion. 

To be more explicit, let us first treat the classical expectation value $\langle \hat{\delta}\rangle_{\rm classical}$ and then include its quantum correction $\langle \hat{\delta}\rangle_{\rm quantum}$. Taking the derivative with respect to $\alpha$ of the classical generating functions, we have
\begin{equation}\label{eq:deltaclassical}
\begin{split}
\langle \hat{\delta}\rangle_{\rm classical} 
&=-\lim_{\alpha\rightarrow0}\partial_{\alpha}\Bigl[\langle e^{-\alpha\delta_+}\rangle -\langle e^{\alpha\delta_-}\rangle\Bigr]_{\rm classical} 
=\lim_{\alpha\rightarrow0}\partial_{\alpha}\langle e^{-\alpha\delta_+}\rangle_{\rm classical} \\
&\approx \frac{2\pi}{\Delta E}\, t \,,
\end{split}
\end{equation}
where the dominant contribution comes from the positive time shift $\delta_+$. In other words, the classical expectation value of the time-shift operator is equivalent to the boundary time $t$ of the TFD state. It grows linearly as expected and can be interpreted as an indicator of the age of the black hole. On the other hand, the quantum correction is derived as
\begin{equation}
\left[\langle e^{-\alpha\delta_+}\rangle -\langle e^{\alpha\delta_-}\rangle\right]_{\rm quantum} \approx 
\frac{2\pi}{\Delta E} \times 
     \begin{cases}
         \dfrac{e^{-\alpha(t+T_{\mt{H}})} (e^{\alpha t}-1)\left( 1+e^{\alpha t}-e^{2\alpha T_{\mt{H}}} \right)}{\alpha^2  T_{\mt{H}} ^2} \,,\quad  t<  T_{\mt{H}} \\[1em]
        \dfrac{-e^{-\alpha(t+T_{\mt{H}})}(e^{\alpha T_{\mt{H}}}-1)^2}{\alpha^2 T_{\mt{H}}^2}  \,, \qquad\qquad\quad \quad \,\,  t> T_{\mt{H}}
     \end{cases} \,.
\end{equation}
Expanding this result around $\alpha\to 0$, we find that the quantum correction contributes a negative term of the same order:
\begin{equation}
\begin{split}
\langle \hat{\delta}\rangle_{\rm quantum}
&=-\lim_{\alpha\rightarrow0}\partial_{\alpha}\Bigl[\langle e^{-\alpha\delta_+}\rangle -\langle e^{\alpha\delta_-}\rangle\Bigr]_{\rm quantum}
\approx -\frac{2\pi}{\Delta E}\, t \,.
\end{split}
\end{equation}
Thus, the cancellation between the classical contribution and the quantum correction yields the vanishing of the regularized time shift\footnote{For comparison, the absolute value of the time shift is given by
\begin{equation}\label{eq:eveabs}
\begin{split}
\langle |\hat{\delta}| \rangle 
&=-\lim_{\alpha\rightarrow0}\partial_{\alpha}\Bigl[\langle e^{-\alpha\delta_+}\rangle +\langle e^{\alpha\delta_-}\rangle\Bigr]
=\langle \hat{\delta}_+ \rangle - \langle \hat{\delta}_- \rangle \\
&\approx \frac{4\bigl(\log (\alpha t)-1 -\gamma \bigr)}{(\Delta E)^2} +  \frac{4\pi}{\alpha^2\,T_{\mt{H}}\,\Delta E} 
+\frac{2\pi}{\Delta E}  \times \begin{cases}
   \dfrac{(t- T_{\mt{H}})^3}{3\,T_{\mt{H}}^2}  \,, & \quad t < T_{\mt{H}} \\[1em]
   0 \,, & \quad t > T_{\mt{H}} \,,
 \end{cases}
\end{split}
\end{equation}
where the first two terms represent the divergences from the classical and quantum parts, respectively.}:
\begin{equation}\label{eq:vanishidelta}
\langle \hat{\delta}\rangle_{\rm reg}
=-\lim_{\alpha\rightarrow0}\partial_{\alpha}\Bigl[\langle e^{-\alpha\delta_+}\rangle -\langle e^{\alpha\delta_-}\rangle\Bigr]
\approx 0 \,.
\end{equation}
One generally expects that the emission of baby universes at late times could effectively reduce the value of the time shift, thereby rendering the black hole younger. However, the result \eqref{eq:vanishidelta} shows that the classical contribution from the disk geometry is exactly canceled by the wormhole (nonperturbative) contributions at all time scales—even in the regime where the classical disk geometry is generically dominant.

\subsubsection{Spectral probes from Fourier transformations}

In a similar manner to the preceding discussion on the generating function $\langle e^{-\alpha \ell} \rangle$ for the geodesic length, we can also use the spectral representation of the generating functions associated with the time shift operator to obtain spectral probes such as the spectral complexity $\mathrm{SC}(t)$ and its generating function $\mathrm{G}_{\mt{SC}}(t)$. The key point is that the sum over the basis $|\delta \rangle$ is nothing but the Fourier transformation. 

Recalling the definition of the probability $P^{\mt{TFD}}$ shown in eq.~\eqref{eq:Pdeltat}, we can decompose the generating function, \eg $\langle e^{-\alpha |\delta|} \rangle$ in terms of \footnote{If we consider $\langle e^{-i \alpha \delta} \rangle$, we will find that the Fourier transformation results in a delta function, \ie 
\begin{equation}
\begin{split}
 \langle e^{-i \alpha \delta }  \rangle \longrightarrow \int_{-\infty}^{+\infty} \, e^{i \delta (E_{12}-\alpha) } d \delta &= 2\pi \delta \left(  E_1 -E_2 -\alpha  \right)\,,
\end{split}
\end{equation}
which only picks up the contribution at $E_1 - E_2 =\alpha$. It will make $\langle e^{-i \alpha \delta} \rangle$ keeping oscillatory in the form of $\cos (\alpha t)$. This is why we do not consider this ``natural'' generating function in this paper.}
\begin{equation}\label{eq:expadeltapm}
 \langle e^{-\alpha |\delta|} \rangle  \equiv  \langle e^{-\alpha \delta_+}\rangle +  \langle e^{\alpha \delta_-} \rangle  \equiv   \frac{e^{2S_0}}{Z^2} \int dE_1\int dE_2 \, e^{-iE_{12}t} \langle  D(E_1) D(E_2) \rangle \int_{-\infty}^{\infty} d\delta \, e^{iE_{12}\delta}   e^{-\alpha |\delta|} \,,  
\end{equation}
where $E_{12} = E_1 -E_2$ is referred to as the energy difference or level spacing. It is obvious that the $\delta$-integral (related to summing over all states $|\delta \rangle$) is equivalent to the (inverse) Fourier transformation between the time shift $\delta$ and level spacing $E_{12}$, namely
\begin{equation}
\int_{-\infty}^{\infty} d\delta \,    e^{-\alpha |\delta|} e^{iE_{12}\delta}= \frac{1}{\alpha-i E_{12}}+\frac{1}{\alpha+i E_{12}}= \frac{2 \alpha }{(E_1-E_2)^2 + \alpha^2 }  \,. 
\end{equation}
Ignoring the normalization constant, the discrete spectral representation of the generating function $ \langle e^{-\alpha |\delta|} \rangle$ takes the form of
\begin{equation}
\langle e^{-\alpha |\delta|} \rangle \quad  \longleftrightarrow \quad  \mathrm{G}_{|\delta|} (\alpha, t) = \sum_{E_1, E_2} \frac{ 2\alpha }{(E_1-E_2)^2 +\alpha^2}   e^{- i(E_1 -E_2)t}\,.  
\end{equation}
Comparing with the generating function $\mathrm{G}_{\mt{SC}}(t)$ (see eq.~\eqref{eq:GSC}) for the spectral complexity, there is no doubt that they are manifestly similar to each other. This similarity also illustrates that both of them show a similar slope-ramp-plateau structure. Correspondingly, this new generating function $ \mathrm{G}_{|\delta|} (\alpha, t)$ also produces the spectral complexity, \ie 
\begin{equation} \label{eq:Cdeltareg}
    - \lim_{\alpha \to 0} \left( \frac{d}{d\alpha} \mathrm{G}_{|\delta|} (\alpha, t)  \right)    =  2 \times \mathrm{SC}(t) +  \text{Divergent Constant}  \,. 
\end{equation}
after ignoring the divergent constant. The subtraction of this divergent part could be achieved by removing the diagonal part by hand. As a result, we can find that the regularized expectation value $\langle | \hat{\delta} | \rangle_{\rm reg}$ is derived as the spectral complexity, up to a time-independent constant, \ie 
\begin{equation}
\langle | \hat{\delta} | \rangle_{\rm reg}  \sim   
\sum_{E_1 \ne E_2} \frac{ - 2 \cos((E_1-E_2)t) }{(E_1-E_2)^2} + \text{Constant} = 2 \times   \mathrm{SC}(t)  \,. 
\end{equation}
The analysis for the two generating functions $\langle e^{-\alpha \delta_+} \rangle$ and $\langle e^{\alpha \delta_-} \rangle $ is similar. Their spectral representations are explicitly given by\footnote{Note that the subleading term here is different from that of $\rm{G}_{|\delta|} (\alpha, t)$ (and $\rm{G}_{\rm{SC}} (\alpha, t)$ defined in eq.~\eqref{eq:GSC} for the spectral complexity):
\begin{equation}
\frac{d}{d\alpha} \mathrm{G}_{|\delta|} (\alpha, t)    = \sum_{E_1, E_2}  e^{- i(E_1 -E_2)t} \left( \frac{ 1}{(E_1-E_2)^2 +\alpha^2}  - \frac{ 2\alpha^2  }{((E_1-E_2)^2 +\alpha^2)^2}   \right) \,,
\end{equation}
where the second term decays in terms of $\alpha^2$ and {\it does not} contribute to the time evolution of the absolute value of time shift $\langle |\hat{\delta} \rangle|$ as derived in the footnote \eqref{eq:eveabs}. To see this difference, we remind the reader of the following Fourier transformations:
\begin{equation}
\int_{-\infty}^{\infty} \frac{ 2\alpha^2 }{\left(\alpha ^2+(E_{12})^2\right)^2} e^{-i E_{12} t} dE_{12}  =     \frac{\pi   (\alpha  t+1)}{ \alpha }e^{-\alpha t} \approx \frac{\pi }{\alpha }-\frac{\alpha }{2}  \pi  t^2 +\mathcal{O}(\alpha^2)  \,,
\end{equation}
and 
\begin{equation}
\int_{-\infty}^{\infty} \frac{  2 \alpha i E_{12} }{\left(\alpha^2+(E_{12})^2\right)^2} e^{-i E_{12} t} dE_{12}  = \pi t  e^{- \alpha t} \approx   \pi t +\mathcal{O}(\alpha)  \,.
\end{equation}
} 
\begin{equation}\label{eq:generationpm}
\left\{ \langle e^{-\alpha \delta_+} \rangle \,, \langle e^{\alpha \delta_-} \rangle  \right\} \quad  \longleftrightarrow \quad  \mathrm{G}_{\delta_\pm} (\alpha, t) = \sum_{E_1, E_2} \frac{ \alpha \pm i (E_1 -E_2) }{(E_1-E_2)^2 +\alpha^2}   e^{- i(E_1 -E_2)t}\,. 
\end{equation}
However, we note that the spectral representation of the positive/negative time shift $\langle \hat{\delta}_\pm \rangle $is not well-defined since the $\alpha \to 0$ limit is singular for its corresponding generating function $\mathrm{G}_{\delta_\pm} (\alpha, t)$. The derivatives of the generating functions are given by 
\begin{equation}
\mp \partial_\alpha \mathrm{G}_{\delta_\pm} (\alpha, t) =
\mp \sum_{E_1, E_2}  e^{- i(E_1 -E_2)t} \left( \frac{ 1}{(E_1-E_2)^2 +\alpha^2}  -  2\alpha \times \frac{ \alpha \pm i (E_1 -E_2) }{((E_1-E_2)^2 +\alpha^2)^2}   \right) \,. 
\end{equation}
Naively, taking $\alpha \to 0$ limit for defining the spectral complexity $\langle \hat{\delta}_\pm \rangle$ 
implies that the second term vanishes and gives rise to $\langle \hat{\delta}_+ \rangle  = -\langle \hat{\delta}_- \rangle$. This looks match with our previous result $\langle \hat{\delta}  \rangle_{\rm reg} \approx 0$.   
However, this is not a correct derivation since the linear $\alpha$ term appearing in the above derivative is non-trivial due to the singularity at $E_1=E_2$ at $\alpha \to 0$. The simplest example is obtained by a limit representation for the Dirac delta function, \eg 
\begin{equation}
\lim_{\alpha\to 0}  \frac{\alpha}{(E_1-E_2)^2 +\alpha^2}  = \pi \, \delta (E_1 -E_2)   \,.
\end{equation}
To understand why we still obtained the vanishing expectation value of time shift $\langle \hat{\delta} \rangle = \langle \hat{\delta}_+ \rangle + \langle \hat{\delta}_- \rangle$. Let us first note that the time-shift expectation value $\langle \hat{\delta} \rangle$ is generated by the following function:  
\begin{equation}
 \langle e^{-\alpha \delta_+}\rangle  - \langle e^{\alpha \delta_-} \rangle    \,\longleftrightarrow \,  \mathrm{G}_{\delta_+} (\alpha, t) - \mathrm{G}_{\delta_+} (\alpha, t) = \sum_{E_1, E_2} \frac{ 2 i (E_1 -E_2) }{(E_1-E_2)^2 +\alpha^2}   e^{- i(E_1 -E_2)t}\,. 
\end{equation}
Obviously, the generating functions contain a pole located at 
\begin{equation}
\text{pole of generating function}: \qquad E_{12} = E_1 -E_2 =  - i \alpha \,,  
\end{equation}  
on the complex plane of $E_{12}$. The universal pole structure plays a crucial role in determining the time evolution of not only generating functions but also quantum complexity measures. We will find that the vanishing of the expectation value of the time shift $\langle \hat{\delta} \rangle$ is due to the cancellation between the residue of this pole from the disconnected spectral correlation and that from the sine kernel, \ie
\begin{equation}
    \begin{split}
    D_{\text{Disk}}(E_i)D_{\rm{Disk}}(E_j)    
   -e^{-2S_0}\frac{\sin^2(\pi e^{S_0}D_{\rm{Disk}}(\bar{E})(E_i-E_j))}{\pi^2(E_i-E_j)^2} \,. 
    \end{split}
\end{equation}
Summing over the (continuous) spectrum in terms of $E_{12}$ is nothing but another Fourier transformation between the energy difference $E_{12}$ and time $t$. For simplicity, we focus on performing the integral with $\int^\infty_{-\infty} dE_{12}$, which captures the leading contributions for a finite energy window.\footnote{For example, one can get  
\begin{equation}\label{eq:approximationintegral}
\int_{-\Delta E}^{\Delta E} \frac{ 2iE_{12}}{\alpha ^2+(E_{12})^2} e^{-i E_{12} t} dE_{12}   \approx    2 \pi  e^{-\alpha t}-\frac{4 \Delta E  \cos ( \Delta E t)}{t (\alpha ^2+\Delta E^2)  }  + \mathcal{O}\left(\frac{1}{t^2}\right)\,.
\end{equation}
The leading term is equivalent to 
\begin{equation}
\int_{-\infty}^{\infty} \frac{ 2i E_{12} }{\alpha ^2+(E_{12})^2} e^{-i E_{12} t} dE_{12}  =     2\pi e^{-\alpha t} \,.
\end{equation}
The subleading correction ignored by the approximation \eqref{eq:approximationintegral} is at the order of $\mathcal{O}(\Delta E)$. 
}
We start from the disconnected spectral correlator whose expansion around $E_1 \sim E_2$ is written a
\begin{equation}
 D_{\mathrm{Disk}}(E_1)D_{\rm{Disk}}(E_2) \approx (D_{\rm{Disk}} (\bar{E}) )^2 +  \frac{(E_{12})^2}{4}\left( D_{\rm{Disk}}'' - D_{\rm{Disk}} D_{\rm{Disk}}' \right)  + \mathcal{O}((E_{12})^4) \,. 
\end{equation}
Noting the pole at $E_{12} = i \alpha$, the classical contribution to the generating function is derived as
\begin{equation}
\int_{-\infty}^{\infty} dE_{12} \, \frac{ 2i E_{12} }{\alpha ^2+(E_{12})^2}  e^{-i E_{12} t} D_{\mathrm{Disk}}(E_1)D_{\rm{Disk}}(E_2)  =   2\pi e^{-\alpha t} (D_{\mathrm{Disk}}(\bar{E}))^2 + \mathcal{O}(\alpha^2)  \,,
\end{equation}
where only the first term contributes after taking the derivative and $\alpha \to 0$. Its derivative yields 
the linear growth of $\langle \hat{\delta} \rangle_{\rm classical}$ as we have derived in eq.~\eqref{eq:deltaclassical}.
Similarly, the quantum part from the sine kernel is   
\begin{equation}
\begin{split}
\int_{-\infty}^{\infty} dE_{12} \, &e^{-i (E_1-E_2)t}    \frac{ i (E_1-E_2)}{\alpha ^2+(E_1-E_2)^2}  \left( e^{-2S_0}\frac{\sin^2(\pi e^{S_0}D_{\rm{Disk}}(\bar{E})(E_i-E_j))}{\pi^2(E_i-E_j)^2} \right)  \\
&=   \frac{2e^{-2S_0} \sinh ^2\left(\pi  \alpha  D_{\rm Disk} e^{S_0}\right)}{\pi  \alpha ^2}e^{-\alpha t} \,.
\end{split}
\end{equation}
Combing the above two pieces from the disconnected spectral correlator and the sine kernel, we can find the precise cancellation for evaluating the time shift $\langle \hat{\delta} \rangle $, \ie 
\begin{equation}
- \lim_{\alpha\to 0} \partial_\alpha (\text{disconnected part} - \text{sine kernel part}) =  \left( 2\pi t  -2\pi t \right) \, (D_{\mathrm{Disk}}(\bar{E}))^2  =0  \,.
\end{equation}
This illustrates the previous finding for the vanishing of the time shift, \ie 
\begin{equation}
\langle \hat{\delta} \rangle_{\rm reg} = 0\,. 
 \end{equation}
In summary, we conclude that the vanishing of the expectation value of the time-shift operator is a direct consequence of the exact cancellation between the classical and quantum contributions. Moreover, this cancellation is encoded in the spectral correlator (see eq.~\eqref{eq:sine}) and the universal pole at $E_{12}=i\alpha$ in the generating function.

\section{Conclusion and Outlook}\label{section:comment}

Understanding the quantum nature of black hole interiors and their connection to quantum chaos has been a central theme in recent developments in quantum gravity. In this work, we have explored how non-perturbative overlaps in JT gravity encode universal signatures of quantum chaos, quantum complexity, and black hole interior dynamics. Our analysis reveals deep connections between quantum gravitational observables and the spectral properties of chaotic quantum systems. Below, we summarize our key results on non-perturbative overlaps in JT gravity and the generating functions of quantum complexity. We then discuss potential generalizations of our findings and highlight open questions for future investigations.

The core of our study has been the overlaps between the TFD state and two distinct classes of states: fixed-length states $|\ell \rangle$ and time-shifted TFD states $|\delta \rangle$. The squared overlaps, \eg $P(\ell, t)$ and $P^{\mt{TFD}}(\delta, t)$, serve as probability distributions that define the expectation values of gravitational operators, thereby probing the non-perturbative structure of black hole spacetime. In the semiclassical limit, these probability distributions exhibit sharp peaks at classical expectation values, reflecting well-understood geometric properties of the two-sided black hole. Using the universal spectral correlation function \eqref{eq:sine} from random matrix theory, we are able to introduce non-perturbative quantum corrections into the squared overlaps. As derived in eq.~\eqref{eq:Ptotal} and illustrated in figures \ref{fig:Ptotal01} and \ref{fig:Ptotal02}, the time evolution of the total probability $P(\ell, t)$ associated with fixed-length states follows the well-known slope-ramp-plateau structure, reminiscent of the spectral form factor in random matrix theory. Similarly, the probability $P^{\mt{TFD}}(\delta, t)$, defined via time-shifted TFD states, exhibits a comparable evolution, as shown in figure \ref{fig:PTFD}, since it is equivalent to the spectral form factor. This universal behavior strongly suggests that the quantum properties of black hole spacetime are governed by the same spectral statistics that underlie quantum chaotic systems. Moreover, it indicates that non-perturbative overlaps provide a novel framework for understanding the late-time behavior of quantum gravitational systems.

Summing over all (infinitely many) fixed-length states in $P(\ell, t)$ allows us to define the expectation value of the geodesic length, which characterizes the volume of the black hole interior. However, this naive expectation is ill-defined due to divergence issues, which arise from the over-completeness of the basis expanded by infinite  fixed-length states $|\ell \rangle$. Unlike the divergent length expectation value, its time derivative remains well-defined. As shown in eq.~\eqref{eq:Lvalues} in section \ref{sec:lengthvalue}, the geodesic length grows linearly with time in the classical limit, consistent with AdS black hole geometry. However, including quantum corrections, the regularized geodesic length $\langle \hat{\ell} \rangle_{\rm gen}$ saturates to a plateau after the Heisenberg time $T_{\mt{H}}$, as given by eq.~\eqref{eq:dLdtquantum}. This finally recovers the expected time evolution of quantum complexity in chaotic systems, in alignment with the complexity=volume conjecture.

Rather than directly defining the expectation value, a key result of this work is the introduction of {\it generating functions of quantum complexity}. The first example, $\langle e^{-\alpha \ell }\rangle$, studied in section \ref{sec:expl}, serves as a regularized version of the length operator, remaining finite for any positive $\alpha$. We explicitly demonstrate in eq.~\eqref{eq:expalphalsum} and figure \ref{fig:expal} that the time evolution of the generating function $\langle e^{-\alpha \ell }\rangle$ exhibits the slope-ramp-plateau structure for a generic $\alpha$. The time evolution of complexity, \eg $\langle \ell \rangle$, is encoded in the generating function in the limit $\alpha T_{\mt{H}} \to 0$. Importantly, we find that the linear ramp disappears in this limit, reflecting the fact that the geodesic length $\langle \hat{\ell} \rangle$ follows a simpler time evolution: an initial linear growth followed by a late-time plateau after the Heisenberg time $T_{\mt{H}}$. 

To further demonstrate that both complexity measures and their generating functions probe the spectral properties of quantum systems, we define the spectral representation of the generating function $\langle e^{-\alpha \ell} \rangle$ in eq.~\eqref{eq:ealdiscrete}. The universal component $\rm{G}_{\mt{SC}}(\alpha, t)$, given by eq.~\eqref{eq:GSC}, plays the role of the generating function for spectral complexity \eqref{eq:SC}. Motivated by these results for fixed-length states $|\ell \rangle$, we extend our investigation to the generating function and expectation value of the time-shift operator $\hat{\delta}$ in section \ref{sec:timeshift}. The time evolution of the corresponding generating functions, \eg $\langle e^{-\alpha |\delta|} \rangle$ and $\langle e^{\mp \alpha \delta_{\pm}} \rangle$, follows the same slope-ramp-plateau structure observed in $\langle e^{-\alpha \ell} \rangle$. However, we find that the classical expectation of the time shift, which predicts linear growth with time, is precisely canceled by quantum corrections. Consequently, the expectation value of the time shift vanishes at all times, namely 
\begin{equation}
\langle \hat{\delta} \rangle_{\rm reg} = 0,
\end{equation}
as derived in eq.~\eqref{eq:vanishidelta}. Noting the dominating contributions origin from the universal pole $E_1 - E_2 = i \alpha$ of the generating function, we then further prove that this ``exact cancellation'' occurs because the residues of the disconnected correlator and the sine kernel contribute oppositely in the limit $\alpha \to 0$.

These results highlight the crucial role of generating functions of quantum complexity in understanding not only the time evolution of complexity but also the quantum nature of black hole interiors. Through their spectral representation, these generating functions encode rich information about the underlying quantum gravitational dynamics, revealing deep connections between quantum chaos and the spectral statistics of quantum gravity. To conclude, we now turn to discuss important subtleties, potential generalizations of our results, and open questions that warrant further exploration.

\subsection*{Complete Basis and Gram-Schmidt Orthogonalization}

In this paper, we explicitly computed the non-perturbative probability distributions $P(\ell,t)$ and $P^{\mt{TFD}}(\delta,t)$ for the fixed-length states $|\ell \rangle$ and time-shifted states $|\delta \rangle$, respectively. However, a naive summation over all states to define the expectation values of the geodesic length operator, $\hat{\ell}:=\sum_{\ell} \ell |\ell\rangle\langle \ell|$, and the time-shift operator, $\hat{\delta}:=\sum_{\delta} \delta |\delta \rangle\langle \delta|$, leads to divergences. This suggests that these infinite states do not constitute a well-defined (complete) basis in the full non-perturbative Hilbert space, and their naive use as eigenstates of physical operators is problematic.

To construct well-defined non-perturbative operators, it is necessary to modify these naive eigenstates by including non-perturbative corrections. A systematic approach is to apply {\it Gram-Schmidt orthogonalization} to the fixed-length and time-shifted states, as proposed by one of the authors in \cite{Miyaji:2024ity}. This procedure ensures that the resulting states form a complete orthonormal basis, thereby resulting in a consistent definition of probability distributions associated with geodesic length, time shift, and the black hole/white hole transition. However, a consequence of this orthogonalization is that the non-perturbative corrections become significant for states with very large geodesic lengths or large time shifts, making their bulk interpretation less straightforward.

\subsection*{Complete Basis from Time-Shifted TFD States}

As demonstrated in section \ref{sec:timeshift}, the expectation value of the time shift operator, obtained by summing over all $|\delta \rangle$ states, suffers from the same type of divergence observed for the geodesic length operator. This suggests that the infinite set of time-shifted TFD states also forms an over-complete basis. On the other hand, we have shown—through multiple perspectives—a surprising result: when quantum corrections are included, the regularized time shift completely vanishes. 

However, one can implement a time-dependent regularization by applying a Gram–Schmidt-type orthogonalization process to the infinite set of $|\delta \rangle$ states, thereby constructing a complete basis that satisfies the completeness relation:
\begin{equation}\label{eq:completetimeshift}
\left( \int P^{\mt{TFD}}(|t-\delta|) \, d \delta \right)_{\text{reg}} = \sum_{\text{complete basis}} P^{\mt{TFD}}(|t-\delta|) = 1  \,. 
\end{equation}
Let us denote the (continuous or discrete) spectrum of the complete basis as
\begin{equation}
\left\{ \delta_{\rm c} \right\} :=\left\{ -\delta_{\rm max}, \cdots , - \delta_i, \cdots, \delta_i \, \cdots\,\delta_{\rm max} \right\} \,,
\end{equation}
where we assume it is symmetric between negative and positive time shifts. The expectation value of the time shift is then given by
\begin{equation}
\begin{split}
\langle \hat{\delta}   \rangle  
= \sum_{\left\{ \delta_{\rm c} \right\}} P^{\mt{TFD}}(|t-\delta|)  \, \delta  &= \sum_{\left\{ \delta_{\rm c} \right\} -t} P^{\mt{TFD}}(|\tilde{\delta}|) \left(\tilde{\delta} + t \right)     \,,\\
&= t- \sum_{\left\{ \delta_{\rm max} -t , \cdots, \delta_{\rm max} +t\right\}} P^{\mt{TFD}}(|\tilde{\delta}|) \,\tilde{\delta} \,,
\end{split}
\end{equation}
where we have used the completeness relation \eqref{eq:completetimeshift} and the symmetry of the complete basis $\left\{ \delta_{\rm c} \right\}$ to derive the second line. It is evident that the correction to the linear time growth arises from the second term, which is sensitive to the maximum time shift $\delta_{\rm max}$. Investigating the behavior of the time shift $\langle \delta \rangle$ by constructing such a complete basis from time-shifted TFD states would be an intriguing direction for future exploration.

\subsection*{The Residual Ramp and the Peak of Complexity}

\begin{figure}[t]
	\centering
\includegraphics[width=5.9in]{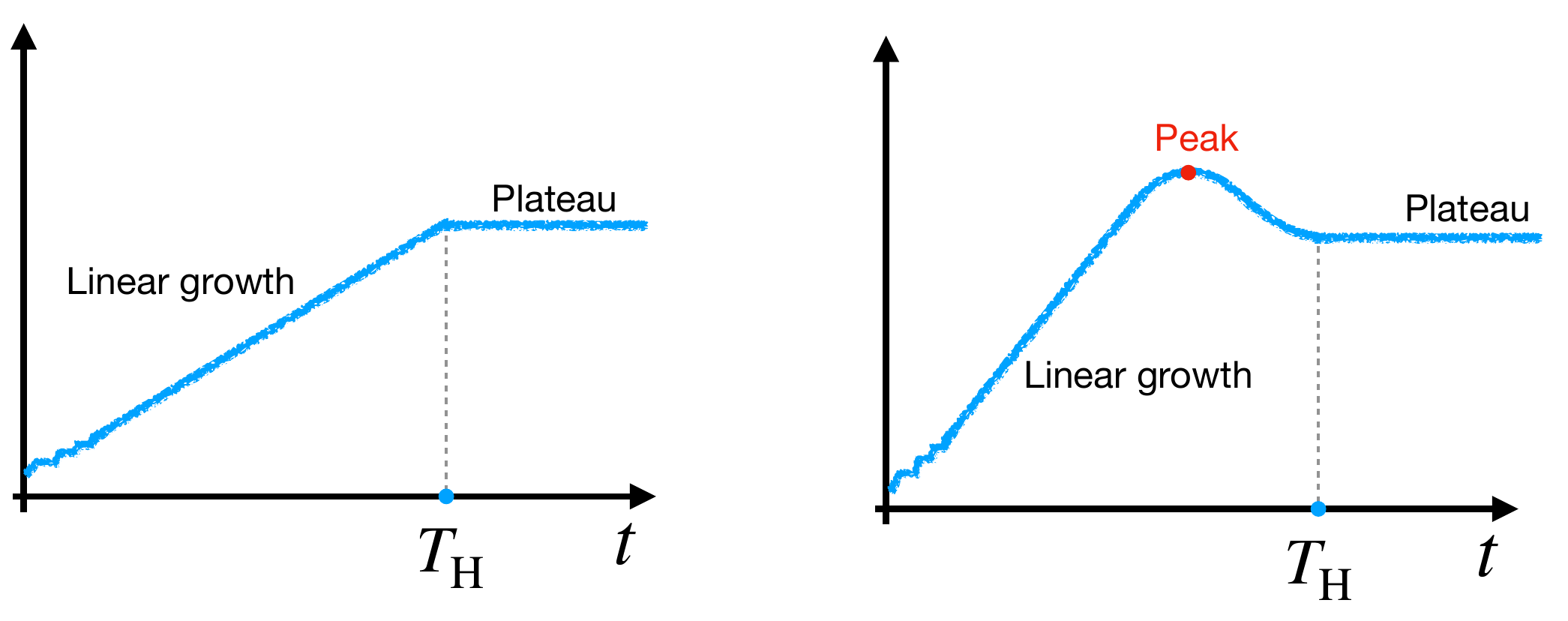}
	\caption{Two typical time evolution pictures for quantum complexity measures of chaotic systems. The appearance of a peak corresponds to the remaining dip in the generating function at the $\alpha \to 0$ limit. The decreasing part of the quantum complexity is thus given by the residual part of the ramp.}
	\label{fig:secondlaw}
\end{figure} 

One of the key findings of this paper is that the universal time evolution of complexity for chaotic systems (as shown in figure \ref{fig:secondlaw}) is governed by the disappearance of the linear ramp in the generating function at the $\alpha \to 0$ limit. In other words, the linear-plateau type evolution of (infinite) complexity measures emerges from the slope-ramp-plateau structure of generating functions in this special limit. By decreasing the parameter $\alpha$, we observe that the transition time from the slope regime to the ramp regime monotonically increases until it reaches the Heisenberg time $T_{\mt{H}}$. 

Denoting this transition time\footnote{For generating functions with a generic $\alpha$, this transition time corresponds to the so-called dip in the slope-dip-ramp-plateau structure. Here, we examine its dependence on the parameter $\alpha$. Since its value varies from $e^{S_0/2}$ to $e^{S_0}$, we refer to it as the transition time to distinguish it from the typical dip.} as $t_{\rm dip}$, we fix it as the point where the time derivative of the generating function changes sign:
\begin{equation}
\left( \frac{d}{dt} \langle e^{-\alpha \ell } \rangle   \right)\bigg|_{t=t_{\rm dip}(\alpha)} =0\,.
\end{equation}
Formally, the typical linear-plateau structure of complexities is equivalent to the following approximate equality:
\begin{equation}\label{eq:tcrt}
\lim_{\alpha T_{\mt{H}} \to 0}  t_{\rm dip}(\alpha) \approxeq  T_{\mt{H}} \,.
\end{equation}
For example, this is illustrated in figure \ref{fig:SYKspecs} for the generating function $\rm{G}_{\mt{SC}}$ in the SYK model. However, an important subtlety arises: the equality \eqref{eq:tcrt} may not strictly hold for certain generating functions. That is, while taking $\alpha \to 0$ eliminates the linear ramp, higher-order corrections can still introduce a small ramp before the plateau. This effect is observed in the generating function $\langle e^{-\alpha \ell } \rangle$ due to the choice of the microcanonical ensemble. Specifically, the transition time from the decreasing region to the plateau cannot fully reach $T_{\mt{H}}$ due to the presence of a $\log(t)$ term in the classical contribution. Performing a detailed analysis in the $\alpha T_{\mt{H}} \to 0$ limit, we find that the generating function takes the form:
\begin{equation}
\lim_{\alpha T_{\mt{H}} \to 0} \langle e^{-\alpha \ell } \rangle  \approx  \text{Constant} + 2  \sqrt{E_0} \alpha \left(\frac{2 \log (2 \alpha  t)}{\pi  \Delta E}+\frac{(T_{\mt{H}}-t)^3}{3T_{\mt{H}}^2}\right) \,.
\end{equation}
Its time derivative is then given by
\begin{equation}
\lim_{\alpha T_{\mt{H}} \to 0}  \frac{d}{dt}\langle e^{-\alpha \ell } \rangle  \approx   2  \sqrt{E_0} \alpha \left(\frac{1}{\pi  \Delta E \, t}- \frac{(T_{\mt{H}}-t)^2}{T_{\mt{H}}^2}\right) \,,
\end{equation}
where the positive term originates from the increasing logarithmic term in $\langle e^{-\alpha \ell } \rangle_{\rm classical}$ \eqref{eq:expalclassical}. The transition time, at which the sign change occurs, is thus given by:
\begin{equation}
\lim_{\alpha \to 0} t_{\rm dip} (\alpha) \approx T_{\mt{H}} - \sqrt{\frac{2T_{\mt{H}}}{\pi \Delta E}}  + \mathcal{O}(1) < T_{\mt{H}} \,.
\end{equation}
This result indicates that a small residual ramp (of width approximately $\sqrt{T_\mt{H}} \sim e^{S_0/2}$) persists in the generating function at the $\alpha \to 0$ limit. From the complexity perspective, this remaining ramp corresponds to the {\it decreasing phase} of complexity, which extends from $t_{\rm crt} (\alpha=0)$ until the system reaches the plateau at the Heisenberg time $T_{\mt{H}}$, as illustrated in the right schematic plot of figure \ref{fig:secondlaw}. The peak appearing in the time evolution of complexity is thus equivalent to the dip of the generating function in the $\alpha \to 0$ limit:
\begin{equation}
\text{Peak time of complexity} = \lim_{\alpha \to 0} t_{\rm dip} (\alpha)  < T_{\mt{H}} \,. 
\end{equation}
For interested readers, we note that a similar analysis applies to the peak observed in the time evolution of Krylov complexity \cite{Balasubramanian:2022tpr,Erdmenger:2023wjg,Camargo:2023eev,Camargo:2024deu,Caputa:2024vrn,Balasubramanian:2024ghv,Baggioli:2024wbz,Bak:2025qgs}. Further investigation into the generating function of Krylov complexity would be an interesting direction for future research.

Another intriguing aspect is that the residual ramp in the generating functions of complexity implies a corresponding decreasing phase in complexity before it saturates. This behavior may suggest a potential violation of the second law of quantum complexity\footnote{The second law of complexity states that, for a generic closed chaotic quantum system, the complexity of its state will {\it most likely} increase over time, analogous to the second law of thermodynamics.} \cite{Brown:2016wib,Brown:2017jil}. Since the appearance of a peak depends on the choice of complexity measures and ensembles, it would be valuable to further investigate the conditions under which a peak appears—equivalently, the conditions under which the inequality $\lim_{\alpha \to 0} t_{\rm dip} (\alpha)  < T_{\mt{H}}$ holds.

\subsection*{Microcanonical Ensemble vs Canonical Ensemble}

In the main text, we focus on the microcanonical ensemble with a fixed narrow energy window $\left[ E_0 - \frac{\Delta E}{2}, E_0 + \frac{\Delta E}{2} \right]$. A natural question is how the corresponding results change for a canonical ensemble with a fixed inverse temperature $\beta$. The time evolution of the geodesic length and its generating function $\langle e^{-\alpha \ell}\rangle$ in the canonical ensemble has been studied in detail in \cite{Iliesiu:2021ari}. Based on the results obtained in this paper, the simplest generalization to the canonical ensemble can be achieved by taking the limit $2E_0 =\Delta E \to \infty$, which corresponds to the case $\beta =0$. However, we find that most conclusions remain similar, particularly those independent of the energy window $\Delta E$. 

Here, we highlight a significant difference arising from the classical contributions: the slope in a microcanonical ensemble decays more slowly than in a canonical ensemble. Using the spectral form factor $P^{\mt{TFD}}(\delta, t) = \text{SFF}(|t-\delta|)$ as an explicit example, the corresponding classical contributions are approximately given by
\begin{equation}\label{eq:micrvscan}
    P_{\rm{classical}}^{\mt{TFD}}(\delta,t)   \approx 
    \begin{cases}
    \dfrac{1}{|\delta -t|^2} \,, \qquad \text{microcanonical ensemble} \\
     \dfrac{1}{|\delta -t|^3} \,, \qquad \text{canonical ensemble} \\
    \end{cases}\,,
\end{equation}
where the microcanonical case is derived in eq.~\eqref{eq:PTFDclassical}, while the canonical result is well known for the GUE in random matrix theory and discussed in Appendix \ref{sec:canonical} for completeness. The key consequence of this difference is the distinct time scales for the dip: approximately $|\delta -t| \sim T_{\mt{H}}^{2/3} \sim e^{2S_0/3}$ in the microcanonical case and $|\delta -t| \sim T_{\mt{H}}^{1/2} \sim e^{S_0/2}$ in the canonical case. This effect is also reflected in the probability distribution $P(\ell, t)$ for fixed-length states, various generating functions, and expectation values. For example, the classical length expectation value $\langle \hat{\ell} \rangle$ in the microcanonical ensemble includes a logarithmic term $\log (t)$ (see eq.~\eqref{eq:applclassical}) because $P_{\rm classical}(\ell, t)$ decays as $\frac{1}{(t_\ell -t)^2}$. 

\begin{figure}[t]
	\centering
\includegraphics[width=3.5in]{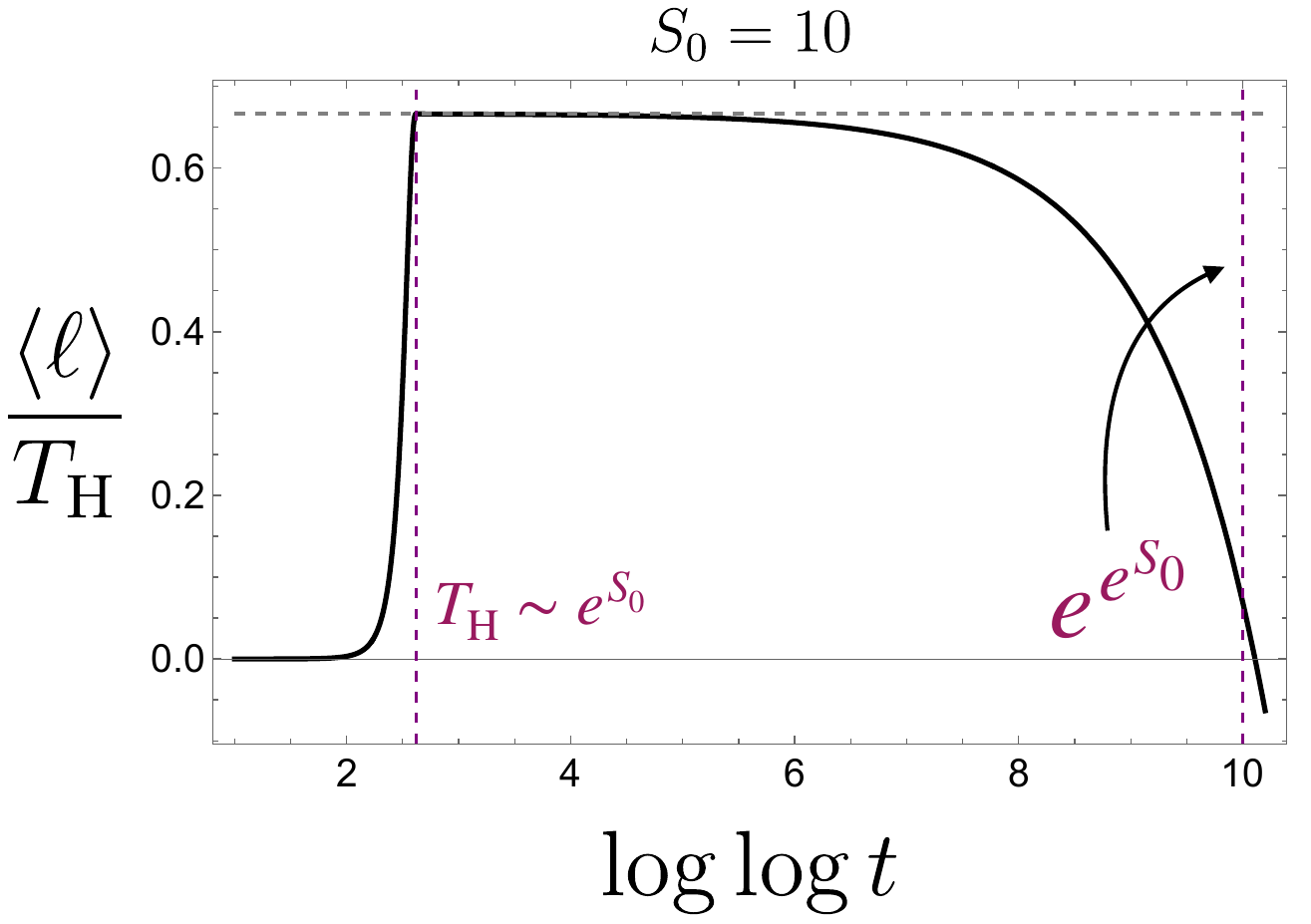}
	\caption{The time evolution of the regularized length expectation value $\langle \ell \rangle_{\rm reg}$, including the $\log (t)$ correction. After a long plateau regime, the length expectation value decreases as time evolves to the order of $e^{e^{S_0}}$.}
	\label{fig:expal03}
\end{figure}

A particularly noteworthy aspect of this logarithmic correction is that it becomes significant only near the recurrence time. As illustrated in figure \ref{fig:expal03}, the time evolution of the regularized expectation value of the length operator $\hat{\ell}$, which includes a subleading negative logarithmic term in eq.~\eqref{eq:Lvalues}, results in a decrease in geodesic length/complexity around $e^{e^{S_0}}$. A cautionary note is warranted here: this example, which is presented in a naive form, is only intended to emphasize the importance of logarithmic corrections. To rigorously derive the decrease in complexity near the recurrence time, a more detailed analysis incorporating all higher-order {\it quantum corrections} is required.

\subsection*{Complexity=Anything and Universal Time Evolution}

According to the complexity=volume proposal, the generating function of the geodesic length, denoted by $\langle e^{-\alpha \ell} \rangle$, is dual to the generating function of quantum complexity in the boundary dual theory. As demonstrated in this paper, there exist infinitely many analogous complexity measures, including the geodesic length (maximal volume in 2D gravity) $\langle \hat{\ell} \rangle$, the spectral complexity $\mathrm{SC}(t)$, and others. These infinite complexity measures neatly illustrate the complexity=anything proposal \cite{Belin:2021bga,Belin:2022xmt,Jorstad:2023kmq}. 

To briefly illustrate this idea, let us consider codimension-one holographic complexity measures as an example. Such codimension-one gravitational observables can be interpreted as generalized volumes, denoted as $V_{\rm gen}$. Analogous to the geodesic length operator defined in eq.~\eqref{eq:loperator}, we can formally define the operator associated with the generalized volume as
\begin{equation}
 \hat{V}_{\rm gen} := \int V_{\rm gen} | V_{\rm gen} \rangle \langle V_{\rm gen} | \, dV_{\rm gen} \,,
\end{equation}
where $| V_{\rm gen} \rangle$ represents states with a fixed generalized volume $V_{\rm gen}$. The information about fixed-volume states is encoded in the wavefunction $\phi_{E}(V_{\rm gen})$ in the eigenenergy basis, \ie 
\begin{equation}\label{eq:Vgenstate}
|V_{\rm gen} \rangle = e^{S_0} \int dE \, D(E) \phi_{E}(V_{\rm gen}) |E\rangle \,.   
\end{equation}
Furthermore, we can define its generating function as
\begin{equation}
 \langle e^{-\alpha V_{\rm gen} } \rangle := \frac{e^{2S_0}}{Z} \int dE_1\int dE_2 \, e^{-iE_{12}t} \langle  D(E_1) D(E_2) \rangle \int_{-\infty}^{\infty} dV_{\rm gen}  \, \phi_{E_1}(V_{\rm gen})\phi^\ast_{E_2}(V_{\rm gen})   e^{-\alpha V_{\rm gen} } \,.
\end{equation}
The universal time evolution of codimension-one holographic complexity is directly determined by the universal pole structure of the corresponding generating function, namely
\begin{equation}
\langle E_1 | \widehat{e^{-\alpha V_{\rm gen} }}  | E_2 \rangle \propto \int_{-\infty}^{\infty} dV_{\rm gen}  \, \phi_{E_1}(V_{\rm gen})\phi^\ast_{E_2}(V_{\rm gen})   e^{-\alpha V_{\rm gen} }  \sim \frac{1}{(\tilde{\alpha} + i E_{12} )(\tilde{\alpha} - i E_{12})} \,.
\end{equation}

From this construction, it is evident that the generalized volume and its generating function serve as spectral probes, similar to the spectral form factor. However, it is important to note that these complexity measures, such as the generalized volume, extract only a limited amount of information about the spectrum. This limitation arises because they only depend on two-point spectral correlators. Nevertheless, it is possible to extend this framework to include codimension-zero holographic complexity measures, such as the action. A similar analysis can be carried out by incorporating three-point spectral correlators. These extensions will be discussed in further detail in the forthcoming companion paper \cite{toappear}.  

\subsection*{More Generalizations and Questions}

Finally, we turn to possible generalizations of our work for future exploration. As mentioned above, it is interesting to explore the generating function of Krylov complexity to better understand its time evolution, which also exhibits the inverted slope-dip-ramp-plateau structure. See \cite{Jian:2020qpp} for explicit calculations on the generating function of Krylov complexity in the SYK model in the large-$q$ limit. In the context of double-scaled SYK (DSSYK), the generating function of the Krylov complexity has also been explicitly calculated in \cite{Xu:2024gfm} and shown to be given by the $6j$ symbol of the quantum group $U_q(\mathfrak{su}(1,1))$. It would be interesting to study the (finite-$N$) quantum corrections of the generating functions in these models. In particular, the Krylov complexity has been shown to be related to the geodesic length in the DSSYK model \cite{Lin:2022rbf,Rabinovici:2023yex,Xu:2024gfm,Heller:2024ldz}. This connection raises an interesting question: how do the quantum complexity generating functions differ between the SYK and DSSYK models? Investigating this difference could provide a geometric perspective to distinguish the holographic dual spacetimes of these models. 
Moreover, in the context of a specified complexity measure, such as Krylov complexity, and their generating functions, it is also intriguing to explore the inverse problem: namely, what is the holographic dual of a given complexity measure? This can be regarded as constructing a holographic dictionary between complexity=anything in bulk spacetime and quantum complexity measures on the boundary.

Furthermore, the study of generating functions can be extended to more general quantum many-body systems, allowing us to explore properties analogous to ``wormhole lengths'' and ``age of black holes'' in non-holographic settings. This broader application can provide new insights into the universality of complexity growth and saturation across different physical systems.

Finally, similar calculations can be made using different ensembles in random matrix theory. Given the known differences in the spectral form factor of different RMT ensembles, such as the three Dyson ensembles (GOE, GUE and GSE) or Altland-Zirnbauer ensembles \cite{Stanford:2019vob}, it is reasonable to expect that the time evolution of complexity measures and their generating functions will exhibit distinct behaviors in other ensembles. Understanding these differences could further refine our understanding of complexity growth and its dependence on spectral correlations.


\acknowledgments
We are grateful to Vijay Balasubramanian, Johanna Erdmenger, Jonathan Karl, Taishi Kawamoto, Soichiro Mori, Dominik Neuenfeld, Kazumi Okuyama, Le-Chen Qu, Andrew Rolph, Masaki Shigemori, Tadashi Takayanagi, Weyne Weng, Jiuci Xu and Zhuo-Yu Xian for discussions and comments. MM is supported by JSPS KAKENHI Grant-in-Aid for Early-Career Scientists (24K17044). KY is supported by JST SPRING, grant number JPMJSP2125, "THERS Make New Standards Program for the Next Generation Researchers." Work at VUB was supported by FWO-
Vlaanderen project G012222N and by the VUB Research Council through the Strategic Research Program High-Energy Physics. This research was supported in part by the International Centre for Theoretical Sciences (ICTS) for the program - Quantum Information, Quantum Field Theory and Gravity (code: ICTS/qftg2024/08).


\appendix

\section{Approximate wavefunction $\psi_E(\ell)$ of fixed-length states}\label{sec:app}
The wavefunction of the fixed-length state, \ie the overlap with the TFD state is derived as 
\begin{equation}
  \psi_{E}(\ell)  = \langle E|\ell\rangle = \sqrt{8 e^{-S_0}} K_{i2\sqrt{E}}(2e^{-\ell/2})\,. 
\end{equation}
The goal of this appendix section is to derive the approximate wavefunction for explicit calculations as shown in the main text. The key ingredient is modified Bessel function of the second kind with a purely imaginary order \cite{dunster1990bessel}, denoted as $K_{i \tau}(z)$ whose integral representation is defined as 
\begin{equation}\label{eq:integral}
\begin{split}
K_{i \tau}(z) &= \int_0^{\infty} \mathrm{e}^{-z \cosh t} \cos( \tau t) \, \mathrm{d} t = \frac{1}{2}\int_{-\infty}^{\infty} \mathrm{e}^{-z \cosh t + i \tau t}  \, \mathrm{d} t \,.\\
\end{split}
\end{equation}
We will only consider real and positive parameters $z, \tau$ in the following due to our interest, \ie 
\begin{equation}
 z= 2 e^{-\ell/2}, \qquad  \tau= 2\sqrt{E} \,.
\end{equation}
One useful trick associated with modified Bessel functions is the so-called Kontorovich–Lebedev transform. The transform and its inversion formula are given by 
\begin{equation}\label{eq:KL01}
f(\tau) = \frac{2 \tau}{\pi^2} \sinh (\pi \tau) \int_0^\infty  \frac{g(z)}{z} K_{i \tau} (z)\, dz\,, 
\end{equation}
and 
\begin{equation}\label{eq:KL02}
g(z) = \int^\infty_0 f(\tau) K_{i \tau}(z) \, d\tau \,.
\end{equation}
These can help us to evaluate many integrals in closed form, \eg 
\begin{equation}
\int^\infty_0 \frac{z^\alpha}{z} K_{i\tau} (z)  \, dz  =  2^{\alpha-2} \Gamma \left(\frac{1}{2}(\alpha - i \tau)\right)\Gamma \left(\frac{1}{2}(\alpha + i \tau)\right) \,,
\end{equation}
and also eq.~\eqref{eq:KLtransformation}
In the following of this Appendix, we focus on deriving the approximate expressions of modified Bessel function by using the method of steepest descent and infinite series expansion, respectively. 

\subsection{Method of steepest descent}
We begin by deforming the $t$-integral and then apply the saddle-point method (method of steepest descent) to obtain the approximate Bessel function and the disk wavefunction. 
Obviously, there are infinite saddle points on the complex $t$-plane. First of all, let us consider the parameter region with 
 $\frac{\tau}{z} > 1$. The infinite saddle points reads
 \begin{equation}
    t_{\ast} = i \pi  \left(\frac{1 }{2} +  2 N \right) \pm \log \left(\sqrt{\frac{\tau ^2}{z^2}-1}+\frac{\tau }{z}\right) \,, \qquad N \in \mathbb{N} \,,
 \end{equation}
which solve $\partial_t (-z \cosh t + i \tau t)=0$. 
It is noteworthy that the dominant contributions to the integral defined in eq.~\eqref{eq:integral} are given by the first two distinct addle points, \ie 
\begin{equation}
     t_\pm \equiv \frac{i \pi }{2} \pm \log \left(\sqrt{\frac{\tau ^2}{z^2}-1}+\frac{\tau }{z}\right) \,,
\end{equation}
Noting 
\begin{equation}
\arg \{ - z\cosh t  \} \big|_{t=t_+} = -\frac{\pi}{2} \,,\qquad  \arg \{ - z\cosh t  \} \big|_{t=t_-} = +\frac{\pi}{2} \,, 
\end{equation}
we can determine the direction of the steepest descent as $\alpha_+ = \frac{3\pi}{4}$ and $\alpha_- = \frac{\pi}{4}$.
The saddle point approximation produces 
\begin{equation}\label{eq:appK}
\begin{split}
    K_{i \tau}(z) &\approx  \left( -e^{-i \frac{3\pi}{4}} \sqrt{\frac{2\pi}{|- z\cosh t_+ |} }\mathrm{e}^{-z \cosh t_+ + i \tau t_+} + e^{i \frac{\pi}{4}} \sqrt{\frac{2\pi}{|- z\cosh t_- |} }\mathrm{e}^{-z \cosh t_- + i \tau t_-}  \right)  \\
    &=\frac{\sqrt{2 \pi } e^{-\frac{1}{2} (\pi  \tau )} }{(\tau ^2-z^2)^{1/4}} \cos \left(-\sqrt{\tau ^2-z^2}+\tau  \log \left(\frac{\tau +\sqrt{\tau ^2-z^2}}{z}\right)-\frac{\pi }{4}\right) \,. 
\end{split} 
\end{equation}
In the closed integral contour, there are still infinite saddle points $t_\ast$ with $N\ge 1$. However, they are subleading corrections since they are suppressed by $e^{i \tau t } \sim e^{- 2 \pi N \tau  } $. 
Substituting the approximate Bessel function \eqref{eq:appK} to the wavefunction of the fixed-length state yields
\begin{equation}
 \psi_{E}(\ell)  \approx  \frac{\sqrt{8 \pi e^{-S_0} } e^{-\pi  \sqrt{E}} \sin \left(-2 e^{-\frac{\ell}{2}} \sqrt{E e^\ell-1}+2 \sqrt{E} \log \left(\sqrt{E e^\ell-1}+\sqrt{E e^\ell}\right)+\frac{\pi }{4}\right)}{e^{-\frac{\ell}{4}} (E e^\ell-1)^{\frac{1}{4}}}\,,   \\
\end{equation}
which is a valid approximation for $E e^\ell >1$. In the regime with $E e^\ell \gg 1$, we can further simply the above result and obtain eq.~\eqref{eq:psiDiskapp}. 

On the other hand, we also need to consider the case with $z> \tau >0$ in order to consider the negative renormalized length $\ell$. Correspondingly, the dominated saddle point is given by 
\begin{equation}
t_0 = i \arcsin \left( \frac{\tau}{z} \right)  \in i\left[0, \frac{\pi}{2}\right]\,. 
\end{equation}
With deforming the integration contour by using the path of steepest descent through this saddle point, we can get the asymptotic dependence of the modified Bessel function as 
\begin{equation}
 K_{i \tau}(z)  \approx \sqrt{\frac{\pi }{2}}\frac{ e^{-\sqrt{z^2-\tau ^2}-\tau  \arcsin\left(\frac{\tau }{z}\right)}}{(z^2-\tau ^2)^{1/4}} \,,
\end{equation}
which is valid in the regime with $z > \tau >0$. This approximation gives rise to a distinct approximate wavefunction
\begin{equation}
 \psi_{E}(\ell)  \approx \frac{\sqrt{2 \pi e^{-S_0}} e^{-2 \left(\sqrt{e^{-\ell}-E}+\sqrt{E} \arcsin \left(\sqrt{E e^{\ell}} \right)\right)}}{(e^{-\ell}-E)^{1/4}}\,, \\
\end{equation}
which is valid when $E e^\ell <1$. However, it is evident that it is doubly exponentially suppressed for $\ell \ll 0$. Consequently, contributions from fixed-length states with negative renormalized length can be ignored. This is a natural expectation, given that the TFD state is evolved along increasing the boundary time, resulting in an increase in the classical geodesic length with time. 

\subsection{Series expansions for modified Bessel functions}
Since the modified Bessel function $K_\nu(z)$ have two independent variables as that of the wavefunction $\psi_E (\ell)$. It is much easier to treat its argument and order parameter individually by using the series expansion. The power series representation of the modified Bessel function is known as 
\begin{equation}
K_\nu(z)=\frac{1}{2}\left(\Gamma(\nu)\left(\frac{z}{2}\right)^{-\nu}\left(\sum_{k=0}^{\infty} \frac{\left(\frac{z}{2}\right)^{2 k}}{(1-\nu)_k k!}\right)+\Gamma(-\nu)\left(\frac{z}{2}\right)^\nu\left(\sum_{k=0}^{\infty} \frac{\left(\frac{z}{2}\right)^{2 k}}{(\nu+1)_k k!}\right)\right)  \,, \quad \nu \notin \mathbb{Z}
\end{equation}
or 
\begin{equation}
\begin{split}
K_v(z) &=\frac{1}{2}\Gamma(v)\left( \frac{z}{2}\right)^{-v}\left(1+\frac{z^2}{4(1-v)}+\frac{z^4}{32(1-v)(2-v)}+O\left(z^6\right)\right)  \\
&+\frac{1}{2}\Gamma(-v)\left(\frac{z}{2}\right)^v\left(1+\frac{z^2}{4(v+1)}+\frac{z^4}{32(v+1)(v+2)}+O\left(z^6\right)\right) \,, \\
\end{split}
\end{equation}
which is valid when $\nu$ is not an integer \footnote{ In the case of $\nu$ being an integer, the sum expansion is different.} From the above series expansion, we can find that the subleading terms can be ignored if 
\begin{equation}
 \frac{z^2}{v} \sim \frac{1}{\sqrt{E}e^\ell} \ll 1 \,.
\end{equation}
For example, we can find that the subleading term after the approximation \eqref{eq:appPsi} is given by  
\begin{equation}
\begin{split}
    \psi_{E}(\ell) - \psi^{(0)} _{E}(\ell)  &\approx \psi_{E}^{(1)}(\ell) \\
    &= -\sqrt{2}e^{-S_0/2} e^{-\ell} \left(e^{i \sqrt{E} \ell} \Gamma \left(2 i \sqrt{E}-1\right)+e^{-i \sqrt{E} \ell} \Gamma \left(-2 i \sqrt{E}-1\right)\right) \\ 
\end{split}
\end{equation}
Using the simpler inequality $ -2 \sqrt{(a+i b)(a-ib)} \le (a+i b) + (a- i b) \le 2 \sqrt{(a+i b)(a-ib)}$, we can find that the subleading term is bounded by 
\begin{equation}
 - \frac{2 \sqrt{\pi} e^{-\ell}e^{-S_0/2}}{{E}^{1/4} \sqrt{(4 E+1) \sinh \left(2 \pi  \sqrt{E}\right)}}  \le  \psi_{E}^{(1)}(\ell)  \le     \frac{2 \sqrt{\pi} e^{-\ell}e^{-S_0/2}}{{E}^{1/4} \sqrt{(4 E+1) \sinh \left(2 \pi  \sqrt{E}\right)}}  \,.
\end{equation}
In the neighborhood of the singularity $z \sim 0$, its asymptotic behavior is thus dominated by   
\begin{equation}\label{eq:zto0}
\begin{split}
z \to 0 : \qquad   K_{v}(z) &\approx  \frac{1}{2} \left(   \Gamma(v)\left(\frac{z}{2}\right)^{-v}+ \Gamma(-v)\left(\frac{z}{2}\right)^v  \right) \,.\\ 
\end{split}
\end{equation}

On the other hand, The power series representation of the modified Bessel functions of a purely imaginary order $K_{i \tau }(z)$ is expressed as \cite{dunster1990bessel}
\begin{equation}
\begin{split}
K_{i \tau}(z) &=-\left(\frac{\tau \pi}{\sinh (\tau \pi)}\right)^{1 / 2} \sum_{s=0}^{\infty} \frac{\left(z^2 / 4\right)^s \sin \left(\nu \ln (z / 2)-\phi_{\tau, s}\right)}{s!\left[\left(\tau^2\right)\left(1^2+\tau^2\right) \cdots\left(s^2+\tau^2\right)\right]^{1 / 2}} \,,
\end{split}
\end{equation}
with 
\begin{equation}
\phi_{\tau, s}=\arg \{\Gamma(1+s+i \tau)\} .
\end{equation}
One can check that the leading term 
\begin{equation}
K_{i \tau}(z)=-\left(\frac{\pi}{\tau \sinh (\tau \pi)}\right)^{1 / 2}\left(\sin \left(\tau \ln (z / 2)-\phi_{\tau, 0}\right)+O\left(z^2\right)\right)\,,
\end{equation}
is the same as eq.~\eqref{eq:zto0} after the analytical continuation with substituting $v \to i \tau$. More explicitly, we can use 
\begin{equation}
 \phi_{\tau, 0}=\arg \{\Gamma(1+i \tau)\}  =   -i \log \left(\sqrt{\frac{\sinh (\pi  \tau )}{\pi  \tau }} \Gamma (i \tau +1)\right) \,, 
\end{equation}
and find  
\begin{equation}\label{eq:Kzero}
\begin{split}
K^{(0)}_{i \tau}(z)&=\left(\frac{\pi}{\tau \sinh (\tau \pi)}\right)^{1 / 2}\sin \left(\tau \ln \left( \frac{2}{z}\right)+ \arg \{\Gamma(1+i \tau)\} \right) \\
&= \frac{1}{2} \left(\Gamma (-i \tau ) \left(\frac{z}{2}\right)^{i \tau }+\Gamma (i \tau ) \left(\frac{2}{z}\right)^{i \tau }\right) \,.
\end{split}
\end{equation}
To recover the leading result \eqref{eq:psiDiskapp} derived from the method of steepest descent, we need to perform the second approximation: a large order $\tau$ expansion for the leading term $K^{(0)}_{i \tau}(z)$. Using the recursion relation: 
\begin{equation}
 \Gamma(1+i \tau) = i \tau \Gamma(i \tau)  \approx  \sqrt{2 \pi } e^{-\frac{i}{12 \tau }-i \tau } (i \tau )^{\frac{1}{2}+i \tau } \,,
\end{equation}
we can approximate the argument function by 
\begin{equation}
  \arg \{\Gamma(1+i \tau)\}  \approx  -\tau -\frac{1}{12 \tau }+\tau  \log (\tau )+\frac{\pi }{4} \,.
\end{equation}
As a result, the large $\tau$ expansion of $K^{(0)}_{i \tau}(z)$ is given by 
\begin{equation}\label{eq:appKtauz}
\begin{split}
K^{(0)}_{i \tau}(z) &\approx 
\sqrt{\frac{2 \pi}{\tau}} e^{-\frac{\pi \tau}{2}} \sin \left(-\tau+\tau \log \left(\frac{2 \tau}{z}\right)+\frac{\pi}{4}\right)\left(1+O\left(\frac{1}{\tau}\right)\right)\,, \\
\end{split}
\end{equation}
which corresponds to the approximate wavefunction shown in eq.~\eqref{eq:psiDiskapp}.

To be more complete, we finally note that the individual asymptotic expansions for either large order (large $E$) or small argument (large $\ell$) is slightly different from eq.\eqref{eq:appKtauz}. Applying the approximation in the limit $v \to i\infty$, one can arrive at 
\begin{equation}\label{eq:appPsi01}
 E \gg 1 : \quad \psi_{E}(\ell) \approx  \frac{\sqrt{8 \pi }e^{-S_0/2} e^{-\pi \sqrt{E}} \sin \left(\frac{e^{-\ell}}{2 \sqrt{E}}+\sqrt{E} (\log (4 E)+ \ell+2)+\frac{\pi }{4}\right)}{E^{\frac{1}{4}}} \,.
\end{equation}
which is similar to the expression as eq.~\eqref{eq:psiDiskapp} by ignoring the terms at the order of $\mathcal{O}(z^2)$. On the other hand, the wavefunction in the large geodesic distance limit is reduced to 
\begin{equation}\label{eq:appPsi}
 e^\ell \gg 1 : \quad \psi_{E}(\ell) \approx \sqrt{2} e^{-S_0/2}\left(\Gamma \left(2 i \sqrt{E}\right) e^{i \sqrt{E} \ell}+\Gamma \left(-2 i \sqrt{E}\right) e^{-i \sqrt{E} \ell}  \right)  \,.
\end{equation}
The corrections resulting from the small $z$ (large $\ell$) approximation are also suppressed by the powers of $E$.

\section{Canonical Ensemble}\label{sec:canonical}
Considering the energy eigenstates of the Hamiltonian, we can define the inner products and completeness relation as 
\begin{equation}
 \langle E | E' \rangle = \frac{\delta(E- E')}{e^{S_0} D_{\rm{Disk}}(E) } \,, \qquad e^{S_0} \int_{0}^{\infty}dE~  D(E) \, |E\rangle\langle E|= \hat{\mathbbm{1}} \,.  
\end{equation}
The fixed length state is thus given by 
\begin{equation}
| \ell_{\rm can} \rangle  : = e^{S_0}\int_{0}^{\infty}dE~  D(E) \psi_{E}(\ell) \, |E\rangle \,,
\end{equation}
where $D(E)$ denotes the density of eigenstates and the wavefunction $\psi_{E}(\ell)$ is given by 
\begin{equation}
 \psi_{E}(\ell_{\rm can}) \equiv  \langle \ell_{\rm can} | E \rangle   =\sqrt{8e^{-S_0}}K_{i2\sqrt{E}}(2e^{-\ell_{\rm can}/2}) \,.
\end{equation}
The fixed-length state defined in the canonical ensemble satisfy the orthogonal condition 
\begin{equation}
\begin{split}
 \langle \ell_{\rm can} | \ell_{\rm can}' \rangle &=   \langle \ell_{\rm can} | \left( e^{S_0} \int_{0}^{\infty}dE~  D_{\rm{Disk}}(E) \, |E\rangle\langle E|   \right)    |\ell_{\rm can}' \rangle  \\
 &=  e^{S_0} \int_{0}^{\infty}dE~  D_{\rm{Disk}}(E) \psi_{E}(\ell_{\rm can}) \psi_{E}(\ell_{\rm can}') \\
 &=\delta (\ell_{\rm can}- \ell_{\rm can}') \,, 
\end{split}
\end{equation}
where the integral is explicitly given by the Kontorovich–Lebedev transform \eqref{eq:KLtransformation}. It thus leads to 
\begin{equation}
\int_{-\infty}^{\infty} \,  \langle  \ell_{\rm can}  | \ell_{\rm can}' \rangle  \, d \ell_{\rm can}  = 1 \,.
\end{equation}
Similarly, one can also derive the completeness relation, \ie 
\begin{equation}
\begin{split}
\left(  \int_{-\infty}^{\infty} d \ell_{\rm can}  \, |  \ell_{\rm can} \rangle   \langle \ell_{\rm can} |  \right)\bigg|_{\mathrm{classical}}=  e^{S_0} \int_{0}^{\infty}dE~  D_{\rm{Disk}}(E) \, |E\rangle\langle E|=\hat{\mathbbm{1}}  \,,
\end{split}
\end{equation}
by using the expansion of $| \ell_{\rm can} \rangle$ and the following explicit integral 
\begin{equation}
  \int_{-\infty}^{\infty} d \ell   \,  \psi_{\rm{Disk}}(E, \ell)  \psi_{\rm{Disk}}(E', \ell) = \frac{\delta(E- E')}{e^{S_0} D_{\rm{Disk}}(E) }  \,.
\end{equation}
The corresponding geodesic length operator reads 
\begin{equation}
\hat{\ell}_{\rm can}    = \int_{-\infty}^{\infty} \,
\ell_{\rm can}\, |\ell_{\rm can} \rangle\langle \ell_{\rm can} |  \, d \ell_{\rm can}  
\,,
\end{equation}
which satisfies the eigen equation: 
\begin{equation}
\hat{\ell}_{\rm can} |\ell_{\rm can}  \rangle = {\ell}_{\rm can} |\ell_{\rm can} \rangle  \,,  
\end{equation}
at the classical level. Including the quantum contributions after the ensemble average, most relations shown above does not hold. For example, the overlaps between two fixed-length states are derived as 
\begin{equation}
\begin{split}
 \langle \ell_{\rm can} | \ell_{\rm can}' \rangle &=   \langle \ell_{\rm can} | \left( e^{S_0} \int_{0}^{\infty}dE~  D_{\rm{Disk}}(E) \, |E\rangle\langle E|   \right)    |\ell_{\rm can}' \rangle  \\
 &=  e^{S_0} \int_{0}^{\infty}dE~ \psi_{E}(\ell_{\rm can}) \psi_{E}(\ell_{\rm can}') * \frac{\langle D(E)D(E) \rangle }{ D_{\rm{Disk}}(E) } \\
 &\ne \delta (\ell_{\rm can}- \ell_{\rm can}') \,. 
\end{split}
\end{equation}
Using the universal spectral two-point function again, the overlap is approximately given by 
\begin{equation}
\begin{split}
 \langle \ell_{\rm can} | \ell_{\rm can}' \rangle &=   
 e^{S_0} \int_{0}^{\infty}dE~ \psi_{E}(\ell_{\rm can}) \psi_{E}(\ell_{\rm can}') * \frac{ D_{\rm{Disk}}^2(E) + e^{-S_0}D_{\rm{Disk}}(E)\delta (E-E)  -  D_{\rm{Disk}}^2(E)}{ D_{\rm{Disk}}(E) } \\
 &=    \int_{0}^{\infty}dE~ \psi_{E}(\ell_{\rm can}) \psi_{E}(\ell_{\rm can}') \delta (0) \to \infty \,.
\end{split}
\end{equation}
This divergence also characterize the violation of the finiteness of the Hilbert space due to the infinite continuous spectrum.

The partition function or the normalization factor
in the canonical ensemble is defined by 
\begin{equation}
  Z (\beta+ i T) \equiv e^{S_0} \int^{\infty}_{0}  D_{\rm{Disk}}(E) e^{- (\beta + i T)E} dE \,.
\end{equation}
Performing the integral directly with using the density of states, one can get 
\begin{equation}
Z (\beta+ i T) Z (\beta- i T)  = e^{2S_0}\frac{e^{\frac{2 \pi ^2 \beta }{\beta ^2+T^2}}}{16 \pi  \left(\beta ^2+T^2\right)^{3/2}}\,,
\end{equation}
with resulting in 
\begin{equation}\label{eq:canZZ}
Z (0+ i T) Z (0- i T)  = \frac{e^{2S_0}}{16 \pi  }\frac{1}{T^3} \,.
\end{equation}
This indicates that the decaying rate of the probability $P^{\mt{TFD}}$ from the classical contribution, \ie the slope regime is described by 
\begin{equation}
P^{\mt{TFD}}_{\rm classical}(\delta, t) \approx  \frac{1}{|\delta -t|^3} \,,
\end{equation}
which decays faster than the one \eqref{eq:PTFDclassical} in the microcanonical ensemble.

The same difference also happens to the classical probability distribution $ P_{\rm classical}(\ell,t)$ of geodesic lengths in canonical ensemble. The definition of $P_{\rm classical}(\ell,t)$ in canonical ensemble is given by
\begin{equation}
    P_{\rm classical}(\ell,t)=\Bigg| \frac{e^{S_0}}{\sqrt{Z}}\int^\infty_0 dE\,e^{-\beta E-iEt}\psi_{E}(\ell)D_{\rm Disk}(E) \Bigg|^2 \,.
\end{equation}
The integral is localized at $E_1,E_2\lesssim1/\beta$. We can obtain analytical expression when $e^{\ell}\lesssim\beta$, in which another approximation \eqref{eq:psinegative} is valid.
Then approximating with small $E_1,E_2$ the integral is evaluated to be
\begin{equation}
    P_{\rm classical}(\ell,t)\approx\frac{4\sqrt{\pi}\beta^{\frac{3}{2}}e^{-\frac{\pi^2}{\beta}}e^{\frac{\ell}{2}-4e^{-\frac{\ell}{2}}}}{8\left(\left(\beta+2e^{\frac{\ell}{2}}\right)^2+4t^2\right)^\frac{3}{2}} \, .
\end{equation}
At large times with $t \gg \beta$, it reduces to
\begin{equation}
     P_{\rm classical}(\ell,t) \propto\frac{1}{t^3} \,, 
\end{equation}
resembling the behavior of SFF in canonical ensemble.



\bibliographystyle{jhep}
\bibliography{references}
\end{document}